\documentclass[a4paper, 11pt, oneside]{other/ThesisCustom}  
\graphicspath{Figures/}  
\usepackage[11pt]{moresize}
\usepackage[square, numbers, comma, sort&compress,round]{natbib}  
\usepackage{url}
\usepackage{verbatim}  
\usepackage{other/vector}  
\usepackage{adjustbox,fancybox,array,caption,pdfpages,appendix,ragged2e}
\usepackage{multirow,setspace}
\usepackage{tabularx}
\usepackage{geometry}
\hypersetup{urlcolor=blue, colorlinks=true}  

\usepackage{multibib}
\newcites{publi}{Publications related to this thesis}
\newcites{paper}{Other publications not related to this thesis}

\usepackage{standalone}
\usepackage[symbol]{footmisc}

\newcommand\bea{\begin{eqnarray}} 
\newcommand\eea{\end{eqnarray}} 
\newcommand\ee{\end{equation}}
\newcommand\be{\begin{equation}}

\newcommand\eeq{\end{equation}}
\newcommand\beq{\begin{equation}}

\def\ba{\begin{array}}
\def\ea{\end{array}}

\def\aa{a_2}
\def\ab{a_1}
\def\ac{a_3}
\def\ad{a_4}
\def\ae{a_5}

\newcommand{\DDt}{\alpha}
\newcommand\bbi{b_{b,\text{in}}}
\newcommand\bci{b_{c,\text{in}}}

\newcommand\GG{G} 

\def\d{\delta}

\def\f{f}
\def\D{D}

\def\tA{\hat{A}}

\newcommand\gammac{\gamma_c}
\newcommand\tbeta{\beta}

\newcommand\wDE{w}

\newcommand\alphaBz{\alpha_{\text{B},0}}
\newcommand\alphaMz{\alpha_{\text{M},0}}

\newcommand\alphaTz{\alpha_{\text{T},0}}

\newcommand\alphaDI{\alpha_{\text{D},I}}
\newcommand\alphaCI{\alpha_{\text{C},I}}
\newcommand\alphaXI{\alpha_{\text{X},I}}
\newcommand\alphaYI{\alpha_{\text{Y},I}}

\newcommand{\fg}{f_{\rm eff}}

\newcommand\alphaSm{\alpha^{\rm eff}_{\text{D,m}}}
\newcommand\alphaDm{\alpha_{\text{D,m}}}
\newcommand\alphaCm{\alpha_{\text{C,m}}}
\newcommand\alphaXm{\alpha_{\text{X,m}}}

\newcommand\alphaYm{\alpha_{\text{Y,m}}}

\newcommand\alphaDc{\alpha_{\text{D},c}}
\newcommand\alphaCc{\alpha_{\text{C},c}}
\newcommand\alphaD{\alpha_{\text{D}}}
\newcommand\alphaC{\alpha_{\text{C}}}
\newcommand\alphaX{\alpha_{\text{X}}}
\newcommand\alphaY{\alpha_{\text{Y}}}
\newcommand\alphaB{\alpha_{\text{B}}}
\newcommand\alphaM{\alpha_{\text{M}}}
\newcommand\alphaK{\alpha_{\text{K}}}
\newcommand\alphaT{\alpha_{\text{T}}}
\newcommand\alphaH{\alpha_{\text{H}}}

\newcommand{\2}{{(2)}}

\newcommand{\aL}{\alpha_{\rm L}}
\newcommand{\aH}{\alpha_{\rm H}}
\newcommand{\aT}{\alpha_{\rm T}}
\newcommand{\bun}{\beta_1}
\newcommand{\bdeux}{\beta_2}
\newcommand{\btrois}{\beta_3}
\newcommand{\aq}{a}  
\newcommand{\A }{\As  } 
\newcommand{\Az}{A}
\newcommand{\Bz}{B}

\newcommand{\CI}{{\cal C}_{\rm I}}
\newcommand{\CII}{{\cal C}_{\rm II}}
\newcommand{\CU}{{\cal C}_{\rm U}}

\newcommand{\tgmn}{{\tilde{g}_{\mu\nu}}}
\newcommand{\gmn}{{g_{\mu\nu}}}
\newcommand{\pmphi}{{\partial_{\mu}\phi}}
\newcommand{\pnphi}{{\partial_{\nu}\phi}}
\newcommand{\tg}{\check{g}}

\newcommand{\NS}{N_{\rm S}}

\newcommand{\omskz}{{\cal E}_1}
\newcommand{\omfkt}{{\cal E}_2}
\newcommand{\omfkz}{{\cal E}_3}

\newcommand{\omtkf}{{\cal E}_4}
\newcommand{\omtkt}{{\cal E}_5}

\newcommand{\tzeta}{{\tilde \zeta}}

\newcommand{\As}{A}

\newcommand{\aone}{{c_{1,0}}}

\newcommand{\bone}{{c_{1,2}}}
\newcommand{\btwo}{{c_{2,2}}}

\newcommand{\ctwo}{{c_{2,4}}}

\newcommand{\dtwo}{{c_{2,6}}}

\newcommand{\MMM}{{\cal M}}
\newcommand{\SSS}{{\cal S}}

\newcommand{\MB}{M_{\rm B}}
\newcommand{\MsqK}{M^2_{\rm K}}

\newcommand{\phimu}{ \phi_{\mu}}
\newcommand{\phinu}{ \phi_{\nu}}

\usepackage{pifont}
\begin{document}

\frontmatter      
\setstretch{1.15}
\thispagestyle{empty}

\vspace*{-2.2cm}

\begin{figure}[htb!]
        \hspace{-2.2cm}
				\begin{minipage}[t]{2cm}
       				 \begin{flushleft}
					\includegraphics[height=3cm]{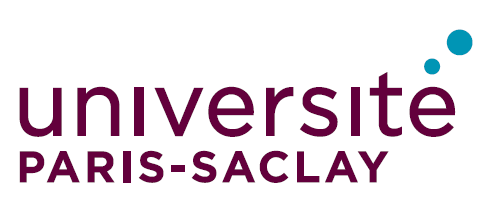}
				\end{flushleft}
				\end{minipage}
	\hspace{5.7cm}
				\begin{minipage}[t]{2cm}
       				 \begin{flushleft}
					\includegraphics[height=2cm]{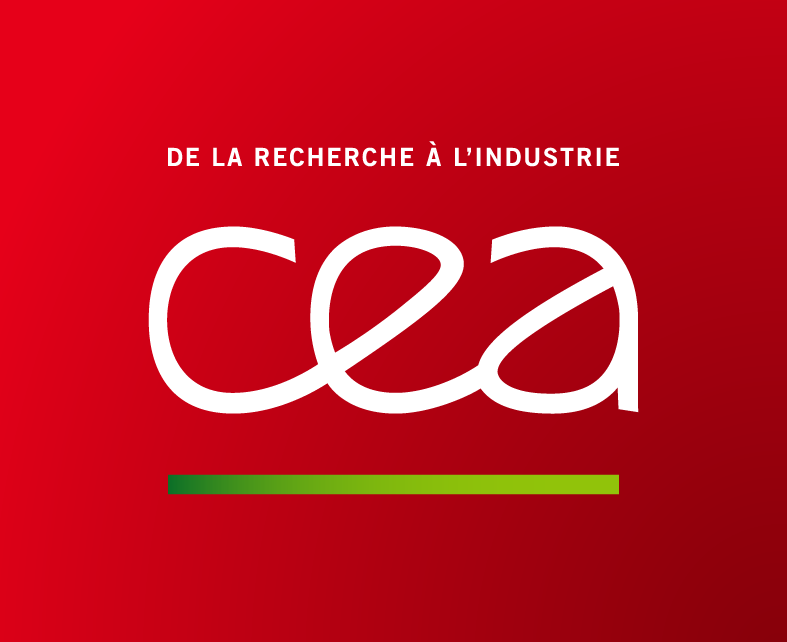}
				\end{flushleft}
				\end{minipage}
	\hspace{2.6cm}
				\begin{minipage}[t]{3cm}
				\begin{flushright}
					\includegraphics[height=3cm]{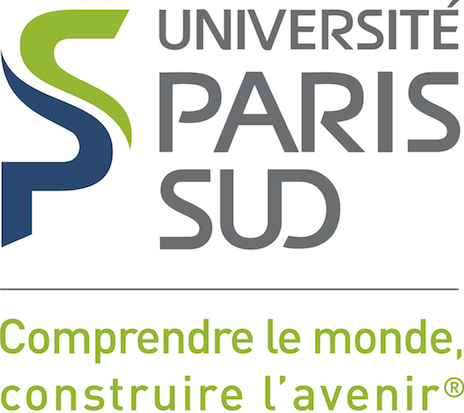} 
				\end{flushright}
				\end{minipage}
\end{figure}

NNT: 2017SACLS163

\begin{center} 
\begin{tabular}{c}
\huge \textsc{Th\`ese de doctorat}\\
    {\huge \textsc{de l'Universit\'e Paris-Saclay}}\\
     {\Large \textsc{Pr\'epare\'e \`a l'Universit\'e Paris Sud}}
		\\
		\\
    {\Large \textsc{Ecole Doctorale 564 - Physique en \^{I}le-de-France}} \\   
		{\large \textsc{Institut de Physique Th\'eorique du CEA Saclay}} \\
		\\
		{\Large \textsc{Discipline : Physique}} 
		\\
		{\large \textsc{Sp\'ecialit\'e : Cosmologie}}\\
    \\
    \\
    \large{Soutenue le 27/06/2017 \`a l'Institut de Physique Th\'eorique du CEA Saclay par} \\
    \\
    \\
    \Huge{\textbf{{Michele Mancarella}}}\\
    \\
		\\
		\\
\renewcommand*{\thefootnote}{\fnsymbol{footnote}}
    \huge\bf{An effective description of dark energy:}\\
		\huge\bf{from theory to phenomenology}\protect\footnotemark\\
    \\
    \\
		\\
		\\
\end{tabular}
\footnotetext[1]{Originally defended with the title ``Tests de coh\'erence de l'Univers et reliques cosmiques''}

\begin{tabular}{p{0cm} p{3.5cm} p{5.4cm} l }
	& \footnotesize\bf{Directeur de th\`ese} : & Filippo Vernizzi & \footnotesize{IPhT - CEA Saclay} \\
	& & &\\
	& \footnotesize\bf\underline{Composition du jury :}& &\\
	& \footnotesize{Pr\'esident du jury} : &Ruth Durrer & \footnotesize{Universit\'e de Gen\`eve}   \\
	& \footnotesize{Rapporteurs} :& Kazuya Koyama & \footnotesize{ICG, University of Portsmouth} \\
  &	& Federico Piazza & \footnotesize{CPT, Aix-Marseille Universit\'e} \\
  & \footnotesize{Examinateurs} : &Ruth Durrer & \footnotesize{Universit\'e de Gen\`eve} \\
  &	& Vanina Ruhlmann-Kleider  & \footnotesize{SPP/IRFU - CEA Saclay} \\
  &    &Philippe Brax  &\footnotesize{IPhT - CEA Saclay} 
\end{tabular}

\end{center}
\renewcommand*{\thefootnote}{\arabic{footnote}}


\setstretch{1.3}  

\fancyhead{}  
\rhead{\thepage}  
\lhead{}  

\clearpage  

\setstretch{1.3}  



\pagestyle{fancy}  

\lhead{\emph{Contents}}  
\tableofcontents  

\lhead{\emph{List of Figures}}  
\listoffigures  

\lhead{\emph{List of Tables}}  
\listoftables  

\setstretch{1.5}  
\clearpage  
\lhead{\emph{Abbreviations}}  
\listofsymbols{ll}  
{
\textbf{GR} & \textbf{G}eneral \textbf{R}elativity \\
\textbf{EOM} & \textbf{E}quation(s) \textbf{O}f \textbf{M}otion \\
\textbf{DOF} & \textbf{D}egree(s) \textbf{O}f \textbf{F}reedom \\
\textbf{FLRW}& \textbf{F}riedmann-\textbf{L}ema\^itre-\textbf{R}oberston-\textbf{W}alker \\
\textbf{ADM}&\textbf{A}rnowitt-\textbf{D}eser-\textbf{M}isner \\
\textbf{WEP} & \textbf{W}eak \textbf{E}quivalence \textbf{P}rinciple\\
\textbf{DHOST} & \textbf{D}egenerate \textbf{H}igher-\textbf{O}rder \textbf{S}calar-\textbf{T}ensor\\
\textbf{EST} & \textbf{E}xtended  \textbf{S}calar-\textbf{T}ensor\\
\textbf{dS} & \textbf{d}e  \textbf{S}itter\\
\textbf{KMM} & \textbf{K}inetic  \textbf{M}atter \textbf{M}ixing

}



\clearpage  
\lhead{\emph{Notations}}  
\listofnomenclature{lll}  
{
$\Phi$ & $00$ part of the metric \\
$\Psi$ & trace of the spatial metric \\
$f_A\equiv \frac{\partial f}{\partial A}$ \\
$\phi_\mu \equiv \nabla_\mu \phi$,\, \\
$\phi_{\mu \nu} \equiv \nabla_\nu \nabla_\mu \phi$,\,  \ldots \\
$X\equiv \phi_\mu \phi^\mu$ \\
$R\equiv {}^{(3)}\!R$ & When not specified, $R$ denotes the three-dimensional Ricci scalar.\\
$\epsilon_{\mu \nu \rho \sigma }$ & Totally antisymmetric Levi-Civita tensor
}
\fancyhead{}  
\rhead{\thepage}  
\lhead{} 
\setstretch{1}  


\addtocontents{toc}{\vspace{2em}}  

\mainmatter	  

\pagestyle{fancy}  


\chapter*{Introduction}
\addcontentsline{toc}{chapter}{Introduction} 
\lhead{\emph{Introduction}}  

In the last decades, a cosmological model that fits observations through a vast range of scales emerged. It goes under the name of $\Lambda$CDM.
At the current state of the art, the six free parameters of $\Lambda$CDM are known~\cite{Ade:2015xua}, but there are challenging problems still open, in particular the cause of the observed accelerated expansion of the universe.
The simplest explanation is a cosmological constant $\Lambda$, but this is not technically natural in quantum field theory, from the point of view of the stability under radiative corrections. Many alternative models have been proposed, where the accelerated expansion is induced by a dynamical field or by a modification of General Relativity (GR). They are generally referred to as ``Dark Energy''.  Independently on any theoretical prejudice, the crucial point is that these alternatives to $\Lambda$CDM are \emph{testable} and it is thus worthwhile to study them.
A very promising way to perform these tests is to consider cosmological perturbations. 
In $\Lambda$CDM, the growth of perturbations is fixed by the value of the cosmological constant. Alternatives to it instead generally come with extra degrees of freedom that give different dynamics. Hence, here is where any deviation from $\Lambda$CDM can become manifest.
Crucially, our knowledge about the growth of inhomogeneities that generate the large scale structure we observe today can still be improved. This is the goal of a number of missions planned for the next decade (such as EUCLID~\cite{Laureijs:2011gra} and LSST~\cite{Abell:2009aa}), that will be able to push current constraints on the growth of structures down by one or two orders of magnitude. This will enable us to actually falsify many models and to shed light on the nature of the acceleration of the universe.

This effort in constraining deviations from $\Lambda$CDM should be supported by an appropriate theoretical insight. On the one hand, one should look for a simple, general and effective way to bridge theory and observations. On the other, we would like to keep control on the viability of the theory and on its agreement with basic principles of physics (such as causality and locality) when exploring the parameter space to fit data. These thesis presents an approach that relies on these two cornerstones. It is called Effective Theory of Dark Energy~\cite{Gubitosi:2012hu,Bloomfield:2012ff,Gleyzes:2013ooa,Bloomfield:2013efa,Piazza:2013coa,Tsujikawa:2014mba,Gleyzes:2014rba}, and amounts to the description of linear cosmological perturbations through all the operators compatible with symmetries in theories where a single scalar degree of freedom is added on top of the usual two helicity-two modes of GR, referred to as ``scalar-tensor theories''.

In Chapter~\ref{chap:DHOST} I will summarise what is currently the most general class of viable scalar-tensor theories. One of the sufficient conditions for a theory to be stable is that the equations of motion (EOM) contain at most two derivatives of the fields.
What has been for a long time considered the most general viable class of scalar-tensor theories, known as Horndeski theories~\cite{Horndeski:1974wa}, relies on this condition.
I will explain that however this is not a \emph{necessary} condition, and Horndeski theories can be extended to more general classes introducing the notion of \emph{degeneracy}, i.e. requiring the existence of constraints in the equations of motion. An example is the case of theories known as ``beyond-Horndeski''~\cite{Gleyzes:2014dya,Gleyzes:2014qga}, that I will discuss.
These paved the way for the discovery of larger classes of theories, known as Degenerate Higher Order Scalar-Tensor (DHOST)~\cite{Langlois:2015cwa}, or Extended Scalar-Tensor (EST)~\cite{Crisostomi:2016czh} theories. In the rest of the thesis, I will develop an effective description of linear cosmological perturbations in this class of theories.

In Chapter~\ref{chap:EFToDE} I will introduce the the Effective Theory of Dark Energy. I will first resume the formalism and show how to construct a very general action for linear cosmological perturbations. Although so far I only mentioned scalar-tensor theories, when dealing with the late universe the presence of matter is of course relevant. This opens the possibility to envisage different interactions between the matter fields and the other sectors. Part of my work has been devoted to include in the effective treatment very general couplings between matter, the metric and the scalar field, which I will illustrate in Chapter~\ref{chap:EFToDE_m}. These include also the possibility that different species couple differently to gravity.\\
A key advantage of the formulation through an action is that we can thoroughly analyse the stability conditions of the theory. I will discuss this aspect in detail in Chapter~\ref{chap:EFToDE_lin}, focussing on the dispersion relations for the propagating scalar modes, underlying the impact that the presence of matter can have and some subtleties arising in the case of DHOST theories. I will show that the formulation in terms of the effective description and the stability analysis allow to substantially reduce the very large number of independent DHOST theories to a few classes that are observationally viable, and that such classes are equivalent to Horndeski and ``beyond-Horndeski'' theories, up to non trivial couplings between matter and the gravitational sector.

Besides developing a solid theoretical understanding of the effective description, a goal of this thesis is also to start investigating the signatures of deviations from $\Lambda$CDM on observables. This is the subject of the last two Chapters.
In Chapter~\ref{chap:Pheno} I will consider the possibility of an interaction in the ``dark sector'', i.e. between dark matter and the dark energy. I will provide an analytical understanding of the effects and analyse the constraining power that future experiments will have on the free parameters of the effective description.
To solve the equations, I will resort to the so-called ``quasi-static limit'' valid for Fourier modes that are well inside the sound horizon of dark energy, These are the relevant modes for Large Scale Structure experiments. In Chapter~\ref{chap:KMM} I will go beyond this assumption and use a Boltzmann code which implements scalar-tensor theories in the effective theory formalism. I will focus on the peculiar observational effects that arise in theories ``beyond-Horndeski'', showing that in this case a frame-independent kinetic mixing between matter and the scalar field arises. I will illustrate and quantify its peculiar effect, namely the weakening of gravity at scales relevant for redshift surveys.

I chose not to include too lengthy calculations in the main text. I rather recall the formulae I need for the discussion, and what I think are the most interesting results, focusing on their physical meaning and their potential impact. The papers I published on the topics I present in the thesis contain the technical details and are included as appendices. I indicate in the text where to find the calculations the reader might be interested in.


\chapter{A general formulation of scalar-tensor theories}\label{chap:DHOST}
\lhead{\emph{A general formulation of scalar-tensor theories}}

\section{Scalar fields and higher derivatives in cosmology}\label{sec:ScalarInCosmo}
This thesis focuses on \emph{scalar-tensor} theories, i.e. theories where a single scalar degree of freedom is added to GR.
These represent the simplest way to try to explain the observed acceleration of the universe with a dynamical field. For this reason, they are widely studied in the context of late time cosmology.

In the last years, an intense activity has been devoted to find more and more general extensions of scalar-tensor theories. In particular, a way to go beyond the most studied ones is to allow derivatives higher than second in the Lagrangian. However, special care must be taken when considering this possibility.\\
Indeed, in presence of more than two derivatives, the fact that one introduces only one scalar field in the Lagrangian is not enough to ensure that the theory will contain a single propagating mode. One degree of freedom is characterised by two initial conditions needed to solve the equations of motion. On the other hand, in the presence of derivatives higher than second, extra initial conditions should be provided. This means that additional propagating degrees of freedom appear. Moreover, these modes are usually associated with instabilities in the system, and the presence of higher derivatives is severely restricted by a powerful theorem that dates back to the work by Ostrogradsky~\cite{Ostrogradsky:1850fid}. 
Due to this theorem, for a long time the requirement of having second order equations of motion was not questioned. Only recently it was realised that stability can be achieved even in theories with equations of motion of order higher than second by considering ``degenerate'' theories, i.e. theories where constraints among the canonical variables are present so that the number of propagating degrees of freedom is reduced and the dangerous modes eliminated.

In this work I will discuss the most general class of theories obtained in this way. This encompasses all the most studied theories such as quintessence~\cite{Ratra:1987rm,Wetterich:1987fm}, k-essence~\cite{ArmendarizPicon:1999rj,ArmendarizPicon:2000ah}, $f(R)$~\cite{Starobinsky:1980te,DeFelice:2010aj}, Horndeski~\cite{Horndeski:1974wa}-Generalised Galileons~\cite{Deffayet:2011gz} and their extensions known as ``beyond Horndeski''~\cite{Gleyzes:2014dya}.

\section{Horndeski theories}

To begin with, I recall the most general theories with second-order dynamics both for the scalar field and the metric, and generalise them later. The most general scalar-tensor theory that obeys this requirement dates back to Horndeski's work~\cite{Horndeski:1974wa}. The key idea is that one can admit higher derivatives in the Lagrangian, provided that its variation gives only second order EOM both for the scalar field and for the metric.
The most general Lagrangian satisfying the above property amounts to the four terms
\be \label{Hornd}
\begin{split}
L_2^{\rm H} & \equiv  G_2(\phi,X)\;, \qquad  
L_3^{\rm H}  \equiv  G_3(\phi, X) \, \Box \phi \;, \\
L_4^{\rm H} & \equiv G_4(\phi,X) \, {}^{(4)}\!R - 2 G_4{}_{,X}(\phi,X) (\Box \phi^2 - \phi^{ \mu \nu} \phi_{ \mu \nu})\;, \\
L_5^{\rm H} & \equiv G_5(\phi,X) \, {}^{(4)}\!G_{\mu \nu} \phi^{\mu \nu}  +  \frac13  G_5{}_{,X} (\phi,X) (\Box \phi^3 - 3 \, \Box \phi \, \phi_{\mu \nu}\phi^{\mu \nu} + 2 \, \phi_{\mu \nu}  \phi^{\mu \sigma} \phi^{\nu}_{\  \sigma})  \,.
\end{split}
\ee
In the above action, I introduced the notations
\be\label{notation}
\phi_{\mu } \equiv \nabla_\mu \phi \, , \quad X\equiv g^{\mu\nu} \phimu \phinu \, , \quad \phi_{\mu \nu} \equiv \nabla_\nu \nabla_\mu \phi \,, \quad \Box \phi \equiv g^{\mu\nu} \nabla_{\mu}\nabla_{\nu} \phi \, .
\ee
The idea at the base of the Horndeski Lagrangians is to find an \emph{antisymmetric} structure of the coefficients such that the terms with more than two derivatives cancel from the EOM.

This is clear if one considers first the case of flat space. This was studied in Ref.~\cite{Nicolis:2008in}. On Minkowski space, the only dynamical degree of freedom is the scalar. Its action is assumed to have the symmetry
\be
\phimu \mapsto \phimu+c_{\mu} \,, \qquad c_{\mu}=\text{const.} \, ,
\ee
which can be seen as a generalisation of the Galilean invariance. Hence the theory is called \emph{Galileon theory}. The most general theory with the above symmetry that gives second-order equations of motion has been shown to consist of the Lagrangians
\be
\begin{split}
L_2^{\rm gal}&=-\frac{1}{2}  (\partial \phi)^2 \, , \qquad
L_3^{\rm gal}=-\frac{1}{2}(\partial \phi)^2 \Box \phi, \\
L_4^{\rm gal} &=-\frac{1}{2} (\partial \phi)^2 \left[(\Box\phi)^2-(\partial_\mu\partial_\nu\phi)^2\right],  \\
L_5^{\rm gal} &= -\frac{1}{4}(\partial \phi)^2 
\bigl[(\Box\phi)^3 -3\Box\phi(\partial_\mu\partial_\nu\phi)^2
+2(\partial_\mu\partial_\nu\phi)^3 \bigr],
\label{galileon}
\end{split}
\ee
where I did not use the notation~\eqref{notation} to stress that in the above action a flat metric $\eta_{\mu\nu}$ is used, and all derivatives are partial derivatives. In particular, $\Box \phi = \eta^{\mu \nu}\partial_{\mu}\partial_{\nu} \phi$.
One can easily prove that the variation of each component contains only second derivatives.

The theory above is very instructive to identify the correct structure of the Lagrangian, and one can already notice that this structure reflects the one of the terms of the Horndeski Lagrangians. To go further, one can try to formulate the Galileon theory in a fully covariant form. If we ``covariantise'' the Galileon Lagrangians by replacing the partial derivatives with covariant derivatives associated to the metric $\gmn$, however, the EOM contain derivatives of the metric of order higher than second~\cite{Deffayet:2009mn,Deffayet:2009wt,Deffayet:2011gz}. To avoid this, it is necessary to introduce some ``gravitational counterterms'', or non-minimal gravitational couplings to $\phi$, in the covariant completion of $L_4$ and $L_5$. One can show that the couterterms are
\be
L_4 \in \frac{1}{8} X^2 \, {}^{(4)}\!R \, , \qquad L_5 \in -\frac{3}{8} X^2 \, {}^{(4)}\!G_{\mu \nu} \phi^{\mu \nu}  \, .
\ee
As a final generalisation, one can promote the coefficients $(\partial \phi)^2$ of the Galileon Lagrangians~\eqref{galileon} (which become $X$ in the covariant version) to functions of $\phi$ and $X$, provided that the tuning of the gravitational counterterms is preserved. The Lagrangian obtained this way is exactly the Horndeski Lagrangian~\eqref{Hornd}~\cite{Deffayet:2009mn}. In particular, the ``rediscovery'' of Horndeski's result that I just sketched gives an idea of why the structure of the quartic and quintic part should be as in~\eqref{Hornd}. It also explains the origin of the tuning between the coefficient of the Ricci scalar and the higher derivative terms for $\phi$ in $L_4^H$, and the same for $L_5^H$.
Any other Lagrangian satisfying the requirement of giving second-order EOM both for the metric and the scalar can be reduced to~\eqref{Hornd} by integration by parts.

\section{Theories ``beyond Horndeski''}\label{BHth}

The condition of second-order equations of motion is indeed a \emph{sufficient} condition, but one can ask whether it is \emph{necessary}. In other words, one can wonder whether higher order derivatives can be introduced in the equations of motion maintaining at the same time the correct number of initial conditions needed to solve the system. We can have an intuition of why this could be the case by considering an example where a healthy higher-order theory is built from a second-order one~\cite{Zumalacarregui:2013pma,Gleyzes:2014rba,Domenech:2015tca}.
Suppose we start with a simple theory belonging to the Horndeski class~\eqref{Hornd}, i.e. the Einstein-Hilbert action plus an action $S_{\phi}$ for the scalar field which I assume to yield second-order equations of motion. I denote with $\tgmn$ the metric that describes the gravitational and scalar field sector. I then add to that a matter Lagrangian constructed with a metric that depends also on the scalar field gradient, which I call $\gmn$:
\be
S=\frac{M_{\text{Pl}}^2 }2 \int d^4x \sqrt{-\tilde{g}}\,{}^{(4)}\!\tilde{R}+S_{\phi}\left[\tgmn, \phi \right]+S_{\rm m}\left[\gmn,\Psi\right]\,, \quad \gmn=C(X)\tgmn \, .\label{pureConf}
\ee
where $\Psi$ denotes the matter fields.
One can rewrite the action in terms of the metric $\gmn$ to which matter couples:
\be
\begin{split}
S=&\frac{M_{\text{Pl}}^2 }2\int d^4x \sqrt{-{g}}\, \left( C^{-1}{}^{(4)}\!{R}+\frac{3}{2C^3}  \partial_{\mu} C\, \partial^{\mu}C \right)+S_{\phi}\left[C^{-1}\gmn, \phi \right]+S_{\rm m}\left[\gmn,\Psi\right]\, \\
=&\frac{M_{\text{Pl}}^2 }2\int d^4x \sqrt{-{g}}\, \left( C^{-1}{}^{(4)}\!{R}+6\, C_X^2 \, \phi^{\mu} \phi_{\mu \rho} \phi^{\rho \nu} \phi_{\nu} \right)+S_{\phi}\left[C^{-1}\gmn, \phi \right]+S_{\rm m}\left[\gmn,\Psi\right] \, .
\end{split}
\ee
The above Lagrangian does not belong to the Horndeski class~\eqref{Hornd}. 
The EOM obtained by the variation of the action with respect to the scalar field contain derivatives of the latter up to fourth order~\cite{Zumalacarregui:2013pma,Bettoni:2015wla}. They read
\be \label{escf}
{\nabla}_{\mu} \left[ \phi^{\mu} \, C^{-3/2} C_{{X}} \left(C^{-1/2} {}^{(4)}\!{R} -6 {\Box} C^{-1/2}\right)\right]=\frac{\delta \mathcal{L}_{\phi}}{\delta \phi} \, .
\ee
However, it was shown in~\cite{Zumalacarregui:2013pma} that the higher derivative terms
can be eliminated by taking the trace of the metric equations:
\be
\left(C^{-1/2} {}^{(4)}\!{R} -6 {\Box} C^{-1/2}\right)\left(C^{-1/2}+{X} C^{-3/2} C_{{X}} \right)=\frac{2}{M_{\text{Pl}}^2} {T},
\ee
where $T={g}^{\mu \nu} \left({T}^{\phi}_{\mu\nu}+{T}^{\rm m}_{\mu\nu}\right)$. The above equation allows to eliminate the higher-derivative terms from~\eqref{escf} by introducing a ``mixing term'' with the energy-momentum tensor~\cite{Bettoni:2015wta}. Explicitly,~\eqref{escf} becomes
\be
{\nabla}_{\mu} \left( \phi^{\mu}\,  \mathcal{T}_K\right)=-\frac{1}{2}\frac{\delta \mathcal{L}_{\phi}}{\delta \phi} \, , \qquad \mathcal{T}_K\equiv -\frac{C^{-3/2} C_{{X}}}{M_{\text{Pl}}^2\left(C^{-1/2}+{X} C^{-3/2} C_{{X}}\right)} T
\ee
This way, it is evident that the scalar field equations actually require only two initial conditions to be solved (provided that $\delta \mathcal{L}/\delta \phi$ and ${\nabla}_{\mu} {T}$ do, which I assume is the case).
 This is an example of a theory that has only one propagating scalar degree of freedom even if formally it does not fit into the Horndeski Lagrangians~\eqref{Hornd}. The reason relies on the fact that a hidden constraint equation is present in the system due to the form of the transformation~\eqref{pureConf}. 
 
Along this line of reasoning, it was realised that the Horndeski class of theories~\eqref{Hornd} can be extended including the two Lagrangians~\cite{Gleyzes:2014dya,Gleyzes:2014qga}
\begin{align}
L_4^{\rm bH}& \equiv F_4(\phi,X) \epsilon^{\mu\nu\rho}_{\ \ \ \ \sigma}\, \epsilon^{\mu'\nu'\rho'\sigma}\phi_{\mu}\phi_{\mu'}\phi_{\nu\nu'}\phi_{\rho\rho'}\;, \label{L4bH} \\
L_5^{\rm bH}&\equiv F_5 (\phi,X) \epsilon^{\mu\nu\rho\sigma}\epsilon^{\mu'\nu'\rho'\sigma'}\phi_{\mu}\phi_{\mu'}\phi_{\nu \nu'}\phi_{\rho\rho'}\phi_{\sigma\sigma'} \label{L5bH}\,,
\end{align}
Remarkably, if we ``naively'' covariantise the quartic and quintic Galileon Lagrangians promoting partial derivatives to covariant derivatives, the resulting Lagrangians belong to the ``beyond Horndeski'' class, i.e. they are of the form~\eqref{L4bH}-\eqref{L5bH}. In curved space, the EOM for the metric contain third order derivatives of the scalar field while the EOM for the scalar field contain third order derivatives of the metric. In flat space, the scalar field dynamics reduces to second order. In curved space it can be shown that the number of propagating scalar degrees of freedom is also one.
To do so, an argument similar to the simple example I discussed above applies to the beyond-Horndeski theories. One can start from a theory containing only $L_4^{\rm bH}$ and find a transformation of the form
\be
\tgmn=\gmn+D(\phi,X)\phimu\phinu
\ee
such that in the final frame the system can be reduced to a second order one. The same can be done starting from $L_5^{\rm bH}$ alone, even though when both $L_4^{\rm bH}$ and $L_5^{\rm bH}$ are present such a transformation cannot be found. Nevertheless, the counting of the propagating degrees of freedom in these theories have been confirmed basing on a hamiltonian analysis.

In the rest of the thesis, I will refer to the theories introduced in this section as beyond Horndeski theories. This choice simply adapts to the name most used in the literature (other names are GLPV or $G^3$).

\section{Degenerate Higher-Order Scalar-Tensor theories}\label{sec:Ostro}

One can ask if theories even more general than the extensions presented above exist. The use of transformations to different frames to find ``hidden constraints'' suggests that one can allow for higher derivatives provided that there exists a way to reduce the number of initial conditions of the system to only two in the scalar sector. To do so in a systematic way, one should recall Ostrogradsky's theorem for higher-derivative Lagrangians.

\subsection{Ostrogradsky's theorem}\label{sec:OstroTh}

Ostrogradsky's result can be formulated in a very simple statement: the Hamiltonian constructed from non-degenerate Lagrangians that depend upon more than one time derivative 
necessarily develops an instability. Let me show this, and clarify the notion of degeneracy, considering a higher-order Lagrangian for the classical canonical variable $\phi(t)$~\cite{Woodard:2006nt,Woodard:2015zca},
\be\label{ostroL}
\mathcal{L}=\frac{a}{2} {\ddot{\phi}}^2+\frac{m}{2}{\dot{\phi}}^2 -\frac{m\omega^2}{2} \phi^2\; .
\ee
It is easy to show that the Hamiltonian is \emph{linear} in the conjugate momentum $P_{\phi}\equiv \partial \mathcal{L}/\partial \dot{\phi}$. We can just promote the ``velocity'' $\dot{\phi}$ to a new canonical variable $Q=\dot{\phi}$. With the aid of a Lagrange multiplier, we have the equivalent Lagrangian
\be
\mathcal{L}=\frac{1}{2}a {\dot{Q}}^2+\frac{1}{2}m{Q}^2-\frac{m\omega^2}{2} \phi^2-\lambda\left(Q-\dot{\phi}\right) \; .
\ee
The conjugate momenta are
\be \label{conMom}
P_{\phi}= \frac{\partial \mathcal{L}}{\partial \dot{\phi}}=\lambda \; , \qquad P=\frac{\partial \mathcal{L}}{\partial \dot{Q}}=a \dot{Q} \, ,
\ee
and, inverting the above relations, we can write the Hamiltonian
\be\label{HamOsc}
\mathcal{H}=P_{\phi} Q +\frac{1}{2 a} P^2-\frac{m}{2}  Q^2 +\frac{m\omega^2}{2} \phi^2\, .
\ee
Already at the classical level, this Hamiltonian reveals instabilities in the system.
Indeed, the presence of the term linear in $P_{\phi}$ makes it unbounded from below, so there exists an open direction in the phase space leading to states with negative energies. In the case of the system~\eqref{ostroL}, this can be seen explicitly. Following~\cite{Woodard:2015zca}, we can write the Euler-Lagrange equations obtained from~\ref{ostroL} and their solution in the form:
\be
\begin{split}
&\frac{a}{m}\ddddot{\phi}+\ddot{\phi}+\omega^2 \phi=0 \, , \\
&\phi(t)=C_+ \cos(k_+ t)+S_+ \sin (k_+ t )+C_- \cos(k_- t)+S_- \sin (k_- t ) \, , \\
\end{split}
\ee
where
\be
\begin{split}
&k_{\pm}=\omega\, {\left[\frac{1\pm\sqrt{1+4\epsilon}}{2\epsilon}\right]}^{1/2} \, , \qquad \epsilon\equiv\frac{ \omega^2 a}{m} \, , \\
& C_+ = \frac{k_-^2 \phi_0 \!+\! \ddot{\phi}_0}{k_-^2 \!-\! k_+^2} \qquad  ,  \qquad
S_+ = \frac{k_-^2 \dot{\phi}_0 \!+\! \dddot{\phi}_0}{k_+ (k_-^2 \!-\! k_+^2)} \; , \\
&C_- = \frac{k_+^2 \phi_0 \!+\! \ddot{\phi}_0}{k_+^2 \!-\! k_-^2} \qquad  ,  \qquad
S_- = \frac{k_+^2 \dot{\phi}_0 \!+\! \dddot{\phi}_0}{k_- (k_+^2 \!-\! k_-^2)} \; ,
\end{split}
\ee
and $\phi_0$, $\dot{\phi}_0$, $\ddot{\phi}_0$, $\dddot{\phi}_0$ are the initial data. One can recast the Hamiltonian~\eqref{HamOsc} in terms of the above constants, and gets
\be
\mathcal{H}= \frac{m}2 \sqrt{1 \!+\! 4 \epsilon} \, k_+^2 (C_+^2 \!+\! S_+^2) - 
\frac{m}2 \sqrt{1 \!+\! 4 \epsilon} \, k_-^2 (C_-^2 \!+\! S_-^2) \; . 
\ee
In this form, one can see explicitly that the mode $k_-$ has negative energy. 
Moreover, the fact that the positive and negative energy modes originate from the same higher derivative dynamical variable implies they are necessarily coupled.
The presence of such a mode allows states with arbitrarily high energies to be excited. To conserve the total energy, it is sufficient to excite other states with the same amplitude and opposite sign.
The propagating mode with negative energy that is present in the system is called the \emph{Ostrogradsky ghost}.

The only crucial assumption to arrive at these results is that we could invert the second equation~\eqref{conMom}, i.e. we were able to express the highest time derivatives in terms of canonical variables. This is the meaning of \emph{non-degeneracy}. In this case, this amounts to require that 
 \be\label{nonDeg}
 \frac{\partial \mathcal{L}}{\partial \ddot{\phi}}\neq 0 \; ,
 \ee
which is to say that the higher-order term cannot be eliminated through integration by parts.
For the case of a single variable, it is clear that the above argument is actually a no-go theorem for higher derivatives, since assuming a degenerate theory amounts to reduce to the standard case $a=0$. 

\subsection{Eliminating the Ostrogradsky instability}\label{sec:ToyModel}
In order to avoid the presence of the Ostrogradsky ghost in a non-trivial way, one must relax the assumptions of the theorem. In particular, this means to consider a degenerate theory with multiple fields. In modified gravity, we can have a situation where a higher-derivative Lagrangian for the additional scalar field is coupled to the scalar degrees of freedom of the metric. 
In the toy-model above, we can couple the higher derivative Lagrangian to $n$ regular canonical variables $q_i(t)$~\cite{Langlois:2015cwa}, ($i=1,...,n$). Keeping only the terms with two or more derivatives, we can write
\be\label{highDerEx}
\mathcal{L}=\frac{1}{2}a {\ddot{\phi}}^2+\frac{m}{2}{\dot{\phi}}^2 +\frac{1}{2}k_{ij}{\dot{q}^i \dot{q}^j}+b_i\, \ddot{\phi} \dot{q}^i\; .
\ee
Note that the ``interaction'' term proportional to $b_i$ generates \emph{third order} equations of motions while the term proportional to $a$ gives fourth order equations.
Reducing again to a second order system with the canonical variable $Q=\dot{\phi}$, we can reformulate the theory as
\beq\label{reformulated toy model}
L=\frac12 a\, \dot Q^2 + \frac12 k_{ij} \dot q^i \dot q^j  +  b_i \dot Q \dot q^i + \frac{m}{2}  Q^2- \lambda (Q-\dot \phi)\,,
\eeq
The inversion of the relation between the canonical variables and their conjugate momenta amounts to the inversion of the kinetic matrix, i.e. the symmetric matrix $M$ describing the part of the Lagrangian quadratic in time derivatives:
\begin{eqnarray}
M = \left(
 \begin{array}{cc}
 a & {b_j} \\
{b_i} & {k_{ij}}
 \end{array}
 \right) \,.
 \label{kinetic_matrix}
 \end{eqnarray}To avoid the presence of the Ostrogradsky ghost we require that this matrix is non-invertible. This is called the \emph{degeneracy condition}, and can be formulated as:
\begin{eqnarray}
\text{det}(M)=\text{det}(k) \left(a-b_i b_j (k^{-1})^{ij} \right)=0 \, \Rightarrow a- b_i \, b_j \, (k^{-1})^{ij} \; = \; 0 \,.  \label{detM} 
\end{eqnarray} 
where I assumed $\text{det}(k)\neq0$.
Imposing the above condition, one can find a null eigenmode that satisfies a constraint equation and reduce the system to a second order one. 
Three cases are possible:
\begin{enumerate}
\item A trivial degeneracy is present: the kinetic matrix has a row and column of zeros. This corresponds to have $a=b_i=0$, i.e. a canonical case with second order equations of motion. 
\item The degeneracy comes only from the coupling term: this corresponds to $a=0$ but $b_i\neq0$. The corresponding theory has third order equations of motion. 
\item Both $a$ and $b_i$ are non vanishing and the degeneracy comes from an interplay between the coupling terms, the higher derivative term and the healthy sector.
\end{enumerate}
Noticeably, we will see that when generalised to the scalar-tensor theories, the first two cases are analogous to the Horndeski and beyond-Horndeski theories introduced above.

The example presented above and introduced in~\cite{Langlois:2015cwa} can be made rigorous by a hamiltonian analysis that actually shows how the degeneracy is associated to the presence of a primary constraint in the theory, that eliminates the Ostrogradsky ghost~\cite{Langlois:2015cwa,Langlois:2015skt,Crisostomi:2016tcp,Crisostomi:2016czh,Motohashi:2016ftl}.
In the next section I will introduce the Lagrangians obtained by this method in the context of scalar-tensor theories of gravity.

\subsection{DHOST theories}\label{sec:subDHOST}

In this section I will show how the ideas presented in the toy models above have been applied to the case of scalar-tensor theories of gravity. I will follow the discussion of Ref.~\cite{Langlois:2015cwa}.

The first step is to write a general action that includes higher-order derivatives for the scalar field. I will consider the case where the Lagrangian can depend quadratically on second order derivatives of the scalar field, $\phi_{\mu\nu}$ (the discussion of the case with Lagrangians cubic in $\phi_{\mu\nu}$ is considerably more involved but conceptually equivalent. It can be found in~\cite{BenAchour:2016fzp}).
The most general action can be written as
 \be \label{action}
S[g,\phi] = \int d^4 x \, \sqrt{- g }
\left[ P(X,\phi) + Q(X,\phi) \Box \phi
+
f_2(X,\phi) \,  {}^{(4)}\!R+ C_\2^{\mu\nu\rho\sigma} \,  \phi_{\mu\nu} \, \phi_{\rho\sigma}
\right]  \;.
\ee

The tensor 
$C_\2$ is the most
general tensor constructed from the metric $g_{\mu\nu}$ and the first derivative of the scalar field 
$\phi_\mu$. I included for completeness the terms in $P(X,\phi)$ and $Q(X,\phi)$, even if these do not contribute to the degeneracy of the Lagrangian\footnote{One could add another term of the form $f_1(X,\phi) \phi^{\mu\nu} \,{}^{(4)}\!R_{\mu\nu}$, but this can be reabsorbed by integration by parts~\cite{BenAchour:2016fzp}.}.
The quadratic terms can be decomposed into the sum of five elementary Lagrangians,
\beq
\label{C2}
 C_\2^{\mu\nu\rho\sigma} \,  \phi_{\mu\nu} \, \phi_{\rho\sigma} =\sum_{A=1}^{5}a_A(X,\phi)\,   L^\2_ A\,,
\eeq
with 
\be
\label{QuadraticL}
\begin{split}
& L^\2_1 = \phi_{\mu \nu} \phi^{\mu \nu} \,, \qquad
L^\2_2 =(\Box \phi)^2 \,, \qquad
L_3^\2 = (\Box \phi) \phi^{\mu} \phi_{\mu \nu} \phi^{\nu} \,,  \\
& L^\2_4 =\phi^{\mu} \phi_{\mu \rho} \phi^{\rho \nu} \phi_{\nu} \,, \qquad
L^\2_5= (\phi^{\mu} \phi_{\mu \nu} \phi^{\nu})^2\,.
\end{split}
\ee

The Ostrogradsky ghost is eliminated choosing the functions $\aq_A$ in the expression (\ref{C2}) so that the corresponding theory is degenerate. As for the toy model of Sec.~\ref{sec:ToyModel}, this is done writing the kinetic matrix for the system and imposing that its determinant vanishes. This requires to separate space and time derivatives in the action. To do so in a very general way, it is convenient to use a 3+1 decomposition of spacetime, that I am going to introduce in the next subsection.

\subsubsection{3+1 decomposition}\label{sec:31}

\begin{figure}[t]
\centering
\includegraphics[width=0.6\textwidth]{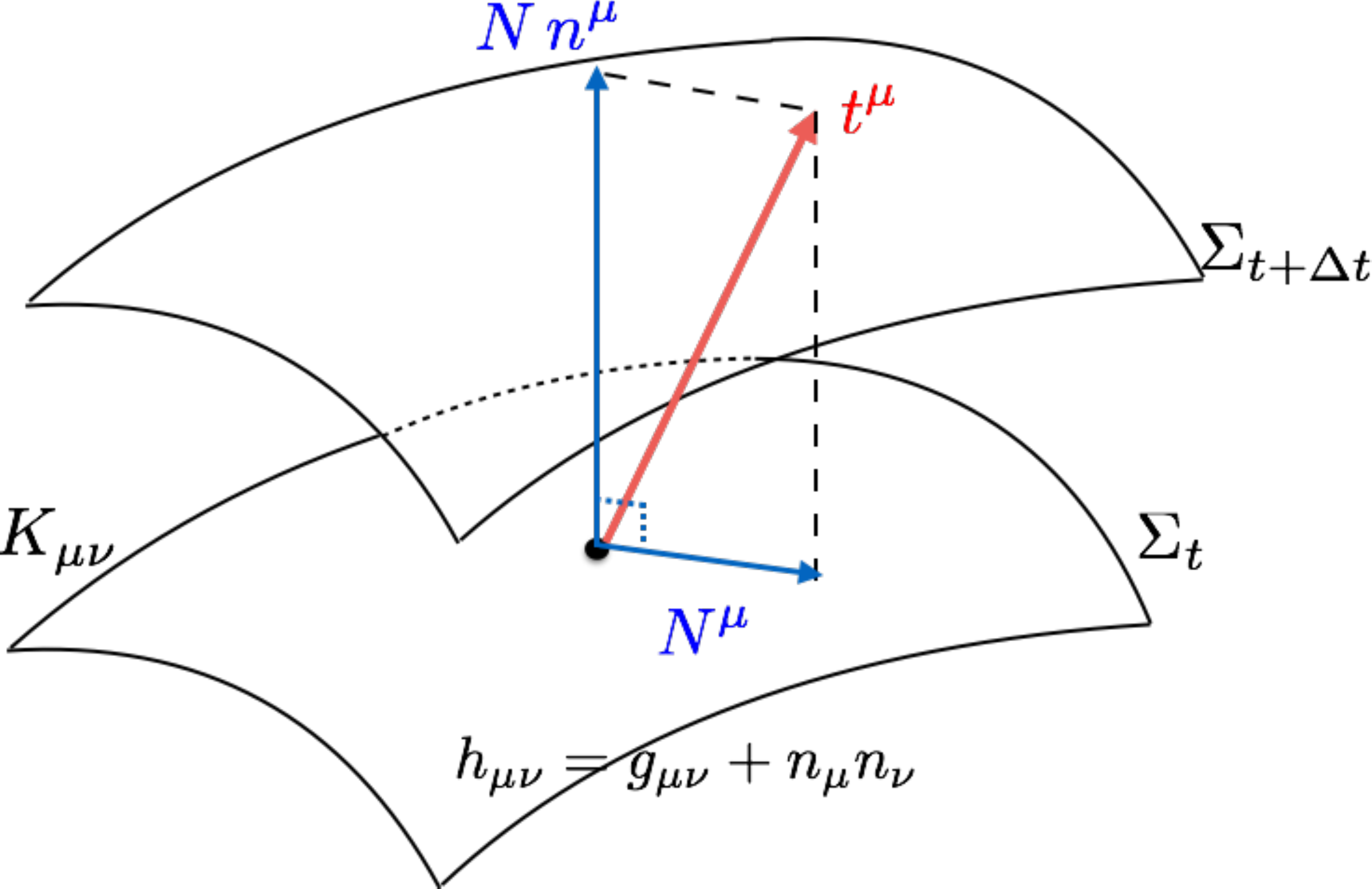}
\caption{Basic geometrical quantities in a 3+1 decomposition.}
\label{fig:hyp}
\end{figure}
Consider a scalar function $t$ such that $t=$const. defines a family of non-intersecting spacelike hypersurfaces $\Sigma_t$.
This is called a foliation of spacetime. So far, $t$ is completely arbitrary. We can define the following quantities~\cite{Menotti:2017cvp}:
\begin{itemize}
\item The unit vector $n^{\mu}$ normal to the hypersurfaces, which is timelike and normalised so that $n_{\mu} n^{\mu}=-1$.
\item The three-dimensional metric induced on the hypersurface $\Sigma_t$:
\be
h_{\mu\nu}=g_{\mu\nu}+n_{\mu} n_{\nu} \; .
\ee
\item The time flow vector $t^{\mu}=\partial / \partial t$ associated to the coordinate $t$. This generates the diffeomorphism which maps $\Sigma_t$ in $\Sigma_{t+\Delta t}$.
It can be decomposed as
\be
t^{\mu}=N n^{\mu}+N^{\mu} \; .
\ee
The above equation defines the \emph{lapse function} $N(\vec{x},t)$ and the \emph{shift vector} $N^{\mu}(\vec{x},t)$ orthogonal to $n^{\mu}$. These, together with other quantities, are illustrated in Fig.~\ref{fig:hyp}.
\item The extrinsic curvature of the hypersurface, which quantifies the properties of the embedding of the 3-surface in a 4-D spacetime through the variation of the normal vector:
\be\label{extrC}
K_{\mu\nu} \equiv h^{\rho}_{\mu} \nabla_\rho n_{\nu}  \, ,
\ee
\end{itemize}
Let me also introduce the normal projection of the vector $A_{\mu}$, 
\beq \label{A}
\A \equiv A_{\mu} n^{\mu}\, ,
\eeq
 which plays an important role for the degeneracy, and the spatial projection
 \be \label{Ahat}
 \hat{A}_\mu \equiv h^{\nu}_{\mu} A_{\nu} \; .
 \ee
Time derivatives are defined as the projection of Lie derivatives with respect to $t^{\mu}$. I will denote them with a ``dot''. In particular,
\be
\dot A \equiv t^{\mu}\nabla_{\mu} A \, .
\ee
Finally, we can construct 3-dimensional covariant derivatives associated to the metric $h_{\mu\nu}$ on the 3-dimesional hypersurface, that I denote with $D_i$.
 
\subsubsection{Degeneracy conditions}
With the use of a 3+1 decomposition, one has the tools to separate space and time derivatives and write the kinetic matrix for the action~\eqref{action}. The procedure is completely analogous to the one followed for the toy model. 

As a first step, one introduces a new variable to reduce the system to a second-order one. So, I define
$
A_{\mu}=\phimu
$
and enforce this property in the action through a Lagrange multiplier $\lambda_{\mu}$. Omitting the terms in $P(X,\phi)$ and $Q(X,\phi)$ that are not relevant for the degeneracy,~\eqref{action} becomes:
\be \label{action2}
S[g,\phi,A_{\mu},\lambda_{\mu}] = \int d^4 x \, \sqrt{- g }
\left[
f_2(X,\phi) \,  {}^{(4)}\!R+ C_\2^{\mu\nu\rho\sigma} \,  \nabla_{\mu} \A_{\nu} \, \nabla_{\rho}\A_{\sigma}+\lambda^{\mu}\left(\phi_{\mu}-\A_{\mu}\right)
\right]  \;.
\ee
Then, one can rewrite~\eqref{action2} in terms of the quantities introduced in~\ref{sec:31}. 
After some manipulations, the kinetic part of the Lagrangian in a 3+1 decomposition can be written in the form \cite{Langlois:2015cwa}
\bea\label{general kin}
L_{\rm kin} = {\cal A}(\phi,X,\A) \dot{\A}^2 + 2 {\cal B}^{\mu\nu}(\phi,X,\A) \dot{\A} K_{\mu\nu} + {\cal K}^{\mu\nu\rho\sigma}(\phi,X,\A) K_{\mu\nu} K_{\rho\sigma}\,,
\eea
where $A$ is the normal projection defined in~\eqref{A}, and $K_{\mu\nu}$ is the extrinsic curvature~\eqref{extrC}. The coefficients are given by
\begin{align}
&\mathcal{A} =\frac{1}{N^2}\left[\ab+\aa-(\ac+\ad)\A^2+
\ae \A^4\right] \,,  \label{calA} \\
 &{\cal B}^{\mu\nu} \, = \, \frac{\A }{2N}\left(2\aa-\ac\A^2+4 f_{2,X}\right) h^{\mu\nu} -\frac{A }{2N}\left(\ac+2\ad-2\ae\A^2\right) \, \hat{A}^{\mu}\,  \hat{A}^{\nu}\,, \label{B}\\
&{\cal K}^{\mu\nu\rho\sigma}\, = \, ( a_1 A^2+f_2) h^{\mu(\rho} h^{\nu)\sigma} + (a_2 A^2-f_2) h^{\mu\nu} h^{\rho\sigma} +\hat{\cal K}^{\mu\nu\rho\sigma} \,. \label{calK}
 \end{align}
The tensor $\hat{\cal K}^{\mu\nu\rho\sigma}$ is given by
\be
\begin{split}
\hat{\cal K}^{\mu\nu\rho\sigma} =&\frac {4f_{2,X}-a_3 A^2}{2} \left(\tA^{\mu}\tA^{\nu} h^{\rho\sigma}+\tA^{\rho} \tA^{\sigma} h^{\mu\nu}\right) -a_1  \left(\tA^{\mu}\tA^{(\rho}h^{\sigma) \nu}+\tA^{\nu}\tA^{(\rho} h^{\sigma)\mu}\right) \\
&+(a_5 A^2-a_4) \tA^{\mu} \tA^{\nu}\tA^{\rho}\tA^{\sigma}\,.
\end{split}
\ee
%
Note that the structure of the Lagrangian~\eqref{general kin} is the same as the one of the toy-model~\eqref{reformulated toy model} with the correspondence (up to a factor of 2)
\be \label{corr}
\begin{split}
&\A \mapsto Q \, ,\quad K_{\mu\nu} \mapsto \dot{q}_i \, , \\
&{\cal A}(\phi,X,\A) \mapsto a \,, \quad {\cal B}^{\mu\nu}(\phi,X,\A) \mapsto b_i\, , \quad {\cal K}^{\mu\nu\rho\sigma}(\phi,X,\A) \mapsto k_{ij} \, .
\end{split}
\ee
So, in this case the role of the ``healthy'' canonical degrees of freedom in the toy-model~\eqref{highDerEx} is played by the degrees of freedom of the metric, contained in the extrinsic curvature $K_{\mu\nu}$ (recall from~\eqref{extrC} that $K_{\mu\nu} \ni \dot{h}_{\mu\nu}$ where $h_{\mu\nu}$ is the spatial part of the metric).

It is possible to cast the determinant of the system~\eqref{general kin} in the form of a polynomial in $\A^2$:
\bea
\label{det}
D(\phi,X,\A^2) \equiv {\cal A} - {\cal K}^{-1}_{\mu\nu\rho\sigma}{\cal B}^{\mu\nu} {\cal B}^{\rho\sigma} = D_0(\phi,X) + \A^2 D_1(\phi,X) + \A^4 D_2(\phi,X)  =0\, ,
\eea
where
\be\label{D012}
\begin{split}
\D_0(\phi,X)&\equiv\ -4 (\aa+\ab) \left[X \f (2\ab+X\ad+4\f_X)-2\f^2-8X^2\f_X^2\right]\,, \\
\D_1(\phi,X)&\equiv\ 4\left[X^2\ab (\ab+3\aa)-2\f^2-4X\f \aa\right]\ad +4 X^2\f(\ab+\aa)\ae +8X\ab^3\\
&-4(\f+4X\f_X-6X\aa)\ab^2 -16(\f+5X \f_X)\ab \aa+4X(3\f-4X \f_X) \ab\ac \\
&-X^2\f \ac^2 +32 \f_X(f+2X f_X) \aa-16\f \f_X \ab-8\f (\f-X\f_X)\ac+48\f \f_X^2 \,, \\
\D_2(\phi,X)&\equiv\ 4\left[ 2\f^2+4X\f \aa-X^2\ab(\ab+3\aa)\right]\ae  + 4\ab^3+4(2\aa-X\ac-4\f_X)\ab^2 \\
&+3X^2 \ab\ac^2-4X\f \ac^2+8 (\f+X\f_X)\ab\ac -32 \f_X \ab\aa+16\f_X^2\ab \\
&+32\f_X^2\aa-16\f\f_X\ac\,.
\end{split}
\ee
The theory is degenerate when the expression~\eqref{det} vanishes for any value of $A$.
This gives the three independent relations 
\bea\label{general deg}
D_0(\phi,X)=0\,, \qquad D_1(\phi,X)=0\,, \qquad D_2(\phi,X)=0 \, ,
\eea
One should solve simultaneously the three equations above to fix three among the functions $f_2$ and $a_A$. 
The theories obtained by imposing the corresponding conditions have been called ``Degenerate Higher-Order Scalar-Tensor'' (DHOST) theories in~\cite{Langlois:2015cwa}, and ``Extended Scalar-Tensor'' (EST) in~\cite{Crisostomi:2016czh}. In the rest of this work, I will use the acronym DHOST. 
These theories have first been identified  at  quadratic order
 in $\phi_{\mu\nu}$
  in \cite{Langlois:2015cwa} and further studied in \cite{Langlois:2015skt,Crisostomi:2016czh,Achour:2016rkg,deRham:2016wji} (see also~\cite{Ezquiaga:2016nqo} for an approach to scalar-tensor theories based on differential forms). 
  The identification of DHOST/EST theories has recently been extended up to cubic order in \cite{BenAchour:2016fzp} where a full classification can be found.

In summary, there exist seven classes of purely quadratic theories and nine classes of purely cubic theories. These quadratic and cubic classes can be combined to yield hybrid theories, involving both quadratic and cubic terms, but all combinations are not possible: only 25 combinations (out of 63)  lead to degenerate theories, often with extra conditions on the free functions in the Lagrangian (see \cite{BenAchour:2016fzp} for details). 
I will however show that at the level of linear perturbations the analysis is greatly simplified, as all the degeneracy conditions of the above classes reduce to only two sets of conditions.

In order not to complicate the discussion, in the main text I will discuss the quadratic case. Let me thus summarise the classes of quadratic DHOST theories: 
\begin{itemize}
\item \emph{Minimally coupled theories}. They correspond to the case $f_2=0$. In this case, the curvature ${}^{(4)}\!R$ disappears from the action. They contain three classes:  $^2$M-I/IIIa, $^2$M-II/IIIb, $^2$M-I/IIIc.
\item \emph{Non-minimally coupled theories}. In this case $f_2\neq0$. There are four classes: $^2$N-I/Ia, $^2$N-II/Ib, $^2$N-III/IIa, $^2$N-IV/IIb. 
\end{itemize}
In each of the above classes, three different functions among $f_2$ and $a_A$ are fixed in terms of the others. The explicit expressions are not relevant for the present discussion and can be found in~\cite{BenAchour:2016fzp}.

Let me finally underline some relevant aspects of the degeneracy and its relation to Horndeski theories and their extension introduced in Sec.~\ref{BHth}.
Basing on the correspondence~\eqref{corr} between the toy model and scalar-tensor theories, the three cases discussed in Sec.~\ref{sec:ToyModel} correspond to:
\begin{enumerate}
\item Horndeski theories. The quartic Horndeski Lagrangian corresponds to the case 
\be\label{Horndas}
f_2=G_4 \, , \quad a_1=-a_2=2G_{4,X} \, , \quad a_3=a_4=a_5=0 \, .
\ee
As can be seen explicitly from~\eqref{A} and~\eqref{B}, this leads to a trivial degeneracy with $\cal{A}$ $={\cal{B}}^{\mu\nu}=0$ and second order equations of motion.
\item Theories ``beyond-Horndeski''. The degeneracy comes from the interaction terms, ${\cal{B}}^{\mu\nu}\neq0$ but $\cal{A}$ $=0$. The theory has third order equations of motion, like the quartic beyond Horndeski Lagrangian~\eqref{L4bH}. This Lagrangian corresponds to the case 
\be \label{Bhas}
a_1=-a_2=X F_4\, \quad a_3=-a_4=2F_4\, , \quad a_5=0 \, .
\ee
\item More general DHOST/EST theories. New classes with $\cal{A}$ $\neq 0$, ${\cal{B}}^{\mu\nu}\neq0$.  
\end{enumerate}

In particular, in Horndeski and beyond Horndeski theories we can see from~\eqref{D012} that the condition $D_0=0$ is always satisfied (since $a_1=-a_2$). Then, one can use the other two conditions to express $a_4$ and $a_5$ in terms of $a_2$ and $a_3$. The requirement to have also $\mathcal{A}=0$ gives $a_5=0$, $a_3+a_4=0$, leaving only two arbitrary functions. This corresponds to the sum of Horndeski and beyond Horndeski quartic Lagrangians. These are contained in the class $^2$N-I/Ia.  

In the following, I will explore the phenomenological properties of the DHOST theories summarised above basing on an effective description. 


\chapter{Effective Theory of Dark Energy}\label{chap:EFToDE}
\lhead{\emph{Effective Theory of Dark Energy}}

\section{An effective description of dark energy}\label{sec:EFTintro}

In Chapter~\ref{chap:DHOST} I introduced a general class of scalar-tensor theories formulated in terms of covariant Lagrangians. 
The ``top-down''  procedure to test these theories amounts to solve the equations for the propagating degrees of freedom, compute the effects on the observables, and try to constrain the free parameters with observations.  Ultimately, we would like to compare the performance of alternative models with $\Lambda$CDM.
The idea of developing an effective description is to find a ``bottom-up'' approach to test linear perturbations in scalar-tensor theories against $\Lambda$CDM, in such a way that we can be agnostic about the underlying fundamental theory. This can be achieved by writing directly a general action for the fluctuations around a time-dependent FLRW background solution in the case where a single scalar degree of freedom is added to GR. 
Let me point out two reasons why an action is important. First, it allows a link with basic principles of physics. In particular, any deviation from $\Lambda$CDM described this way will be automatically consistent with locality, causality and unitarity~\cite{Lewandowski:2016yce,Endlich:2017tqa}. Second, an action allows a systematic study of the stability of the theory. 

The basic idea of the effective description is the following.
$\Lambda$CDM is based on GR and thus has an invariance under coordinate transformations, $x^{\mu} \mapsto \tilde x ^{\mu} =\tilde x ^{\mu}(x^{\nu})$. 
In the alternatives to $\Lambda$CDM I am considering, the acceleration is caused by an additional scalar field $\phi(\vec{x},t)$. In cosmology, due to homogeneity and isotropy this field acquires a time-dependent background value $\bar{\phi}(t)$.
This \emph{spontaneously} breaks the time reparametrisation invariance. So, it makes sense to describe deviations from $\Lambda$CDM assuming spontaneous breaking of the time diffeomorphisms. 
Analogy with spontaneously broken gauge theories suggests that there will be massless excitations (Goldstone modes) describing the low-energy dynamics. These are the fluctuations of the additional scalar degree of freedom, $\delta \phi(\vec{x},t)=\phi(\vec{x},t)-\bar{\phi}(t)$.
They have to transform linearly under the \emph{unbroken} symmetries, i.e. space translations and rotations.
We can thus construct the most general action compatible with this residual symmetry, allowing operators that break time diffeomorphism invariance. The coefficients of these operators will be functions of time and can be constrained by observations. I will present a formulation where these parameters are chosen to represent deviations from $\Lambda$CDM. Remarkably, the large classes of theories introduced in Chapter~\ref{chap:DHOST} reduce to a limited number of free functions in the effective description, as I will discuss.
For any covariant theory, one can compute its free functions in the effective description and then compute the observables. On the other hand, it is also possible to assume a parametrisation for these functions and constrain them directly. This is the true advantage of the effective description. In both cases, the computation of the observable deviations from $\Lambda$CDM, or the implementation of the equations in numerical codes, can be done once and for all.
I will show how to make the connection between covariant theories and the effective description in Sec.~\ref{sec:AlphaInterpr}, after reviewing how to write the action in Sec.~\ref{sec:EFTaction}.

\section{Generalities. Building the action}\label{sec:EFTaction}
Let me now show how to write an action based on the above ideas. We are looking for an action invariant under spatial diffeomorphisms but not under time ones. Thus, we need to separate space and time components and to identify the allowed operators. The natural framework to do so is the 3+1 decomposition, introduced in Sec~\ref{sec:31} exactly for the same purpose. In addition to this, there are two additional steps that one can make.
\begin{itemize}
\item \emph{Unitary gauge.} The fact that the scalar field has a background value $\bar{\phi}(t)$ defines a preferred foliation of spacetime, given by the hypersurfaces of constant $\phi$. In a cosmological context, the usual assumption is that the scalar field gradient is \emph{spacelike}, $\phimu\phi^{\mu}<0$, so these hypersurfaces are spacelike. 
To adapt to this preferred foliation, we can choose the background value of the scalar field as a ``clock'', such that constant time hypersurfaces correspond to constant $\phi$ ones. This choice of the time coordinate is called the \emph{unitary gauge}. We have:
\be 
\begin{split}
&\phi(t,\vec x)=\bar \phi(t)+\delta \phi(t,\vec x) \, , \\
&\phi(t,\vec x)=\bar \phi(t) \Leftrightarrow  \delta \phi(t,\vec x)=0 \qquad\text{(Unitary gauge)}\;.
\end{split}
\end{equation}
After the gauge fixing, we are left with the symmetry $x^{i} \mapsto \tilde x ^{i} =\tilde x ^{i}(x^{\nu})$, which is exactly the unbroken part of the general coordinate invariance of GR.
The scalar degree of freedom appears now in the metric: for example, the kinetic term $X=g^{\mu\nu}\phimu \phinu$ becomes just $g^{00} \dot{\phi}^2=-\dot{\phi}^2/N^2$. At linear level, its contribution will be encoded in the expansion of the metric element $g^{00} $, or equivalently of the lapse function $N$. \\
The geometrical quantities on the hypersurfaces on constant $\phi$/constant time are those introduced in Sec~\ref{sec:31}. These geometrical quantities will now be related to $\phi$. In particular, the normal to the hypersurface is proportional to the gradient of the scalar field:
\be
\begin{split}
n_{\mu}&=-\frac{\partial_{\mu} \phi}{\sqrt{-X}} \; , \quad X\equiv g^{\mu\nu} \partial_{\mu} \phi\partial_{\nu} \phi \, .
\end{split}
\ee

\item \emph{ADM coordinates}. In Sec~\ref{sec:31} I introduced the geometrical quantities in the 3+1 decomposition without referring to a specific coordinate system. To further simplify the problem, one can choose a coordinate system that adapts to the preferred foliation. 
A priori, the coordinate $t$ is completely arbitrary, but it is possible to construct a coordinate system that uses it as the time coordinate.
This coordinates are $(t,x^i)$, $i=1,2,3$, where $x^i$ are the spatial coordinates on the hypersurface $\Sigma_t$. This is called the ADM~\cite{Arnowitt:1962hi} coordinate system. The line element can be written as\footnote{I use latin indeces ($i$, $j$, ...) for the spatial parts.}:
 \beq
 \label{ADM}
 ds^2=-N^2 dt^2 +h_{ij} (dx^i+N^i dt)(dx^j+N^jdt)\,.
 \eeq
In the above equation, I used the lapse function $N(\vec{x},t)$, the shift vector $N^i(\vec{x},t)$, and the \emph{spatial metric} $h_{ij}$ introduced in Sec~\ref{sec:31}. In particular, the latter measures distances between points on every hypersurface, $d \ell^2=h_{ij }dx^i dx^j$.
In the ADM coordinate system, the relevant geometrical quantities have the following form~\cite{Menotti:2017cvp}:
\begin{itemize}
\item Normal unit vector:
\be
\quad n^{\mu}=\left(\frac{1}{N}, \frac{N^i}{N}\right) \, , \quad n_{\mu}=(-N,0,0,0) \, .
\ee
\item Extrinsic curvature:
\be
K_{ij} =\frac{1}{2N} \big(\dot h_{ij} - D_i N_j - D_j N_i \big) \, ,
\ee
where $D_i$ denotes the covariant derivative associated to the metric $h_{ij}$, and a dot a derivative with respect to the time $t$.
\end{itemize}
Le me introduce two other quantities that are useful to characterise the 3-dimensional surfaces:
\begin{itemize}
\item Tangent to the hypersurface: 
\be
 a_{\mu}=n^{\nu}\nabla_{\nu}n_{\mu}=\left(0,\frac{\partial_i N}{N} \right)\, .
\ee
\item The intrinsic curvature on the hypersurface. It is quantified by the three-dimendional Ricci tensor on the hypersurface, $R_{ij}$. In 3-D, this contains as much information as the Riemann tensor.
\end{itemize}

\end{itemize}
We now have all the elements to write down a very general action. 
In particular, we can include 
any time-dependent operator, tensors with free zero indexes, namely $g^{00}=-1/N^2$,\footnote{This actually exhausts the possibilities, since the shift vector and the extrinsic curvature do not have by definition components with $0$ indexes.} and diff-invariant combinations of tensors with spatial indexes such as the extrinsic curvature $K_{ij}$ and the intrinsic curvature $R_{ij}$. Spatial indices are lowered and raised with the spatial metric $h_{ij}$ or its inverse $h^{ij}$, respectively. We can take covariant derivatives $D_i$ associated with the three-dimensional spatial metric and time derivatives $\partial_0$ that I will denote by a dot. 
The building blocks of the action are constructed with the geometrical elements that characterise the  hypersurfaces written in unitary gauge and in ADM coordinates~\cite{Creminelli:2006xe,Cheung:2007st}.
The gravitational action is generically of the form\footnote{The ``acceleration'' vector $a_i$ is not included explicitly since it can be obtained by taking spatial derivatives of the lapse. 
The shift vector $N^i$ should enter in the diff-invariant combination $\dot N-N^i \partial_i N$, but this term reduces to $\dot{\delta N}$ at linear order. }
 \be
\label{g_action}
S_{\rm g}=\int d^4 x \sqrt{-g}\,  \mathcal{L}(N,  K_{ij}, R_{ij}, D_i , \partial_0 ;t) \;.
\ee

The above form is very general. To have an intuition, let me consider the Einstein-Hilbert action
\be
S_{\rm GR}=\int d^4x \sqrt{-g}\, \frac{M_{\text{Pl}}^2 }2\, {}^{(4)}\!R\,.
\ee
One can use the Gauss-Codazzi relation
\be
\label{GC1}
{}^{(4)}\! R =  K_{\mu \nu} K^{\mu \nu}- K^2 + R  + 2 \nabla_\mu (K n^\mu - n^\rho \nabla_\rho n^\mu ) \;,
\ee
to rewrite it in 3+1. The Lagrangian reads
 \be
 \label{pureGR}
\mathcal{L}_{\text{GR}}=\frac{M_{\text{Pl}}^2}2\left[ K_{ij}K^{ij}-K^2+R \right]\,, 
\ee
and is of the form~\eqref{g_action}. \\
Another example worth to mention is a quintessence field added to GR~\cite{Ratra:1987rm,Wetterich:1987fm}, 
\be\label{quint}
S=\int d^4x \sqrt{-g}\, \Big[ \mathcal{L}_{GR}+\frac12 \phi \Box \phi -V(\phi)\Big]\,,
\ee
where $ \mathcal{L}_{GR}$ is the Einstein-Hilbert Langrangian density.
In 3+1 we have the Lagrangian
\be\label{quint31}
\mathcal{L}(t,N)=\mathcal{L}_{GR}+\frac{\dot\phi^2(t)}{2N^2} -V\big(\phi(t)\big) \,.
\ee
So, we added to the GR Lagrangian~\eqref{pureGR} a dependence on the lapse function $N$.\\
More general actions would introduce more complicated terms but can be always reduced to the form~\eqref{g_action}. A complete discussion on how to write general covariant Lagrangians in the 3+1 form has been provided in~\cite{Gleyzes:2013ooa} for the Horndeski case. In Appendix A of~\cite{Langlois:2017mxy} one can find the full DHOST Lagrangians (up to cubic order) in 3+1. 
I will discuss in the main text the details at the level of linear perturbations. 

Even if I included them in the most general case, special care must be taken with time derivatives. In the following discussion, I concentrate on the scalar sector. 
In the effective description, the tensor $K_{ij}$ contains one time derivative of the metric, so any operator quadratic in it will already yield two time derivatives, which correspond to the presence of one propagating degree of freedom. 
Hence, not taking time derivatives of the three-dimensional tensors listed above is enough to ensure the presence of a single propagating degree of freedom; additional conditions should then be imposed on its action in order to avoid that it is itself a ghost and that it contains gradient instabilities. I will describe the physical meaning of these requirements in the following\footnote{Note that we must impose conditions on the action for the propagating degree of freedom after solving the constraints of the theory. Imposing conditions on the initial action can lead to too restrictive conditions.}.\\
This however does not represent the most general case; indeed, we saw in Chapter~\ref{chap:DHOST} that suitable degeneracy conditions can be imposed on an action to eliminate unwanted degrees of freedom\footnote{In effective theories, higher time derivatives are indeed allowed provided that they are suppressed below the cutoff scale, so that the ghosts are out of the domain of validity of the theory. Here I shall consider them on the same footing as the other operators.}. This can be done also in the effective description: in particular, we can allow for time derivatives of the lapse function $N$, but find degeneracy conditions to ensure that only one DOF propagates. In particular, the presence of $\dot N$ is the unitary gauge analogue of introducing a kinetic term for the ``velocity'' $A_{\mu}=\phimu$ introduced in Sec.~\ref{sec:subDHOST}, and the degeneracy conditions needed in the effective description are the analogue of those obtained in the covariant formulation of DHOST theories. 
I will explain in detail the relation between the degeneracy conditions obtained at the covariant and linear level.
\\The case of spatial derivatives is different. Some of the operators built with the above tensors alone can lead to higher order spatial derivatives unless their relative coefficients are appropriately tuned.
In the following, I will consider the most general action for cosmological perturbations that contains operators with at most two derivatives in the effective description. After solving the constraints, this action can contain higher order spatial derivatives. However, I will show that when imposing the appropriate degeneracy conditions obtained at the covariant level, the theory will be free of higher spatial derivatives as well\footnote{Differently from time derivatives, higher spatial derivatives are not necessarily suppressed and may dominate the dispersion relation, such as in the Ghost Condensate theory~\cite{ArkaniHamed:2003uy}. In this case, higher spatial gradients become relevant, even if they begin operating at very short distances~\cite{Creminelli:2008wc,ArkaniHamed:2005gu}, typically shorter than the cosmological ones. Another case where higher order spatial derivatives are present are models that explicitly break Lorentz invariance, such as Horava gravity and its extensions~\cite{Blas:2010hb}. Often, these models are formulated directly in the unitary gauge, but their generalisation to arbitrary gauges could contain additional propagating degrees of freedom. I will show how the effective description encompasses such models.}.

\subsection{Background evolution}
As far as the effective description is concerned, the background evolution is fully encoded in one free function of time $H(t)$ that can be obtained solving the equations of motion in a specific model. In a model independent approach, one can just fix it so to reproduce the observed background expansion history. 
Note that to completely characterise the gravity and dark energy sector we shall also provide a constant, i.e. the fractional matter density today $\Omega_{\rm m, 0}$, since we could trade some dark matter with a suitable amount of time-dependent dark energy keeping the measurements of the background unchanged~\cite{Kunz:2007rk}.\\

Here I will just recall how to obtain the background equations in the effective formalism. 
On a spatially flat FLRW spacetime,  the line element takes the form
\be
ds^2=-\bar N ^2(t) dt^2 +a^2(t) \delta_{ij} dx^i dx^j\,.
\ee
Among the tensors that enter in the action~\eqref{g_action}, the intrinsic curvature tensor of the constant time hypersurfaces vanishes, i.e. $R_{ij}=0$, and the components of the extrinsic curvature tensor are given by $K^i_j=H\delta^i_j$, where $H \equiv \dot{a}/(a \bar{N})$ is the Hubble parameter.

\be
\label{defH}
K^i_j=H\delta^i_j \, , \quad H \equiv \frac{\dot a}{\bar{N} a} \,,
\ee
where $H$ is the Hubble parameter. 
Note that we must retain the background value of the lapse, $\bar N(t)$, since the variation of the action with respect to it gives the
 first Friedmann equation. The homogeneous Lagrangian in~\eqref{g_action}, is a function of $\bar N(t)$, $a(t)$ and of time only.
\be
\bar {\mathcal{L}}(a, \dot a, \bar N)\equiv  \mathcal{L}\left[K^i_j=\frac{\dot a}{\bar N a}\,\delta^i_j, \bar N(t), \dot {\bar N}(t)\right] \;.
\ee
Adding matter minimally coupled to the metric $g_{\mu\nu}$\footnote{The variation of the corresponding action with respect to the metric defines the energy-momentum tensor, 
\be
\delta S_{\rm m}=\frac12\int d^4x \sqrt{-g}\,  T^{\mu\nu} \, \delta g_{\mu\nu}\,.
\ee
In a FLRW spacetime, this reduces  to 
\be
\label{linearmat}
\delta \bar S_{\rm m}=\int d^4x  \bar N a^3\left(-\rho_{\rm m}\frac{\delta N}{\bar N}+3 p_{\rm m}\frac{\delta a}{a}\right) \;.
\ee
}, the variation of the total homogeneous action $\bar{S}=\bar{S}_g+\bar{S}_{\rm m}$ with respect to $N$ and $a$ yields, respectively, the first and second Friedmann equations~\cite{Gleyzes:2013ooa,Gleyzes:2014rba}-\citepubli{Gleyzes:2015pma}:
\be \label{Fried}
\begin{split}
\bar{\mathcal{L}}+\bar{N} \mathcal{L}_N-3H \mathcal{F}-\frac{1}{\bar{N} a^3}\frac{d}{dt}\left(\bar{N} a^3 \mathcal{L}_{\dot N}\right)=&\rho_{\rm m}\, ,\\
\bar{\mathcal{L}}-3H\mathcal{F}-\frac{\dot {\mathcal{F}}}{\bar{N}}=&-p_{\rm m}\, ,
\end{split}
\ee
where
\be
\label{F}
\left(\frac{\partial \mathcal{L}}{\partial K_{ij}}\right)_{\rm bg}\equiv \mathcal{F}a^{-2}\delta^{ij}\,, \quad \mathcal{L}_N=\left(\frac{\partial \mathcal{L}}{\partial N}\right)_{\rm bg}\, , \quad \mathcal{L}_{\dot{N}}=\left(\frac{\partial \mathcal{L}}{\partial \dot{N}}\right)_{\rm bg}\, .
\ee
Again, the above equations are very general but one can recover the well-known cases by doing the calculation. For example,  in GR we have:
\be 
 \frac{\partial \mathcal{L}_{\rm GR}}{\partial K^i_j}=M_{\rm Pl}^2\left(K^j_i-K \d^j_i\right),
 \ee
 which, after substituting $K^i_j=H \d^i_j$, yields, 
 \be
 \mathcal{F}_{\rm GR}=-2 M_{\rm Pl}^2 H\,,
 \ee
 whereas $\bar{\mathcal{L}}_{\rm GR}=-3 M_{\rm Pl}^2 H^2$ and $\mathcal{L}_N=\mathcal{L}_{\dot N}=0$. With these expressions, one recovers the usual Friedmann equations.

\section{Linear perturbations}
To study linear perturbations, one needs to expand the action at second order around the homogeneous background.
Fixing the background gauge $\bar N=1$,  these are
\be\label{pert}
\delta N = N - 1 \,, \qquad \delta K_{ij} = K_{ij} - H h_{ij}\, , \qquad R_{ij} \, ,
\ee
where $R_{ij}$ is already a perturbation since its background value vanishes.\\
Let me resume the idea of the procedure, without entering into too lengthy calculations. The expansion of the Lagrangian $\mathcal{L}$ up to quadratic order is of the form
\be
\label{L_up_to_2}
\mathcal{L}(N, K^i_j, R^i_j,\dots)=\bar {\mathcal{L}}+\mathcal{L}_N\d N+\frac{\partial \mathcal{L}}{\partial K^i_j} \d K^i_j+\frac{\partial \mathcal{L}}{\partial R^i_j} \d R^i_j+
\mathcal{L}^{(2)}+\dots .
\ee
The first order part cancels upon use of the background equations~\eqref{Fried}. The quadratic part is given by
\be
\label{L_quad}
\begin{split}
\mathcal{L}^{(2)}
  = & \frac12 \mathcal{L}_{NN} \d N^2+\frac12\frac{\partial^2 \mathcal{L}}{\partial K^i_j\partial K^k_l} \d K^i_j \d K^k_l+\frac12\frac{\partial^2 \mathcal{L}}{\partial R^i_j\partial R^k_l} \d R^i_j \d R^k_l+
 \cr
 & + \frac{\partial^2 \mathcal{L}}{\partial K^i_j\, \partial R^k_l}\d K^i_j \d R^k_l+
 \frac{\partial^2 \mathcal{L}}{\partial N\partial K^i_j} \d N\d K^i_j+\frac{\partial^2 \mathcal{L}}{\partial N \partial R^i_j} \d N\d R^i_j+\dots \;,
 \end{split}
 \ee
where the dots indicate all the other possible terms. The partial derivatives are evaluated on the background and the notation $ \mathcal{L}_{NN} $ indicates the second derivative with respect to the lapse. One can further simplify the second order action by integration by parts and using the background equations of motion. The details can be found in~\cite{Gleyzes:2014rba}. So far, I have not imposed yet any constraint on the form of the action. In particular, the final expression can in principle contain higher spatial derivatives and time derivatives of the lapse function that signal the presence of an additional scalar degree of freedom, as I pointed out previously. \\
In this work, I will study systematically Lagrangians including at most two time or space derivatives in perturbations\footnote{This means that I will not include operators such as $\delta K R$, $R^2$, $R_i^j \,R_j^i$, $R \dot{\delta N}$ that contain three derivatives. Derivatives of the extrinsic curvature are not allowed for the same reason.
Note that this procedure doesn't imply that the action for the propagating degree of freedom won't contain higher order space derivatives. For example, the operators $\delta K_{ij}\delta K^{ij}$ and $\delta K^2$ indeed generate higher order gradients without a tuning of their relative coefficient. }. 
Imposing this requirement, it is possible to find combinations of the coefficients of the expansion~\eqref{L_quad} such that the quadratic action can be written in  the form 
\be
\begin{split}
\label{SBAction0}
& S^{(2)} = \int d^3x \,  dt \,  a^3  \frac{M^2}2\bigg\{ \delta K_{ij }\delta K^{ij}- \left(1+\frac23\aL\right)\delta K^2  +(1+\alphaT) \bigg( R \frac{\delta \sqrt{h}}{a^3} + \delta_2 R \bigg)\\
&  + H^2\alphaK \delta N^2+4 H \alphaB \delta K \delta N+ ({1+\alphaH}) R  \delta N   +  4 \bun  \delta K  {\delta \dot N }   + \bdeux  {\delta \dot N}^2 +  \frac{\btrois}{a^2}(\partial_i \delta N )^2   
\bigg\} \; ,
\end{split}
\ee
where $\delta_2$ denotes taking the expansion at second order in perturbations. I will show that the action~\eqref{SBAction0} describes the linear perturbations of all the DHOST theories and show the connection between their covariant formulation and the above expression for the action. 
The coefficients appearing in the action~\eqref{SBAction0} correspond to distinct physical effects. They are functions of time, since the scalar field has a nontrivial background evolution and the action has been built to respect space diffeomorphisms only. The definition of these functions is such that they parametrise \emph{deviations} from $\Lambda$CDM, which corresponds to set them to zero\footnote{An exception to this is given by the \emph{cuscuton}~\cite{Afshordi:2006ad,Afshordi:2007yx}, where the cosmological background evolution is modified but no new degrees of freedom appear in perturbations.}. In this case, the scalar sector does not contain propagating degrees of freedom and the above action reduces to the description of the two degrees of freedom in the tensorial sector. 
The functions $\alphaK$, $\alphaB$, $\alphaT$ \cite{Bellini:2014fua}, together with the variation of the  effective Planck mass squared  $M^2$,
\be
\label{alphaM}
\alphaM\equiv \frac{d \ln M^2}{ d \ln a}\,,
\ee
are sufficient to cover linear perturbations in Horndeski theories, $\alpha_{\rm H}$ \cite{Gleyzes:2014rba} corresponds to their extension ``beyond Horndeski''~\cite{Gleyzes:2014dya,Gleyzes:2014qga}, while $\aL, \bun, \bdeux, \btrois$~\citepubli{Langlois:2017mxy} appear in the DHOST theories, or in Lorentz breaking theories.

The time dependent functions are defined so to be independent of the background expansion history \cite{Bellini:2014fua,Gleyzes:2014rba}.
Any of the models introduced in Chapter~\ref{chap:DHOST} can be cast in the above form using the ADM decomposition in unitary gauge. The most general ``dictionary'' can be found in~\citepubli{Langlois:2017mxy}. 
The fact that the action can be organised in powers of the perturbations and allows a clear separation from the background is one of the most powerful features of the effective description in the ADM formalism in unitary gauge with respect to a covariant effective approach \`a la Weinberg~\cite{Weinberg:2008hq,Park:2010cw,Bloomfield:2011np}. In the latter case, adding a new operator would correspond to a change in the background as well and would lead to study the model again from the beginning. A second point is that the relative importance of different operators in the covariant language can be studied only around a specific background. \\

\section{Effects on linear perturbations}\label{sec:AlphaInterpr}

As I said, the functions $\alpha_A$, $\beta_A$ correspond to distinct physical effects that I shall briefly recall in this section. Table~\ref{tabAlphas} summarises their presence in different scalar-tensor theories.

\renewcommand{\arraystretch}{1.4}
\begin{table}[t]
\small
\begin{center}
\begin{adjustbox}{max width=\textwidth}
\begin{tabular}{|c||c|c|c|c|c|c|c|c|c|c|}
  \hline
  &$\alphaK$& $\alphaB$&$\alphaM$&$\alphaT$& $\alphaH$ & $\aL$ &$\bun$ & $\bdeux$ & $\btrois$ \\
  \hline\hline
$\Lambda$CDM   &  &  &   & &&&&&\\
 \hline
  Quintessence~\cite{Ratra:1987rm,Wetterich:1987fm}, \emph{k}-essence~\cite{ArmendarizPicon:1999rj,ArmendarizPicon:2000ah} & \ding{56}&&&&&&&&
 \\
  \hline
  Kinetic Gravity Braiding-Cubic Galileon~\cite{Deffayet:2010qz,Kobayashi:2010cm,Pujolas:2011he}  &\ding{56}&\ding{56}&&&&&&&
  \\
  \hline 
  Galileon Cosmology~\cite{Chow:2009fm}, Brans-Dicke~\cite{Brans:1961sx},  f(R)~\cite{Starobinsky:1980te,DeFelice:2010aj} &\ding{56}&\ding{56}&\ding{56}&&&&&&
  \\
   \hline
   Horndeski~\cite{Horndeski:1974wa}-Generalized Galileons~\cite{Deffayet:2011gz}&\ding{56}&\ding{56}&\ding{56}&\ding{56}&&&&&
   \\
 \hline
   Beyond Horndeski~\cite{Gleyzes:2014dya}&\ding{56}&\ding{56}&\ding{56}&\ding{56}&\ding{56}&&&&
   \\
   \hline
   \hline
   Ia DHOST~\cite{Langlois:2015cwa} ${}^{\star}$&\ding{56}&\ding{56}&\ding{56}&\ding{56}&\ding{56}&&\ding{56}&\ding{56}&\ding{56}\\
    \hline
   IIa DHOST~\cite{Langlois:2015cwa}${}^{\star}$&\ding{56}&\ding{56}&\ding{56}&\ding{56}&\ding{56}&\ding{56}&\ding{56}&\ding{56}&\ding{56}
   \\
\hline
  Horava gravity~\cite{Horava:2009uw}${}^{\star\star}$&&&&&&\ding{56}&&&
   \\
\hline
  Healthy extensions of Horava gravity~\cite{Blas:2010hb} ${}^{\star\star}$&\ding{56}&&&&&\ding{56}&&&\ding{56}
   \\
\hline
   Chronometric theories~\cite{Blas:2010hb,Blas:2009yd}  ${}^{\star\star}$&&&&\ding{56}&\ding{56}&\ding{56}&&&\ding{56}
   \\
\hline

  \end{tabular}
\end{adjustbox}
\end{center}
\caption[Free functions of the effective description and their appearance in different modified gravity and dark energy theories.]{Free functions of the effective description and their appearance in different modified gravity and dark energy theories. Theories  marked by ${}^{\star}$ require degeneracy conditions to be imposed on the coefficients in order to avoid the propagation of extra degrees of freedom.  
Theories marked by ${}^{\star\star}$ are formulated directly in the unitary gauge and their covariantization either introduces a ghost or requires to restrict the space of solutions. }
\label{tabAlphas}
\end{table}

\begin{itemize}

\item\emph{General Relativity}.
As a preliminary example, consider the Einstein-Hilbert action in 3+1 given in Eqn.~\eqref{pureGR}
It easy to show that, when expanded at quadratic order in perturbations, it has the structure of~\eqref{SBAction0} with $M=M_{\text{Pl}}=\rm const.$ and all the $\alpha_i$, $\beta_i$ set to zero.

\item \emph{Kineticity} $ \alphaK$. This function arises directly from the most standard kinetic term for the scalar field and it is the typical contribution of the scalar field in basic models where dark energy has a perfect fluid energy-momentum tensor. It encodes the kinetic energy of the scalar field which in unitary gauge remains hidden in the metric. Lagrangians that depend only on the scalar field and its gradient, $L(\phi, \phimu)$, lead only to this term.\\
This is the cas of quintessence, written in 3+1 in Eqn.~\eqref{quint31}.
The potential $V(\phi(t))$ doesn't depend on any of the tensors appearing in the action for linear perturbations; indeed, in unitary gauge it is a background quantity, $V=V(t)$, and it's fully fixed by the Friedmann equation once $H(t)$ and $\Omega_{\rm m,0}$ are given. This shows the effectiveness of this parametrisation in splitting background and perturbations. Differently from GR, as we saw, the Lagrangian~\eqref{quint31} has a dependence on the lapse function. This gives the non-vanishing function
\be
\alphaK=\frac{\dot {\bar{\phi}}^2}{H^2 M_\text{Pl}^2}
\ee

\item \emph{Kinetic braiding} $ \alphaB$. In the Lagrangian~\ref{quint} the operator $\Box \phi$ has the coefficient $\phi$, which amounts to a standard kinetic term. Let's now promote $\phi$ to a function $G_3(\phi ,X)$~\cite{Deffayet:2010qz,Kobayashi:2010cm,Pujolas:2011he}:
\be
\label{KB}
S=\int d^4x \sqrt{-g}\, \Big[ \frac{M_{\text{Pl}}^2 }2\,{}^{(4)}\!R+G_3(\phi,X)\Box \phi\Big]\,,
\ee

Since the operator $\Box$ contains covariant derivatives, the dependence of  $G_3$ on $X$ will lead to the presence of terms of the type $\partial g\,\partial \phi$, i.e. a kinetic coupling between the scalar and the metric. Hence the name of kinetic braiding. In unitary gauge, these are encoded in the operator $\delta N \delta K$ and lead to a non vanishing $\alphaB$ and $\alphaK$. In particular,
\be
\alphaB=-\frac{G_{3X}}{H M_{\rm Pl}^2}\, .
\ee

\item \emph{Planck mass running rate} $ \alphaM$. 
The time evolution of the Planck mass can be seen as a time-dependend conformal rescaling of the metric. In the absence of matter, it would be re-absorbed by a conformal transformation; when matter is present, however, this comes at the price of introducing a non-minimal coupling with the matter fields, as I will discuss in Chapter~\ref{chap:EFToDE_m}. The simplest example is given by Brans-Dicke theories~\cite{Brans:1961sx}, where the action can be written in the form:
\be
\label{BD}
S_{\rm BD}=\int d^4x \sqrt{-g}\, \, \Big[\phi \,{}^{(4)}\!R-\frac{\omega_{\rm BD}}{\phi}\phi_{\mu} \phi^{\mu}  +V(\phi)\Big]\,.
\ee

This leads to the following non-vanishing functions in the effective action:

\be
\alphaM=\frac{d \ln \phi}{d \ln a}\, , \quad \alphaK=\omega_{\rm BD}\alphaM^2\,,\quad \alphaB=\frac{\alphaM}{2}\, .
\ee

\item \emph{Tensor speed excess} $ \alphaT$. When constructing Lagrangians for the additional scalar degree of freedom, the allowed terms can lead also to modifications of the tensorial part of the action\footnote{Note that the tensor sector is affected also from $\alphaM$ through additional friction.}. The time kinetic term for the gravitons comes form the extrinsic curvature $K_{ij}$, while the spatial part is encoded in the intrinsic curvature $R$. Detuning the two from the GR relation can lead to a propagation speed for the gravitons different from that of light. 
The simplest example where this happens is the quartic Galileon~\cite{Deffayet:2011gz}, given by the Lagrangian $L_4$ in Eqn.~\eqref{Hornd}. 
Assuming for simplicity that $G_4$ is a function of $X$ only, it is useful to write explicitly the corresponding expression in unitary gauge to explicitly see the detuning:
\be
L_4=G_4 R+(2X G_{4,X}-G_4)(K^2-K_{ij}K^{ij})\, .
\ee
This gives non-vanishing functions $\alphaK$, $\alphaB$, $\alphaM$ and 
\be
\alphaT=-2\frac{G_{4,X}}{G_4+2G_{4,X}}\, .
\ee

\item \emph{Kinetic mixing with matter} $ \alphaH$. Besides a mixing between the gravitational scalar degree of freedom and $\phi$, when dealing with the late universe we should also take into account the presence of matter. The presence of the function $\alphaH$ leads to a situation where the propagating scalar modes in the presence of matter are mixed states of the latter and $\phi$.\footnote{This can be interpreted also as a particular type of \emph{disformal} coupling to matter in the frame where the matter fields are non minimally coupled. The mixing is however a physical effect independent of the frame. I will analyse this in detail in Sec.~\ref{KMMsec} } This effect arises in the theories ``beyond Horndeski''. 
An explicit example is given by the Lagrangian $L_4^{\rm bH}$ in Eqn.~\eqref{L4bH}.
This generates non vanishing $\alphaK$, $\alphaB$, $\alphaT$, $\alphaM$, as well as
\be
\alphaH=\alphaT=\frac{F_4}{1-F_4} \, .
\ee

\end{itemize}
The above five free functions and the corresponding operators do not explicitly introduce time derivatives of the lapse function in the action~\eqref{SBAction0} nor spatial derivatives of order higher than two.
 For the remaining functions, both of these two cases are in general realised and we need to impose degeneracy conditions to avoid instabilities.
The additional free functions are the following:
\begin{itemize}

\item \emph{Lorentz breaking} $\aL$. It corresponds to a detuning of the extrinsic curvature terms. Its presence is reminiscent of the fact that the two terms $K_{ij}K^{ij}$ and $K^2$ are \emph{separately} invariant under space diffs, while a full time and space diff invariance would require $\aL=0$. As such, this function is typical of theories that already in their original formulation assume a preferred time slicing, such as Horava gravity~\cite{Horava:2009uw} and its extensions~\cite{Blas:2010hb}\footnote{In the context of covariant theories, it can appear only together with other operators that would ensure full diff invariance to the action. }. For example, Horava's model in the low energy limit has the Lagrangian:\footnote{This model however includes a scalar mode that features instability and strong coupling problems~\cite{Blas:2009yd}.}
\be
L=\frac{M_{\rm Pl}^2}{2} \left[ R+K_{ij}K^{ij}-\lambda K^2\right] \, ,
\ee
which gives 
$\aL=3(\lambda-1)/2$.
In Lorentz-breaking theories, it gives rise to nonlinear dispertion relations of the form $\omega^2=A k^2+B k^4$. In covariant theories, I will show that the conditions to ensure the absence of additional degrees of freedom at the covariant level prevents to obtain a dispertion relation of this form. 

\item \emph{Acceleration} $\btrois$. This coefficient comes from the operator that can be built with the acceleration vector $a_i=\partial_i N /N$ at the nonlinear level. It also provides extra spatial derivatives to the action and it appears, for example, in healthy extensions of Horava gravity~\cite{Blas:2009qj}. For example, the healthy extension of Horava's ``non-projectable'' model has the Lagrangian

\be
L=\frac{M_{\rm Pl}^2}{2} \left[ R+K_{ij}K^{ij}-\lambda K^2-\alpha \frac{D_i N\,D^i N}{N^2}\right] \, ,
\ee
which simply gives
\be
\aL=3(\lambda-1)/2\, , \qquad \btrois = \alpha \; .
\ee

\item \emph{Phantom kineticity} $\bdeux$ . This function parametrises the pure kinetic term for the additional degree of freedom that appears allowing for time derivatives of the lapse function. As such, it is the analogue of the kineticity $\alphaK$.

\item \emph{Phantom kinetic mixing} $\bun$. When this function is non vanishing, the propagating scalar mode is a mixing of the metric perturbations and the lapse perturbations. Thus, it represents a generalisation of the kinetic braiding to the additional degree of freedom in higher-order theories.

\end{itemize} 
When allowing the functions $\bun$, $\bdeux$ to be nonzero, we are adding a propagating mode already at the level of the linear action in unitary gauge, while adding $\aL$ and $\btrois$ induces higher spatial derivatives. To eliminate higher derivatives and/or the additional propagating degree of freedom, the four former functions should obey degeneracy conditions. In particular, there are no viable theories where  $\bun$ and $\bdeux$ can enter separately (see discussion in Sec.~\ref{HOThDOFs}).



\chapter{Field redefinitions and coupling to matter}\label{chap:EFToDE_m}
\lhead{\emph{Field redefinitions and coupling to matter}}


In the late universe, the action~\eqref{SBAction0} should be supplemented by an action describing the matter sector.
This is relevant for the effective description: in general, there can be some arbitrariness in the choice of the metric used to describe the gravitational sector. Indeed, we are always allowed to perform ``field redefinitions'' such that the structure of the action remains unchanged. 
In the presence of matter, we have to take into account that the coupling between the matter fields and gravity changes as well. Suppose we start with a
 \emph{minimal} coupling of the matter fields to the metric, which simply amounts to choose the same metric $\gmn$ to describe the matter and the gravitational sector. In this case, test particles follow by definition the geodesics of the ``gravitational'' metric. After a field redefinition in the gravitational sector $\tgmn=\tgmn(\gmn)$, however, these geodesics will be those of a metric ($\gmn$) \emph{different} than the one used to describe the gravitational sector ($\tgmn$). Our description of the physics in the two frames would therefore be different.
In this Chapter, I will discuss how to include the coupling to matter in the effective theory introduced in the previous Chapter.
\section{Physics in different frames}
As an illustrative example, let me consider the simple case of Brans-Dicke theories~\eqref{BD} with matter minimally coupled to the metric $\gmn$,
\be
\label{ST}
S=S_{\rm BD}\left[\gmn, \phi\right]+S_{\rm m}\left[ \gmn, \Psi\right]\,.
\ee
Here, $\Psi$ denotes the matter fields. In the frame defined by $\gmn$ there is no direct interaction between the scalar field and the matter fields. This is usually called the \emph{Jordan frame}. By construction, in this frame the matter energy-momentum tensor is conserved, $\nabla_{\mu} T^{\mu\nu}=0$. On the other hand, the dynamical equations for the metric has a form different form the usual Einstein equations. Schematically, we can write
\be
 G_{\mu\nu}+\Delta G_{\mu\nu}=8\pi G T^{\rm (m)}_{\mu\nu}\, ,
\ee  
where $\Delta G_{\mu\nu}$ encodes the modification to the Einstein equations due to the presence of the scalar field.
It is well known that one can make a conformal transformation that depends on $\phi$,
\be\label{confBD}\tgmn=\phi \gmn \, , \ee 
such that the new metric $\tgmn$ obeys the usual Einstein equations. 
Indeed, making a field redefinition $\phi \mapsto \psi(\phi)$ to canonically normalise $\phi$, we can re-cast the action in the form~\cite{Clifton:2011jh}
\be
\label{BDE}
S_{\rm BD}=\int d^4x \sqrt{-\tilde{g}}\, \, \Big[ {}^{(4)}\!\tilde{R}-\frac12\tilde{\nabla}_{\mu} \psi\tilde{\nabla}^{\mu} \psi +\tilde{V}(\psi)\Big]\,+S_{\rm m}\left[\phi^{-1}\tgmn, \Psi\right] \, .
\ee
By comparison with the gravitational action~\eqref{BD}, we see that its structure has been preserved by the transformation~\ref{confBD}. With this I mean that no operators different from those present in the original action are generated (in this case ${}^{(4)}\!R$, $\partial_{\mu}\phi\partial^{\mu}\phi$ and $V(\phi)$), but only their coefficients changed. \\
In the frame defined by $\tgmn$, the contribution of the scalar field is encoded in the energy-momentum tensor, as if $\psi$ was an ordinary matter field:
\be
\tilde{G}_{\mu\nu}=8\pi G \left(\tilde{T}^{\rm (m)}_{\mu\nu}+\tilde{T}^{\phi}_{\mu\nu}\right)\, .
\ee
Clearly, if it wasn't for the presence of matter, by inspection of the actions~\eqref{BD} and~\eqref{BDE} we would conclude that Brans-Dicke theories are equivalent to General Relativity with a scalar field in a potential. 
However, since the metric that couples to the matter fields also transforms, this is not the case.
Note that in the second frame the covariant conservation of the energy-momentum tensor with respect to the new metric will not hold separately for the contributions of matter and of the scalar field\footnote{In the case of a purely conformal coupling, radiation fields are an exception, since their action is conformally invariant. For these we shall need a more general version of the transformation that I will discuss later.}. We will rather find an interaction of the form
\be
\tilde{\nabla}_{\mu} \tilde{T}^{\mu\nu}_{\rm (m)}\propto \tilde{\nabla}^{\nu} \phi \, .
\ee
This second frame is known as the \emph{Einstein} frame. 
It is clear that due to the arbitrariness of the transformation~\eqref{confBD} we can find infinite reference frames among which the description of the physics can change. The two above stand out for the clean different interpretation: either we have matter following geodesics of the gravitational metric, which however is not descrided by the Einstein-Hilbert Lagrangian, or we have a gravitational sector described by the same equations as General Relativity but with matter interacting in a non trivial way with the metric. In this second case, the scalar mediates an additional force (``fifth force'').  
The advantage of using the Jordan frame to derive predictions is that only the gravitational sector is non-standard; thus, one does not need to care about modifications of non-gravitational forces, which would otherwise greatly complicate the analysis. In the following, I will adopt this strategy.
\section{Disformally related frames and non-minimal couplings}
More general actions would of course require more general transformations to play the role of conformal transformations in Brans-Dicke actions. Remarkably, transformations that preserve the structure of the action exist for all the classes of theories that I shall consider in this work. These are the so-called \emph{disformal transformations}~\cite{Bekenstein:1992pj}, that generalise~\eqref{confBD} and can be written in their most general form as~\cite{Bekenstein:1992pj}
\be\label{cdgen}
\tgmn=C(\phi,X)\gmn+D(\phi,X)\pmphi \pnphi \, .
\ee
There are two main differences with respect to a $\phi$-dependent conformal transformation. First, the new metric $\tgmn$ is now allowed to depend on the gradient of the scalar field, thus changing the lightcones\footnote{For this reason, also radiation fields are affected by a disformal transformation, contrarily to the case of a purely conformal one.}. Second, the functions $C$ and $D$ themselves can depend not only on the value of the field but also on the metric through the kinetic term $X$. Due to this second property, a dependence on $X$ of the functions  $C$ and $D$ can lead to the introduction of higher order derivatives. In unitary gauge, in particular, the dependence of $C$ on $X$ (thus on $N$) corresponds to the introduction of time derivatives of the lapse function in the theory. I will proceed by increasing complexity and consider three cases:
\begin{enumerate}

\item \emph{Horndeski}+\emph{$\phi$-dependent conformal/disformal transformation}~\cite{Bettoni:2013diz}. Horndeski theories are defined by the requirement that the equations of motion are at most second order. The transformation that preserves their structure should therefore not generate higher derivatives. This kind of transformation is given by a conformal/disformal transformation of the form~\eqref{cdgen} where the functions $C$ and $D$ depend on $\phi$ only:
\be\label{cdgenH}
\tgmn=C(\phi)\gmn+D(\phi)\pmphi \pnphi \, .
\ee

\item \emph{Beyond Horndeski}+\emph{$\phi$-dependent conformal}+\emph{$\phi$ and $X$-dependent disformal transformation}~\cite{Gleyzes:2014qga}. In the case of theories ``beyond Horndeski'', the covariant action can allow for higher order derivatives. This reflects into the fact that this class of theories is invariant under a more general class of transformations, where the disformal factor depends on $X$, while the conformal one - $C$ - does not:
\be\label{cdgenKMM}
\tgmn=C(\phi)\gmn+D(\phi,X)\pmphi \pnphi \, .
\ee
In unitary gauge, this corresponds to avoid the introduction of time derivatives of the lapse in the theory.

\item\emph{DHOST}+\emph{$\phi$ and X-dependent conformal/disformal transformation}~\cite{Achour:2016rkg}. The case of the most general healthy class of theories correspond to the transformation~\eqref{cdgen}, where both the functions are allowed to depend on $\phi$ and $X$.
The presence of time derivatives of the lapse in unitary gauge is not a problem, since they are already present in the original action. Crucially, the degeneracy conditions that have to be imposed to get rid of the extra degree of freedom are preserved by the transformation~\cite{Achour:2016rkg}-\citepubli{Langlois:2017mxy}. This means that all the classes of quadratic DHOST theories introduced in Sec.~\ref{sec:subDHOST} are preserved by conformal and disformal transformations\footnote{An explicit proof for the cubic case has not been done yet, but one can expect that it is the case.}. In particular, this means that every theory in class Ia is completely equivalent to a Horndeski+beyond Horndeski theory in vacuum. In the presence of matter, it is equivalent to a Horndeski+beyond Horndeski theory with matter exhibiting a coupling of the form~\eqref{cdgen}.
\end{enumerate}

Let me observe that for the theories above and their corresponding transformations, we can always define a \emph{Jordan frame} while due to the complexity of the gravitational sector in general we are not able to find an \emph{Einstein frame} where the gravitational Lagrangian reduces to the Einstein-Hilbert one as in scalar-tensor theories.

So far, I considered matter minimally coupled to the metric. In this case I showed that the description of the physics in two conformally-disformally related frames is not the same due to the presence of matter.
To establish a correspondence between two frames that are equivalent, one must allow for a non-minimal coupling of matter to the metric. 
For each of the above classes, we have the freedom to couple matter to the most general metric of the form~\eqref{cdgen} that preserves the structure of the class.
\subsection{Violations of equivalence principle and interacting dark energy}
A coupling of the form~\eqref{cdgen} preserves the Weak Equivalence Principle (WEP) if we assume that the functions $C(\phi,X)$, $D(\phi,X)$ are the same for all matter species. Indeed, to be even more general, we can relax this assumption and allow different species to have different conformal and disformal couplings. Of course, the universality of couplings is very well tested on Solar System scales for standard matter such as baryons and photons, as well as the weakness of fifth force effects on these species~\cite{Will:2014xja,Brax:2014vva}. On the contrary, on cosmological scales and for other species such as Cold Dark Matter or neutrinos the constraints are far less stringent~\cite{Ade:2015rim} and it is interesting to consider this possibility which can be included in the effective description. This allows in particular to include all models where dark energy and dark matter can interact, known as ``Interacting dark energy''. These are usually restricted to the case where the scalar field has a quintessence-like action, while in Chapter~\ref{chap:Pheno} I will consider theories belonging to the Horndeski class~\citepubli{Gleyzes:2015pma} and a conformal-disformal coupling that violates the WEP.  \\
\section{Counting parameters}\label{sec:CountPar}
Taking into account the redundancy associated to field redefinitions and the possibility that different species have distinct conformal/disformal couplings of the form~\eqref{cdgen} to the gravitational metric, let me summarise the kind of couplings allowed in different theories and the number of free functions needed to fully characterise the dynamics of linear perturbations in the gravitational and matter sectors. I assume that $\NS$ species are present, labelled by an index $I$, $I=1,...,\NS$. For each species $I$, I denote the corresponding metric by  $\tg^{(I)}_{\mu \nu}$ and I call this the Jordan frame metric associated with this species. 
For each species I also introduce the conformal-disformal coupling
\be
\begin{split}\label{disf_unit_I2_WEP}
\tg^{(I)}_{\mu \nu} =& C^{(\phi)}_I(\phi,X) g_{\mu \nu}  + D^{(\phi)}_I(\phi, X) \partial_\mu \phi \, \partial_\nu \phi \;.
\end{split}
\ee
($C_I^{(\phi)}>0$ in order to preserve the Lorentzian signature of the Jordan-frame metric of the species $I$.).
In unitary gauge, the metric in eq.~\eqref{disf_unit_I2_WEP}  reads
\be\begin{split}
\label{disf_unit_I}
&\tg^{(I)}_{\mu \nu} = C_I(t,N) g_{\mu \nu}  + D_I(t , N) \delta_\mu^0 \delta_\nu^0 \;, \\
&C_I(t,N) =  C_I^{(\phi)} \big( \phi( t) , -\dot \phi( t)^2/N^2 \big)\, , \quad D_I(t, N) =   \dot{ \phi}^2 (t) D_I^{(\phi)} \big( \phi( t) , -\dot \phi( t)^2/N^2 \big)\,.
\end{split}
\ee

To be concrete, I will assume an action to describe the matter sector. For simplicity, I assume that each matter species can be described by a perfect fluid with vanishing vorticity. One can write an action in terms of derivatively coupled scalar fields $ \sigma_I$ with Lagrangians~\cite{ArmendarizPicon:2000dh,ArmendarizPicon:2000ah,Boubekeur:2008kn}:
\be
\label{Lagmatkess}
L_I \Big(   \tg^{(I)}_{\mu \nu}, \sigma_I \Big) \equiv P_I(Y_I ) \;, \qquad Y_I \equiv  \tg_{(I)}^{\mu \nu} \partial_\mu \sigma_I \partial_\nu \sigma_I \;.
\ee
The total action is given by
\be
S_{\rm m} = \sum_I^{N_S} S_I \;, \qquad S_I =  \int d^4 x \sqrt{- \tg^{(I)} }\,  L_I \Big(   \tg^{(I)}_{\mu \nu}, \sigma_I \Big) \; .
\ee

One can split  each scalar field {$\sigma_I$} into a background value and its perturbations, $\sigma_I = \bar{\sigma}_I (t) + \delta \sigma_I (t, \mathbf{ x}) $, and expand to second-order the action $S_{\rm m}$. The explicit calculation can be found in the case of Horndeski theories in Appendix B of ~\cite{Gleyzes:2015pma}, in the case of beyond Horndeski in Appendix A of~\cite{DAmico:2016ntq}, and in the case of DHOST in class Ia in Appendix D of~\cite{Langlois:2017mxy} (in the latter case, for a single matter species only).

What we found is that, for each species, in the most general case we can fully characterise the coupling of the matter sector at the level of linear perturbations by four functions. Two of them characterise the dependence on $D$ and $C$ on the scalar field and were introduced in~\citepubli{Gleyzes:2015pma}. The remaining two characterise the dependence of $D$ and $C$ on the gradient of the scalar field and were introduced in~\citepubli{DAmico:2016ntq} and~\citepubli{Langlois:2017mxy} respectively. Their explicit definitions are:
\be
\label{defalphasM}
\begin{split}
&\alphaCI \equiv \frac{\dot \phi}{2 H C_I } \frac{\partial C_I}{\partial \phi} \; , \quad \alphaYI \equiv - \frac{X}{C_I} \frac{\partial C_I}{\partial X} \, , \quad  \alphaDI \equiv - \frac{D_I}{ D_I+ C_I/X}\, , \quad \alphaXI \equiv - \frac{X^2}{ C_I } \frac{\partial D_I}{\partial X}\; .
 \end{split}
\ee

As I explained before, one is always allowed to perform field redefinitions in both the gravitational and matter sectors. 
Both the scalar-tensor parameters in the action~\eqref{SBAction0} and the four matter parameters~\eqref{defalphasM} transform under a general conformal-disformal transformation of the form~\eqref{cdgen}. Analogously to the matter case~\eqref{disf_unit_I}, this transformation can be written in unitary gauge and parametrised by four functions of time~\citepubli{Gleyzes:2015pma,DAmico:2016ntq,Langlois:2017mxy}
\be
\label{defalphas2}
\begin{split}
&\alphaC \equiv \frac{\dot \phi}{2 H C } \frac{\partial C}{\partial \phi} \; , \quad \alphaY \equiv - \frac{X}{C} \frac{\partial C}{\partial X} \, , \quad  \alphaD \equiv - \frac{D}{ D+ C/X}\, , \quad \alphaX \equiv - \frac{X^2}{ C } \frac{\partial D}{\partial X}\; .
 \end{split}
\ee
This freedom to can be used to reduce the total number of the free functions of the theory. For example, in the most general case of DHOST theories, the above transformation has four free functions that can be chosen so to eliminate four among the functions~\eqref{defalphasM} in the matter sector and the $\alpha$ and $\beta$ functions in the gravitational sector. Let me consider again cases $1,2,3$ separately. Table~\ref{tabCD} summarises the different possibilities.
\begin{enumerate}
\item  \emph{Horndeski}+\emph{$\phi$-dependent conformal/disformal transformation}~\citepubli{Gleyzes:2015pma}.
Horndeski theories are characterised by four free functions in the gravitational sector, $\alphaK$, $\alphaB$, $\alphaM$, $\alphaT$. The coupling to matter adds two functions for each matter species ($2 \NS$) in the matter sector, $\alphaCI$ and $\alphaDI$. The structure of the action is invariant under a transformation with non vanishing $\alphaC$ and $\alphaD$. In summary, $4+2\NS-2=2(\NS+1)$ free functions fully describe this case. The explicit transformations of these functions under a change of frame can be found in ~\cite{Gleyzes:2015pma}.

\item \emph{Beyond Horndeski}+\emph{$\phi$-dependent conformal}+\emph{$\phi$ and $X$-dependent disformal transformation}~\citepubli{DAmico:2016ntq}. 
In this case, we need five free functions to describe the gravitational sector - $\alphaK$, $\alphaB$, $\alphaM$, $\alphaT$, $\alphaH$ - supplemented by three functions for each matter species ($3 \NS$) in the matter sector - $\alphaCI$, $\alphaDI$, $\alphaXI$. The structure of the action is invariant under a transformation with non vanishing $\alphaC$, $\alphaD$, $\alphaX$. We thus have a total of $5+3\NS-3$ free functions. The explicit transformations of these functions under a change of frame can be found in~\cite{DAmico:2016ntq}.

\item\emph{DHOST}+\emph{$\phi$ and X-dependent conformal/disformal transformation}~\citepubli{Langlois:2017mxy}.
In the most general case, we have nine free functions in the gravitational sector. Among these, $\alphaK$, $\alphaB$, $\alphaM$, $\alphaT$, $\alphaH$ are arbitrary. On the contrary, the remaining four, $\aL$, $\bun$, $\bdeux$, $\btrois$, are subject to three degeneracy constraints, as explained in Sec.~\ref{sec:subDHOST}. As for matter, we have four free functions for each matter species ($4 \NS$), $\alphaCI$, $\alphaDI$, $\alphaXI$, $\alphaYI$. The structure of the action is invariant under a transformation with non vanishing $\alphaC$, $\alphaD$, $\alphaX$, $\alphaY$. In total, $9-3+4\NS-4=2(1+2\NS)$ functions are free. The explicit transformations of these functions under a change of frame can be found in~\cite{Langlois:2017mxy}.

\end{enumerate}
In particular, one can use the arbitrariness in the choice of the gravitational metric $g_{\mu\nu}$ to choose one particular matter species, say $I_*$, to be minimally coupled to it, in which case we have $C^{(\phi)}_{I_*}=1$ and $D^{(\phi)}_{I_*}=0$. This defines the gravitational metric as its Jordan metric.
Observables and physically relevant combinations of the parameters such as the degeneracy conditions are left invariant under the transformations above.

\begin{table}[t] 
\centering
\small
\hspace*{-2cm}
\begin{tabular}{|c||c|c|c|c|c|}
  \hline
  Theory & Brans-Dicke & Horndeski  & Beyond Horndeski & DHOST \\
 \hline\hline
\vtop{\hbox{\strut Free functions,}\hbox{\strut gravitational sector}} & $\alphaK$, $\alphaM=2\alphaB$ & $\alphaK$, $\alphaB$, $\alphaM$, $\alphaT$  & $\alphaK$, $\alphaB$, $\alphaM$, $\alphaT$, $\alphaH$ & \vtop{\hbox{\strut $\alphaK$, $\alphaB$, $\alphaM$, $\alphaT$, $\alphaH$, }\hbox{\strut one among \{$\aL$, $\bdeux$, $\bun$, $\btrois$\}}}  
\\
\hline
Coupling& $\tgmn=C(\phi)\gmn$&\vtop{\hbox{\strut $\tgmn=C(\phi)\gmn$}\hbox{\strut $+D(\phi)\pmphi \pnphi$}}  & \vtop{\hbox{\strut $\tgmn=C(\phi)\gmn$}\hbox{\strut $+D(\phi,X)\pmphi \pnphi$}}  & \vtop{\hbox{\strut $\tgmn=C(\phi,X)\gmn$}\hbox{\strut $+D(\phi,X)\pmphi \pnphi$}} 
\\
  \hline
\vtop{\hbox{\strut Free functions,}\hbox{\strut matter sector}}& $\alphaCI$ &$ \alphaCI, \alphaDI$   &$ \alphaCI, \alphaDI, \alphaXI$    & $ \alphaCI, \alphaDI, \alphaXI, \alphaYI$
\\
\hline
 \# of free parameters &  $1+2\NS$& $2(\NS+1)$  &  $2+3\NS$& $2(1+2\NS)$
\\
\hline
  \end{tabular}
  \hspace*{-2cm}
\caption{Transformations that preserve the structure of different classes of theories, the corresponding possible non-minimal couplings with matter and the number of physically relevant free functions.}
\label{tabCD}
\end{table}

\renewcommand{\arraystretch}{1.4}


\chapter{Propagating degrees of freedom and stability}\label{chap:EFToDE_lin}
\lhead{\emph{Propagating degrees of freedom and stability}}

So far, I showed how to describe within the effective formalism the gravitational sector of scalar-tensor theories and how to add very general couplings to the matter fields. Let me now proceed to analyse the behaviour of the propagating degrees of freedom. 

This analysis reveals one of the main advantages of an effective description based on an action. Indeed, even if a theory contains the expected number of dynamical fields, one should check that the propagating degrees of freedom comply with some basic physical principles. In particular, one should require that the theory is stable. In this Chapter, I recall first the stability conditions to impose on the action of the propagating degrees of freedom, and then proceed to discuss perturbations in scalar-tensor theories. 

I will concentrate on the scalar sector, but it is relevant to show the effect of  modifications of gravity also on the tensorial sector, which I will recall first. In particular, this will lead to put stringent constraints on DHOST theories.

\section{Ghosts and gradient instabilities}
Here I recall the conditions to be imposed to have a healthy theory and give a physical intuition. In Sec.~\ref{sec:OstroTh} I described the pathology associated to the presence of higher derivatives in the Lagrangian. Here I discuss the case of second-order theories. Consider a scalar field described by the Lagrangian density\footnote{In general, we could add a mass term and consider the case of negative mass, often called a \emph{tachyonic instability}. In gravity, however, under very general conditions the scalar perturbation is massless~\cite{Weinberg:2003sw}, and so will be in the present work, as I will show.\label{zetaCOns}}
\be
\mathcal L=\frac12 \left[\mathcal A \dot \varphi^2-\mathcal B {(\partial_i \varphi)}^2 \right] \, .
\ee
In terms of the conjugate momentum 
\be
\pi=\frac{\partial \mathcal{L}}{\partial \dot \varphi}=\mathcal{A} \dot \varphi \, ,
\ee
the Hamiltonian density is
\be
\mathcal{H}=\frac12\left[{\mathcal A}^{-1} \pi^2+\mathcal B {(\partial_i \varphi)} ^2\right]\, .
\ee
This is bounded from below if $sign({\mathcal A})=sign({\mathcal B})=1$ , bounded from above if $sign({\mathcal A})=sign({\mathcal B})=-1$ and indefinite if $sign({\mathcal A})\neq sign({\mathcal B})$. A Fourier mode obeys the equation of motion
\be\label{eomphi}
\ddot \varphi= -c_{s}^2 k^2 \varphi  \, , \qquad c_s^2\equiv\frac{\mathcal B}{\mathcal A}\, .
\ee
So, in the latter case where the Hamiltonian is indefinite, $c_s^2\leq0$, this mode admits an exponentially growing solution, $\varphi\propto e^{|c_s|k t}$, so this case must be discarded. This case is called a \emph{gradient instability}.
At the classical level, the other two cases are equivalent, since they lead to the same equations of motion~\eqref{eomphi} with a stable oscillatory solution with $c_s^2\geq0$
\footnote{At the quantum level the situation is different. It can be shown that in the case $\mathcal A \leq 0$, during the quantisation procedure we are forced to choose between violation of unitarity or propagation of negative energies forward in time~\cite{Cline:2003gs}. Since the first picture is unviable, we have to admit particles with negative energies in the spectrum, destabilising the vacuum that can quickly decay in states of positive and negative energy.\label{fnQuantum}}.

However, it can be shown that a field with $\mathcal A \leq 0$ is pathological at the classical level. This require to consider interactions with other fields. In gravity, this will always be the case, since the fluctuations of the scalar $\phi$ are coupled to the metric sector and to matter. This is also the main difference with an Ostrogradsky ghost, whose typical signature is a linear dependence of the Hamiltonian on one of the conjugate momenta and an instability will necessarily show up.
Let's couple the field $\varphi$ to another healthy scalar $\chi$~\cite{Carroll:2003st},  
\be\label{oscInt}
\mathcal L=\frac12 \left[\mathcal A \dot \varphi^2-\mathcal B {(\partial_i \varphi)}^2 +\dot {\chi }^2- c_{\chi}^2{(\partial_i \chi)}^2\right]+\lambda \varphi^2 \chi^2 \, .
\ee
When $\mathcal A >0$, since the total energy $E=E_{\varphi}+E_{\chi}$ is conserved and $E_{\varphi}\geq0$, $E_{\chi}\geq0$, the classical phase space for each of the two oscillators is bounded. 
On the contrary, if $\mathcal A \leq0$ the classical phase space is not bounded by the requirement that the total energy is conserved: if $E_{\varphi}\leq0$, similarly to the case of the Ostrogradsky ghost, a configuration can exist where $\varphi$ is arbitrarily excited towards negative energies as far as $\chi$ compensates this with an equal excitement towards positive energies.
If $\lambda=0$, the energy is separately conserved and this state can't be reached. As soon as we switch on interactions we can however reach the configuration with arbitrarily large excitations and constant total energy.

To summarise, if $\mathcal A \leq0$, the field $\varphi$ is called a \emph{ghost} field and such an arbitrarily excited state can appear in the spectrum leading to instabilities. In the following, I will impose the requirement of not having a ghost in the theory. Then the exponentially growing solution of~\eqref{eomphi} is avoided imposing also $\mathcal B\geq0$. If this condition is not realised, one has a \emph{gradient instability} in the theory.

\section{Tensor modes}


Tensor modes correspond to perturbations of the spatial metric, defined as (focussing only on the tensorial part)
\be\label{tenspert}
h_{ij} = a^2(t) \left(\delta_{ij} + \gamma_{ij}\right) \;,
\ee
with $\gamma_{ij}$ traceless and divergence-free,  $\gamma_{ii}=0 = \partial_i \gamma_{ij}$. 
Using these properties and the expansion~\eqref{tenspert}, one has
\be
\begin{split}\label{KRgravitons}
&\d K^i_j=\frac{1}{2\bar N}\dot\gamma^i_{\, j} \, ,\\
&\d_2 R=\frac{1}{a^2}\left(\gamma^{ij}\partial^2 \gamma_{ij}+\frac34 \partial_k \gamma_{ij}\partial^k \gamma^{ij}-\frac12 \partial_k \gamma_{ij} \partial^j \gamma^{ik}\right)\;.
\end{split}
\ee
The quadratic action for tensor perturbations is
\be\label{gravitonAc}
S_{\gamma}^{(2)} =\int d^3x dt \, a^3  \frac{M^2}{8} \left[  \dot{\gamma}_{ij}^2 -\frac{c_T^2}{a^2}(\partial_k \gamma_{ij})^2 \right] \;, \quad c_T^2\equiv 1+\alpha_T \, .
\ee
Absence of ghosts and gradient instabilities gives the two conditions
\be
M^2\geq0\, ,\qquad c_T^2\equiv 1+\alpha_T \geq0 \, .
\ee
As I anticipated in Sec.~\ref{sec:AlphaInterpr}, the action~\eqref{gravitonAc} shows that the function $\alphaT$ parametrises deviations of the speed of propagation of gravitons from the speed of light. Note also that the presence of the time-dependent Planck mass $M(t)$ provides an extra friction term in the equations of motion, given by $\alphaM$. Explicitly,
\be
\ddot \gamma_{ij} + H(3+\alpha_M) \dot \gamma_{ij} - (1+\alpha_T) \frac{\nabla^2}{a^2} \gamma_{ij} = \frac{2}{M^2} 
\left(T_{ij} -  \frac{1}3 T \delta_{ij} \right)^{TT} \;,
\ee
where $(T_{ij} -  T \delta_{ij}/3)^{TT}$ is the transverse-traceless projection of the anisotropic matter stress tensor.

\section{Scalar modes}
From now on I shall discuss the scalar sector. I will proceed by considering each of the three cases introduced in the previous Chapter separately.
Le me consider the action~\eqref{SBAction0}. In unitary gauge
the scalar modes can be described 
by the 
metric
perturbations  
\cite{Maldacena:2002vr}
\be
N=1+\delta N, \quad N^i=\delta^{ij}\partial_j \psi\,, \quad {h}_{ij}=a^2(t) e^{2\zeta}\,\delta_{ij}\,. \label{metric_ADM_pert}
\ee
This gives:
\be
\label{dhdK}
\begin{split}
&\delta \sqrt{h} = 3 a^3 \zeta\,, \qquad  \delta K^i_{\ j}=\left(\dot\zeta-H\delta N\right)\delta^i_j-\frac{1}{a^2}\delta^{ik}\partial_{k}\partial_j\psi \;, \\
&\delta_1 R_{ij}
=  - \delta_{ij} \partial^2 \zeta -  \partial_i \partial_j \zeta \;, \qquad \delta_2 R=  -\frac{2}{a^2}\left[(\partial\zeta)^2-4\zeta\partial^2\zeta\right]\,.
\end{split}
\ee

In this case, the situation is more involved than for tensors. 
Using the above expressions, one obtains a lengthy Lagrangian in terms of three scalar fields $\delta N$, $\psi$ and $\zeta$. Eventually, since I'm discussing theories with a single scalar degree of freedom, only one of those fields will be dynamical. This means that the other two satisfy constraint equations that can be used to eliminate them from the action. 

Let me start from the case of Horndeski and beyond Horndeski theories. Here, the Lagrangian does not depend on the time derivatives of the lapse and of the shift. The only dynamical variable is the perturbation $\zeta$ which is contained into the spatial metric $h_{ij}$. Hence, variation of the action with respect to the two fields $\delta N$ and $\psi$ yields constraint equations that correspond to the Hamiltonian constraint and to the scalar part of the momentum constraint. In particular, the latter can be used to replace $\delta N$ in terms of $\dot \zeta$ and the matter fields. The detailed calculation can be found in the Appendices for different cases; later in the text I will point to the references in more detail. As an example, let me recall the constraint equation in absence of matter. This reads
\begin{equation}\label{constrH}
\delta N=\frac{\dot \zeta}{H \left(1+\alphaB\right)} \, .
\end{equation}
When $\alphaB=0$, one has the standard GR expression. When $\alphaB\neq0$, upon use of the constraint the term $\delta N \delta K $ in the action~\eqref{SBAction0} gives a contribution proportional to $\dot{\zeta}^2$, as can be seen using the explicit expression~\eqref{dhdK}. So, the mixing between the gravitational and scalar fluctuations contributes in this case to the kinetic energy of the scalar degree of freedom, whence the name \emph{kinetic braiding}.
Once the constraint is used, one ends with an action for the scalar degree of freedom $\zeta$ and the matter fields only. It is on this action that the stability conditions must be imposed.

In the case of DHOST theories, the situation is more subtle. Time derivatives of the lapse function are present, which means that in principle two degrees of freedom could propagate in the scalar sector and that only one constraint is present in the action.
On the other hand, we know that one of the two propagating degrees of freedom is reminiscent of an Ostrogradsky ghost, and that we can impose degeneracy conditions to avoid its presence. Once we do so, I will show that one can find another constraint equation and find again an action for a single degree of freedom.

\section{Horndeski theories}\label{sec:HorndDOF}
In this case,  $\alphaH=\aL=\bun=\bdeux=\btrois=0$. As I explained, we can use the scalar part of the momentum constraint, Eqn.~\eqref{constrH}, to eliminate $\delta N$ in favour of $\zeta$. 
In absence of matter, we get\footnote{As anticipated in footnote~\ref{zetaCOns}, the action for $\zeta$ doesn't have a mass term. }
\be
\label{actionzetaCI}
S^{(2)}= \int d^3 x\,  dt\,  a^3   \frac{M^2 }{2} \frac{\alphaK+6\alphaB^2}{(1+\alphaB)^2}\bigg[     \dot{\zeta}^2 - c_{s,0}^2  \frac{(\partial_i \zeta)^2}{a^2}  \bigg] \;.
\ee
where\footnote{A subscript $0$ will always denote a quantity defined in absence of matter.}
\be
c_{s,0}^2 =\frac{(1+\alphaB)^2}{\alphaK+6\alphaB^2}\bigg\{2 (1+\alphaT)-\frac{2}{a M^2 }\frac{d}{dt}\bigg[\frac{a M^2}{H(1+\alphaB)}\bigg] \bigg\} \, .
\ee
Absence of ghost and gradient instabilities require respectively
\be \label{stabHorndNoM}
\alphaK+6\alphaB^2\geq0 \, , \qquad c_{s,0}^2 \geq 0 \, .
\ee
The above action illustrates how the operators in the action~\eqref{SBAction0} contribute to the scalar dynamics. In particular, it is clear that the kineticity $\alphaK$ and the kinetic braiding $\alphaB$ give a kinetic energy to this degree of freedom. \\
Let me generalise the previous case adding a coupling to the matter fields as described in~\ref{chap:EFToDE_m}. The matter sector is described by the $\NS$ funtions $\alphaCI,\alphaDI$, $I=1,...,\NS$ introduced in~\eqref{defalphasM}. Combining the quadratic action for matter with eq.~\eqref{actionzetaCI}, one can extract a quadratic action that governs the dynamics of the gravitational scalar degree of freedom and the matter ones. One has to solve the constraint that will now depend on the matter fields as well.
The explicit calculation can be found in Appendix B of ~\cite{Gleyzes:2015pma}. The absence of ghosts is guaranteed by the positivity of the matrix in front of the kinetic terms. This condition is given by
\be
\alpha \equiv \alphaK + 6 \alphaB^2 +3 \sum_I \alphaDI \, \Omega_I \geq0\;, \qquad \rho_I+(1+\alphaDI)p_I \geq 0 \, ,
\label{alpha_def}
\ee
where $\rho_I$ and $p_I$ are the energy density and pressure respectively, and $\Omega_I\equiv \rho_I/(3M^2H^2)$ is the density contrast.
The first condition generalises the first inequality in~\eqref{stabHorndNoM}. We see that a \emph{disformal} coupling to the matter fields, parametrised by $\alpha_{D,I}$, affects the kinetic energy of the gravitational degree of freedom and the no-ghost condition (see also \cite{Bruneton:2007si,Zumalacarregui:2012us}). For the matter sector,  the second condition in~\eqref{alpha_def} corresponds to the Null Energy Condition \cite{Hawking:1973uf} in the frame of $g_{\mu \nu}$: 
in the Jordan frame of each species $I$, this can be expressed in terms of the energy density and pressure by $\check{\rho}_I+\check p_I \geq 0 $ (the  symbol $\check{}$ denotes Jordan-frame quantities). The explicit transformations between the two frames can be found in Appendix A of ~\cite{Gleyzes:2015pma}.\\
The propagating degrees of freedom are the matter ones, with sound speeds squared $c_{s,I}^2$, and the gravitational one, with sound speed
\be\label{csHornd}
c_{s}^2=\frac{\alphaK+6\alphaB^2}{\alpha}c_{s,0}^2-\frac{3}{\alpha}
\sum_I  \Big[ 1+ (1+\alphaDI) w_I \Big] \Omega_I \, ,\ee
where $w_I\equiv p_I/\rho_I$ is the equation of state.
Absence of gradient instabilities requires
\be
c_{s}^2\geq0\, ,\qquad \check c_{s,I}^2=c_{s,I}^2(1+\alphaDI)\geq0 \, .
\ee
To summarise, in the case of Horndeski theories with a conformal-disformal coupling that depends on $\phi$ only, the propagating degrees of freedom of the scalar/gravitational sector remain decoupled from matter, but the presence of matter fields alters the kinetic energy of the former one and its sound speed. For matter, stability conditions and fluid quantities take their standard form in the Jordan frame.

\section{Beyond Horndeski and Kinetic Matter Mixing}\label{KMMsec}
In theories beyond Horndeski the operator $\alphaH$ is added on top of the four others characterising the Horndeski class. The analysis of the propagating mode proceeds exactly as in the Horndeski case.
In absence of matter, the physics is not qualitatively different from the Horndeski case; the no-ghost conditions are not affected by $\alphaH$ while the expression of the sound speed is slightly modified\footnote{It has been recently argued that even in vacuum, this can have relevant effects in the context of spatially flat FRW solutions which are geodesically complete without facing gradient instabilities~\cite{Creminelli:2016zwa,Cai:2017dyi}.},
\be\label{bareSSBH}
c_{s,0}^2 =\frac{(1+\alphaB)^2}{\alphaK+6\alphaB^2}\bigg\{2 (1+\alphaT)-\frac{2}{a M^2 }\frac{d}{dt}\bigg[\frac{a M^2(1+\alphaH)}{H(1+\alphaB)}\bigg] \bigg\} \, .
\ee
A genuinely new physical effect emerges in this case when matter is added. In this case, the transformation preserving the structure of the action includes a dependence of the disformal factor $D$ on the gradient of the field, so we can extend the coupling to matter to include this case~\citepubli{DAmico:2016ntq}. This adds one function $\alphaXI$ (see eqn.~\eqref{defalphasM}) for each matter species; here, I will restrict to the case where only one species is present, and a subscript $\rm m$ will denote the matter quantities. A generalisation to the case where multiple matter species are present is discussed in Appendix A of~\cite{DAmico:2016ntq}.
The no-ghost conditions are affected by the dependence of the disformal factor on $X$: we get
\be
\alpha \equiv \alphaK + 6 \alphaB^2 +3 \alphaSm \, \Omega_{\rm m} \geq0\; , \label{alpha_def_bH}
\ee
\be
\label{alphaSI}
\alphaSm \equiv \alphaDm(1+\alphaXm)^2+\alphaXm(2+\alphaXm)+ \frac{1}{2 C_{\rm m}} \frac{\partial^2  D_{\rm m}}{\partial N^2} \;.
\ee
The qualitatively new phenomenon emerges when considering the propagating degrees of freedom. Requiring that the determinant of the kinetic matrix vanishes, we get a dispersion relation of the form
\begin{equation}
( \omega^2 -  c_s^2 k^2) (\omega^2 - c_{\rm m}^2 k^2)  =  \lambda^2 c_s^2    \, \omega^2 k^2    \, , \label{km}
\end{equation}
where 
\be\label{SSBH}
c_{s}^2  \equiv \frac{ \alphaK + 6 \alphaB^2}{\alpha }c_{s,0}^2 -\frac{3  \big[1+w_{\rm m}(1+\alphaDm)\big] \Omega_{\rm m}  } {\alpha}\, \big(1+\alphaH\big)^2 \; ,
\ee
and the parameter $\lambda^2$ on the right-hand side is defined as
\be
\lambda^2 \equiv \frac{3}{\alpha c_s^2}     \Big[1+(1+\alphaDm)w_{\rm m} \Big] \Omega_{\rm m} \, (\alphaH-\alphaXm)^2  \; .
\label{gpar}
\ee
When $\lambda\neq0$, the two non-trivial solutions of the system are \emph{not} given by the scalar and matter degrees of freedom, $\omega^2=k^2 c_s^2$ and $\omega^2=k^2 c_{\rm m}^2$. They are rather mixed states of matter and the scalar propagating at speeds $c_{\pm}^2$ that can be found solving~\eqref{km}. These two must satisfy the stability conditions
\be \label{cpmgeq0}
c_{\pm}^2\geq0\, .
\ee
Thus, in this case the scalar affects also the sound speed of matter. Since the latter is defined as $\delta p_{\rm m}=c_{\rm m} \delta \rho_{\rm m}$, we can think of it as an additional source of pressure. This feature is particularly surprising if one thinks that in general we are able to decouple the gravitational sector from matter by going at sufficiently short distances - this is the Jeans phenomenon. 
 As can be seen from the dispersion relation~\eqref{km}, the amount of mixing is quantified by the parameter $\lambda$, given by a combination of the beyond-Horndeski function $\alphaH$ and the $X$-dependent part of the disformal coupling $\alphaXm$. This feature is physically very interesting for the interpretation of the effect.
Let me recall explicitly the transformation of the relevant parameters under a change of the metric of the form~\eqref{cdgenKMM} (the complete transformation of the other functions can be found in Sec. (2.3) of~\cite{DAmico:2016ntq}):
\be
\label{alphaDC_change}
\begin{split}
\tilde{\alpha}_{\rm T}&=(1+\alphaT)(1+\alphaD)-1\; , \\
\tilde{\alpha}_{\rm H}&= \frac{\alphaH - \alphaX}{1+\alphaX} \;,\\
\tilde{\alpha}_{\text{D,m}} & = \frac{\alphaDm - \alphaD}{1+\alphaD} \;,  \\
\tilde{\alpha}_{\text{X,m}}  &= \frac{\alphaXm - \alphaX}{1+\alphaX} \; .
\end{split}
\ee
Remarkably, one can start with a theory where matter is disformally coupled and the beyond Horndesky parameter $\alphaH$ is absent, and find a transformation that sets $\alphaXm$ to zero and at the same time generates a non vanishing $\alphaH$ without changing the propagation speed of the gravitons. The inverse is also true: a non vanishing beyond Horndeski parameter can be eliminated generating an X-dependent disformal coupling. 
As an example, consider the disformal coupling of matter $D_{\rm m} = -(X+\dot {\bar \phi}^2(t) )/ \dot {\bar \phi}^2(t)$. 
In the  absence of a conformal coupling, this yields $\alpha_{\rm D,m}=0$ and $\alpha_{\rm X,m}=1$, since $\bar{X}=-\dot {\bar \phi}^2(t)$. Thus,  the transformation to Jordan frame leaves $\alphaT$ (and hence the speed of gravitons) unchanged.

Using the remaining transformation between the two frames, moreover, the parameter  $\lambda$ can be shown to be \emph{frame-independent }, thus probing that the kinetic mixing between matter and the scalar is a truly physical effect. I will call this \emph{Kinetic Matter Mixing} (KMM) and show that it has rather unique observational effects in Chapter~\ref{chap:KMM}.  In the Jordan frame, where the coupling is minimal, KMM is encoded in the beyond Horndeski parameter $\alphaH$. As we saw, we can also find a frame where $\alphaH$ is vanishing and matter has a disformal coupling to the metric that depends on the derivative of the field.

\section{Higher-order theories}\label{HOThDOFs}

Theories that further generalise the previous cases require a more thorough investigation. As we saw, the action with $\bun=\bdeux=\btrois=0$ does not explicitly include derivatives of the lapse perturbation $\delta N$, while setting also $\aL$ to zero prevents to have higher spatial derivatives of $\zeta$ in the final action. 
If we want to cover linear perturbations of DHOST theories, we have to introduce the four above operators altogether. In fact, one could start from the covariant formulation of DHOST, Eqn.~\eqref{action}, and work out the action for linear perturbations in unitary gauge. This is given by an expression of the form~\eqref{SBAction0}, and the explicit calculation can be found in Sec.~2.2 of~\cite{Langlois:2017mxy}.
The functions $\alpha$ and $\beta$ appearing in~\eqref{action} are given by combinations of the functions $a_A$, $f_2$ and $f_{2,X}$ evaluated on the FLRW background. Explicitly, for quadratic DHOST we have (the cubic case is discussed in Appendix A of~\cite{Langlois:2017mxy}):
\beq
\label{effective}
\begin{split}
\frac{M^2}{2}=&\, f_2-\aq_1 X\,, \qquad \frac{M^2}{2}(1+\alphaT)= f_2\,, \qquad  \frac{M^2}{2}(1+\alphaH)=f_2-2X f_{2X} \,,
\\
\frac{M^2}{2}\left(1+\frac23\aL\right)=&\, f_2+\aq_2 X\,,\qquad \frac{M^2}{2} \bdeux=-X \left(\aq_1+\aq_2+(\aq_3+\aq_4) X+\aq_5X^2\right)\,,
\\
2 M^2\bun=&\, X (4f_{2X}+2\aq_2+\aq_3X) \,, \quad \frac{M^2}{2} \beta_3=-X(4f_{2X}-2\aq_1-\aq_4 X)\,.
\end{split}
\eeq
The expressions above already allow to draw some conclusions about the viability of some classes of quadratic DHOST theories. 

\begin{itemize}
\item \emph{No propagating gravitons}. If $\aq_1=f_2/X$, one sees immediately that $M=0$. This time-dependent Planck mass is defined as the normalisation of the action for the gravitons, i.e. the coefficient of their kinetic term, as can be seen from~\eqref{SBAction0} and~\eqref{gravitonAc}. Thus the theory does not contain tensorial degrees of freedom and should be discarded. There are three classes of theories with this feature: Ib, IIb and IIIc.
\item \emph{No spatial gradient for the gravitons}. This is the case if $f_2=0$, since the spatial curvature $R$ disappears~\cite{deRham:2016wji} (see the action~\eqref{action}), and so does the gradient term for $\gamma_{ij}$, as can be seen from Eqn.~\eqref{KRgravitons}. This means that the propagation speed for gravitational waves is zero, or equivalently, $\aT=-1$. This happens in classes IIIa  and IIIb. Note that these also verify the  property $\aH=-1$.
\end{itemize}

Therefore from a phenomenological point of view, the remaining classes, Ia and IIa, appear to be the most interesting. In Sec.~\ref{sec:CountPar} I pointed out that theories in class Ia are equivalent to Horndeski+beyond Horndeski with matter conformally and disformally coupled. Theories in class IIa are instead a genuinely new class.

Let me now discuss the degeneracy conditions found in Sec.~\ref{sec:subDHOST} at the covariant level. These are translated into conditions on the functions $\alpha$ and $\beta$ through Eqn.~\eqref{effective}. 
We found that the 
fully nonlinear 
degeneracy conditions boil down to two sets of very simple conditions for the free functions $\alpha$ and $\beta$ appearing in the quadratic
perturbative
 action. 
Depending on the DHOST theory under consideration, these  satisfy either
\beq
\label{Ia}
\CI:\qquad \aL=0\,, \qquad \bdeux=-6\bun^2\,,\qquad   \btrois=-2\bun\left[2(1+\alphaH)+\bun (1+\alphaT)\right]\,,
\eeq
or the set of conditions
\beq
\label{IIa}
\CII:\qquad \bun=- (1+\aL)\frac{1+\alphaH}{1+\alphaT}\,, \quad \bdeux=-6(1+\aL) \frac{(1+\alphaH)^2 }{(1+\alphaT)^2}\,,\quad \beta_3=2\frac{(1+\alphaH)^2}{1+\alphaT}\,,
\eeq
where I assumed that $\aT\neq -1$ in the latter case (otherwise\footnote{As I already pointed out, a model for which $\aT=-1$ is very peculiar since the speed of gravitational waves vanishes.} one should use a regular version of the conditions obtained by multiplying both sides of the equalities by the denominator of the right hand side). In particular, theories in class Ia satisfy $\CI$ while theories in class IIa satisfy $\CII$.
It is immediate  to see that both sets of conditions share the common condition
\beq
\label{C_U}
\CU:\qquad (1+\aL)\bdeux=-6\bun^2\,,
\eeq
which plays a special role in the unitary gauge, as we will see later. I summarise the situation for the quadratic DHOST theories in Table \ref{table_DHOST}.

One can also recover directly the conditions $\CI$ and $\CII$ by rewriting the three degeneracy conditions~\eqref{D012} in terms of the seven parameters  $M^2$, $\aL$, $\aH$, $\aT$ and $\beta_A$ inverting the equations~\eqref{effective}, as we show in Appendix B of~\cite{Langlois:2017mxy}.
We also generalised the discussion presented in this section to DHOST theories up to cubic order. It can be found in Appendix A of~\cite{Langlois:2017mxy}.


\begin{table}
\centering
\begin{tabular}{ | c || c | c | c | }
 		\hline
 		 Subclass (see  \cite{BenAchour:2016fzp}) & $\#$ free functions & Degeneracy & Remarks\\
 		 \hline \hline
 		 \, {$^2$N-I}/Ia \, $$ & 3  &  I & H, bH \& conf-disf transf \\
 		 \, { $^2$N-II}/Ib \, $$ & 3  & 0 & \\
 		 \, { $^2$N-III}/IIa \, $$ & 3  & II & \\
 		 \, { $^2$N-IV}/IIb \, $$ & 3  &  0 & \\
 		 \hline \hline
 		 \, { $^2$M-I}/IIIa \, $$ & 3  & II & $\aT=\aH=-1$ \\
 		 \, {$^2$M-II}/IIIb \, $$ & 3  &  II & $\aL=\aT=\aH=-1$\\
 		 \, { $^2$M-III}/IIIc \, $$ & 4  & 0 &\\
 		 \hline
 	\end{tabular}
 \caption[Subclasses of DHOST theories]{Subclasses of DHOST theories, using the classification of Ref.~\cite{BenAchour:2016fzp}.  Second column: number of free functions among $f_2$, $\aq_A$. In the degeneracy column,  $0$ stands for $M^2=0$, i.e. there are no tensor modes.}
 \label{table_DHOST}
 \end{table}

\subsection{Propagating degrees of freedom on Minkowski space}\label{sec:MinkDHOST}
It is instructive to consider the Minkowski limit first, as it encodes all the relevant physical information that can be later generalised. In a cosmological context, this is equivalent to consider modes with frequencies and wave numbers much higher than the cosmological ones. In this case, all the functions $\alpha$'s and $\beta$'s, as well as $M^2$, are constants, while we can redefine the functions $\alphaK$ and $\alphaB$ by $\MsqK=H^2\alphaK$,  $\MB=H \alphaB$ and then take the limit $a=1$, $H=0$.
As in this case plane waves are eigenfunctions of the system, we can find a dispersion relation simply  considering perturbations of the form ${(N(t, {\bf x}),\zeta(t, {\bf x}),\psi(t, {\bf x}))}^{\dagger}=e^{-i\omega t+i{\bf k}\cdot{\bf x}} (N(\omega, {\bf k}),\zeta(\omega, {\bf k}),\psi(\omega, {\bf k}))^{\dagger}$, and requiring that the determinant of the resulting quadratic Lagrangian vanishes. This yields 
\be\label{dispMink}
\begin{split}
\omskz\,  \omega^4  +\big(\omfkt k^2+ \omfkz \big)\omega^2 + \omtkf k^4+ \omtkt k^2 =0\; ,
\end{split}
\ee
with the  coefficients
\be
\begin{split}
\omskz = & \ 3 \big[(1+\aL)  \bdeux+ 6 \bun^2\big]\; ,\\
\omfkt = & \  6  \big[ 2 (1+\alphaH)+  (1+\alphaT)\bun\big]\bun +\aL  (1+\aT) \bdeux+3 (1+\aL) \btrois\;, \\
\omfkz = &\    3 \big[  (1+\aL)\MsqK+6\MB^2 \big] \; ,\\
\omtkf = & - \aL \big[2 (1+\alphaH)^2 - (1+\alphaT) \btrois \big]\; ,\\
\omtkt = & \  (1+\alphaT) \left( \aL \MsqK+6 \MB^2 \right)\;.\\
\end{split}
\ee

In the general case, the dispersion relation is a quartic polynomial in $\omega$ with only even powers, 
which means that there are two solutions for $\omega^2$, corresponding to the presence of two scalar modes, as expected.
In particular, the two parameters $\bun$ and $\bdeux$ contribute to the highest order coefficient in $\omega$, which is consistent with their interpretation of a ``kinetic'' and ``braiding'' contribution I gave in Sec.~\ref{sec:AlphaInterpr}. Interestingly, 
the structure of the coefficient $\omskz$ is the same as that of $\omfkz$ with $\bun$ and $\bdeux$  playing the role of $\alphaB$ and $\alphaK$, respectively (reminding that $M_{\rm B} \equiv H \alphaB$ and $M_{\rm K}^2 \equiv H^2 \alphaK$).
Note also that the highest term in spatial derivatives disappears when $\aL=0$.
\\
If the condition $\omskz=0$ is satisfied, then only a single scalar mode remains. This amounts to impose the condition $\CU$ in Eqn.~\eqref{C_U}.
It is also instructive to look for cases where~\ref{dispMink} can be reduced to a standard linear dispersion relation of the form  $\omega^2 = c_s^2\,k^2$. 
This can be achieved by setting $\omfkt=0$ and $\omtkf =0$. 
Solving the above conditions, we obtain that they are equivalent to impose either $\CI$~\eqref{Ia} or $\CII$ \eqref{IIa}.
In both cases, 
the dispersion relation takes the  very simple form 
\be
\label{disprelMin}
\omega^2- c_s^2 k^2=0, \qquad c_s^2\equiv - \frac13 \frac {(1+\alphaT)(6 \MB^2 + \aL \MsqK)}{\MsqK (1+\aL)+6\MB^2}\, .
\ee
To summarise, we found that requiring to have a standard dispersion relation for one propagating mode in unitary gauge leads to impose the same degeneracy conditions found at the covariant level appropriately expressed in terms of the free functions of the effective description. However, the fact that a single scalar mode remains when $\CU$ is imposed is valid only for linear perturbations in unitary gauge; if this is not the case, one should expect the presence of an additional propagating mode that doesn't show up here. To ensure the the extra mode is absent at any level, one has to impose the full degeneracy conditions $\CI$ or $\CII$.


\subsection{Unitary gauge analysis in cosmology and gradient instablilities}\label{sec:CosmonoMatterDHOST}

Here I generalise the discussion to the cosmological case. The details of the calculations are in Sec.~4 and Appendix D of~\cite{Langlois:2017mxy}.
Differently from the Horndeski and beyond Horndeski cases, when $\aL\neq 0$, the action contains terms quadratic in $\partial \psi$,\footnote{I will discuss here this more general case; the case $\aL= 0$ can be obtained at the end taking the smooth limit $\aL \rightarrow 0$.} where $\psi$ has been defined in Eqn.~\eqref{metric_ADM_pert}, $N^i=\delta^{ij}\partial_j \psi$.
The scalar component of the momentum constraint becomes then a linear equation in $\partial \psi$ and we should use it to solve for $\psi$ (rather than $\delta N$ as in the case $\aL=0$). The remaining action in general describes two propagating degrees of freedom, with a kinetic part in the variables $(\dot \zeta$, $\dot{\delta N})$ described by the matrix 
\be
\mathcal{M} = \,
\begin{pmatrix}
6 (1+\aL)  & -6\bun \\
-6 \bun& 6\bun^2+\aL \bdeux\\
\end{pmatrix} \; .
\ee
If the above matrix has vanishing determinant, we can find a null eigenmode that is not a propagating degree of freedom. This amounts to impose the condition
\be
0 = \det [ {\mathcal M} ] = 36 \bun^2 +6 \bdeux \, (1+\aL) \;   \quad  \Rightarrow \quad \bdeux =   -\frac{ 6 \bun^2}{1+\aL} \quad (\CU) \, ,\label{deg_cond}
\ee
which, not surprisingly, is the same found from the dispersion relation on Minkowski~\eqref{C_U}.
The action is diagonalised by the transformation
\be\label{tzetadef}
\tzeta=\zeta-\frac{\bun}{1+\aL} \delta N \; ,
\ee
which represents the propagating degree of freedom in this case. Varying the action with respect to $\delta N$ yields now another constraint,
\be
\label{momaz}
\delta N=\frac{\dot \tzeta}{H(1+\alphaB)-\dot \bun} \; ,
\ee
which generalises equation~\eqref{constrH} and can be used to integrate out $\delta N$. After a spatial Fourier transform, the final action has the form ($\hat k\equiv k/a$):
\be \label{Sx}
\begin{split}
S=& \frac{1}{2(2\pi)^3}\int \,dt\,d^3k   \frac{a^3M^2}{\MMM_{22}+\hat{k}^2 \SSS_{22} }  \bigg[ \Big (\aone+\bone \hat{k}^2\Big)   \, \dot{\tzeta}_{\mathbf{k}} \, \dot{\tzeta}_{\mathbf{-k}} + \frac{\btwo \hat{k}^2+\ctwo \hat{k}^4+\dtwo \hat{k}^6}{\MMM_{22}+\hat{k}^2 \SSS_{22}}  \, \tzeta_{\mathbf{k}}\,\tzeta_{\mathbf{-k}} \bigg] \; .
\end{split}
\ee
The explicit expression for the coefficients is not important for the present discussion and can be found (including also the matter contributions) in Appendix D of~\cite{Langlois:2017mxy}. 

The above action describes a scalar field with a dispersion relation $\omega^2 = \omega^2(k^2)$ that is in general a  rational  function of $k^2$.
This generalises the case of flat space analysed in the previous section to the cosmological context.
Again, we can look for cases where the dispersion relation has the standard form $\omega^2 = c_s^2\,k^2$. As in the case of Minkowski, using the explicit form of the coefficients $c_{i,j}$, one finds two solutions corresponding to the cases $\CI$~\eqref{Ia} and $\CII$~\eqref{IIa}. In this case the action takes the usual form,
\be
\label{actionzetaCI}
S= \int d^3 x\,  dt\,  a^3   \frac{M^2 }{2} \bigg[  {\Az}_{\tzeta}   \dot{\tzeta}^2 +  {\Bz}_{\tzeta} \frac{(\partial_i \tzeta)^2}{a^2}  \bigg] \;.
\ee
The explicit form of the coefficients ${\Az}_{\tzeta}$, ${\Bz}_{\tzeta}$ can be found in Eqns.~(4.9)-(4.10) of~\cite{Langlois:2017mxy} for theories  $\CI$, and in Eqn.~(4.26) for theories $\CII$.
Absence of instabilities requires that the  coefficients in the action satisfy
\be\label{stabCII}
{\Az}_{\tzeta} \ge 0 \;, \qquad {\Bz}_{\tzeta} \le 0 \; .
\ee
A very important result follows from the above conditions for theories $\CII$~\eqref{IIa}. The explicit expression of $ {\Bz}_{\tzeta}$ reads
\be
B_{\tzeta} = 2 (1+\alphaT)\; \qquad (\CII) \;.
\ee
According to the stability condition~\eqref{stabCII} for the scalar mode, the above expression should be \emph{negative}. On the other hand, $(1+\alphaT)$ corresponds to the square of the propagation speed of gravitons $c_T^ 2$, defined in Eqn.~\eqref{gravitonAc}. This quantity should therefore be \emph{positive} to guarantee stability in the tensorial sector. It follows that \emph{theories satisfying the condition $\CII$ necessarily develop a linear gradient instability either in the scalar or in the tensor sector}. We can thus conclude that these theories are unviable.  

Let me finally point out another important result. The action~\ref{Sx} implies that $\tzeta$ is conserved in the long wavelength limit, i.e.~$\dot \tzeta \approx 0$ for $k\ll aH$. The constraint equation~\eqref{momaz} implies that $\delta N$ vanishes in the same limit. It follows from the definition~\eqref{tzetadef} that $\zeta$  is conserved on large scales, 
\be
\dot \zeta \approx 0 \qquad  (k \ll aH)  \;.
\ee

\subsection{Including matter}\label{sec:DHOSTmatter}

DHOST theories have their structure preserved by a conformal-disformal transformation of the form~\eqref{cdgen}, where both the conformal and disformal factors depend on the scalar field and its gradient. This is the third case considered in Sec.~\ref{sec:CountPar}. We can thus couple matter to a metric of the form~\eqref{cdgen}. 
Assuming no violations of the WEP, the matter sector is characterised by all the four functions~\eqref{defalphasM}, $\alphaCm$, $\alphaDm$, $\alphaXm$, $\alphaYm$.  
The calculation in this case proceeds exactly as in the case without matter that I summarised in Sec.~\ref{sec:CosmonoMatterDHOST}, but this time one has to take into account the matter fields as well. The explicit calculation can be found in Appendix D of~\cite{Langlois:2017mxy}. At the end, one gets an action analogous to~\eqref{Sx} (equation D.16 of~\cite{Langlois:2017mxy}) with coefficients that are in general ratios of polynomials in $k^2$, and only imposing the ``full'' degeneracy conditions $\CI$ and $\CII$ a local form of the coefficients is recovered. 

Let me comment on theories satisfying the conditions $\CI$. One gets a dispersion relation of the form~\eqref{km}, where the sound speed in presence and absence of matter appropriately generalise the ones introduced in Eqns.~\eqref{bareSSBH}-\eqref{SSBH} (see equations D.23-D.25 of~\cite{Langlois:2017mxy}). Hence we don't have qualitatively new physical phenomena; we find again a mixing between the scalar and matter propagating modes, with a frame-invariant parameter $\lambda$ quantifying such mixing, introduced in Eqn.~D.24 of~\cite{Langlois:2017mxy}. This result is in agreement with the fact that we can re-map theories satisfying $\CI$ into theories belonging to the Horndeski and beyond-Horndeski class with a conformal-disformal transformation.

As for theories satisfying the conditions $\CII$, the gradient instability found in Sec.~\ref{sec:CosmonoMatterDHOST} is not cured by the presence of matter.

Another very interesting result follows from the computation of the Poisson equation in the Newtonian limit. This is obtained proceeding as in Sec.~\ref{sec:MinkDHOST} and further taking the limit $\omega=0$. For completeness, one can add a test particle of mass $m$ which is minimally coupled to the metric. 
The $3\times3$ kinetic matrix for the variables $\d N$, $\zeta$ and $\psi$ yields three equations that can be combined to get a generalised Poisson equation. In terms of the gravitational potential $\Phi$, this reads (for details, see Sec.~3.2 of~\cite{Langlois:2017mxy}):
\beq
\label{Poisson}
M^2\left[2\frac{(1+\alphaH)^2}{1+\alphaT}-\btrois\right]\nabla^2\Phi + M^2\left(\MsqK+6\frac{\MB^2}{\aL}\right)\Phi =m\,  \d^{(3)}({\bf x})\,,
\eeq
where $\nabla^2\equiv \delta^{ij}\partial_i\partial_j$ denotes the Laplacian and $\d^{(3)}({\bf x})$ is the three dimensional delta function.
The coefficient in front of $\nabla^2\Phi$ in the above equation corresponds to $(4 \pi G_{\rm N})^{-1}$, where $G_{\rm N}$ is the effective Newton constant. 
For  DHOST theories with $\aL\neq 0$ (such as those satisfying $\CII$), we see immediately that the coefficient  in front of the Laplacian in the Poisson equation vanishes, because of (\ref{IIa}), which means that  the effective Newton constant in the linear regime is infinite. Hence, besides developing gradient instabilities, theories satisfying $\CII$ seem also to fail in recovering a viable Newtonian limit (even if this result should be checked in the nonlinear regime and around a non trivial background.)\\
If instead $\aL= 0$, one obtains the generalized Poisson equation
\beq
M^2\left[2\frac{(1+\alphaH)^2}{1+\alphaT}-\btrois\right]\nabla^2 \Phi +M^2 \MsqK \Phi =m\,  \d^{(3)}({\bf x})\,.
\eeq
For DHOST theories that satisfy the conditions $\CI$ but not $\CII$, 
one thus gets a finite Newton constant $G_{\rm N}$
in the linear regime
\beq
\label{GNewton}
8 \pi G_{\rm N}= \frac{1}{M^2} \left[\frac{(1+\alphaH)^2}{1+\alphaT}-\frac{\btrois}{2} \right]^{-1}\,.
\eeq

In conclusion, we found that among the very large number of DHOST theories, only those satisfying the conditions $\CI$ (that are related to Horndeski and beyond Horndeski via conformal or disformal transformations) are phenomenologically viable.
%


\chapter{Phenomenology of Interacting Dark Energy}\label{chap:Pheno}
\lhead{\emph{Phenomenology of Interacting Dark Energy}} 

In this Chapter, I consider a model belonging to the first case analysed in Chapter~\ref{chap:EFToDE_m}, i.e. a gravitational sector described by a Horndeski theory with CDM having a conformal-disformal coupling that depends on the scalar field only, Eqn.~\ref{cdgenH}. I introduce the relevant equations to be solved, the minimal set of parameters needed to fully describe linear perturbations in the quasi-static approximation, and present Fisher matrix forecasts for the constraining power of future surveys for those~\citepubli{Gleyzes:2015rua}.

To discuss the phenomenology, it is convenient to use a gauge where a more direct connection to the physics can be made. 
One can leave the unitary gauge description introduced previously, by ``covariantizing'' the action. This can be done explicitly by performing a time reparametrization of the form
\be
t \to \phi = t+ \pi(t, \mathbf{x}) \;, \label{stuek}
\ee
where the unitary time $t$ becomes a four-dimensional scalar field $\phi$. I denote by $\pi$ the fluctuation of $\phi$. 
By substituting the above transformation into the total action  $S = S_{\rm g} + S_{\rm m} $, one obtains an action that depends on the scalar field $\phi$ and an arbitrary metric $g_{\mu\nu}$. \\
To study cosmological perturbations, I then fix the Newtonian gauge with only scalar perturbations, i.e.,
\be
\label{metric_Newtonian}
ds^2 = - (1+2 \Phi) dt^2  + a^2(t) (1-2 \Psi) \delta_{ij}   dx^i dx^j \;.
\ee
As for matter, in this gauge the scalar part of the stress-energy tensor for each species, at linear order, is 
\begin{align}
T_{(I)}{}^0_{\   0} &\equiv -  (\rho_I+ \delta \rho_I) \;, \label{se1}\\
T_{(I)}{}^0_{\   i} & \equiv \rho_I(1 + w_I)  \partial_i  v_I = - a^2 T_{(I)}{}^i_{\ 0}\;, \label{se2} \\
T_{(I)}{}^i_{\   j} &\equiv (\rho_I w_I  +  \delta p_I) \delta^i_j + \left( \partial^i \partial_j - \frac13 \delta^i_j \partial^2 \right) \sigma_I \label{se3}\;,
\end{align}
where  $\delta \rho_I $ and $\delta p_I$ are the  energy density and pressure perturbations, $v_I$ is the 3-velocity potential and $\sigma_I$ is the  
anisotropic stress potential for the species $I$. In the following, I use the density contrast  $\delta_I \equiv \delta \rho_I /\rho_I$ and consider species with vanishing anisotropic stress.

In principle, the gravitational action contains five non-independent scalar equations: the $(0,0)$, $(0,i)$, $(ii)$ and traceless components of the Einstein equations and the equation for the scalar field $\phi$. These can be combined to yield two independent equations for the metric potentials $\Phi$ and $\Psi$, sourced by the matter perturbations. The first one is a second order differential equation for $\Psi$, while the second is a constraint equation relating $\Phi$ and $\Psi$. 
The corresponding full equations have been derived first in~\cite{Bellini:2014fua} and then in~\cite{Gleyzes:2014rba} for Horndeski theories and extended to the case of $\alphaH$ in~\cite{Lombriser:2015cla}. Their expression in the case of the model treated in this Chapter can be found in Sec.~4.1 of ~\cite{Gleyzes:2015pma}.\\
To close the system, one needs to specify the evolution equations for the matter perturbations. These are derived from the invariance of the matter action under arbitrary diffeomorphisms; if matter is minimally coupled, this yields the usual conservation equation for the energy-momentum tensor, while if a non minimal coupling is present there can be an exchange of energy between matter and the scalar field. I will study a concrete case in the rest of this Chapter. Using the decomposition~\eqref{se1}-\eqref{se3} and assuming vanishing anisotropic stress, the equations of matter are two first order equations for the density contrast $\delta_I$ (continuity equation) and for the velocity potential $v_I$ (Euler equation).

\section{Model and main equations}
In the late universe, the only relevant matter species are CDM and baryons.
Here, I consider the case where CDM admits a non trivial coupling to the metric while the baryons are minimally coupled, and assume without loss of generality that the metric $g_{\mu\nu}$ corresponds to this frame\footnote{If not, one just needs to apply a metric transformation to reach this frame. The transformations of all the relevant quantities can be found in Sec. (2.5) of ~\cite{Gleyzes:2015pma}.}. The gravitational sector is described by an action belonging to the Horndeski class, i.e. Eqn.~\eqref{SBAction0} with $\alphaH=\aL=\bun=\bdeux=\btrois=0$.
The coupling of CDM to gravity and dark energy  is characterised by the effective metric 
\be\label{disf_unit_I2}
 \check{g}^{(c)}_{\mu\nu}\equiv C_c(\phi) g_{\mu\nu}+D_c(\phi) \partial_\mu \phi \partial_\nu \phi\,,
\ee
from which we have the conformal and disformal parameters introduced in~\eqref{defalphasM}:
\be
\label{defalphac}
 \alpha_{\text{C},c} \equiv \frac{\dot C_c}{2 H  C_c} \, , \qquad \alpha_{\text{D},c}\equiv \frac{D_c}{ C_c-D_c}\,.
 \ee
 In the following, I will also call the coupling~\eqref{disf_unit_I2} ``non minimal coupling'' to distinguish it from that of the baryons.
In summary, linear perturbations are characterised by the six free functions
\be
\alphaK , \, \alphaB, \,  \alphaM , \, \alphaT, \,  \alpha_{\text{C},c} , \,  \alpha_{\text{D},c} \, .
\ee

The  equations of motion for the metric are obtained by varying the total action (after having applied the time reparametrisation~\eqref{stuek}) with respect to $g_{\mu\nu}$,
\be
\label{Einstein_eqs}
\frac{\delta S}{\delta g_{\mu \nu}} =0 \;,
\ee
which provides the generalised Einstein equations. Their explicit form in Newtonian gauge can be found in Appendix C of ~\cite{Gleyzes:2015pma}. \\
Since baryons are minimally coupled, their evolution is just given by the standard conservation equation
\be
\nabla_\mu T_{(b)}{}_{ \ \nu}^\mu =0  \label{evol_matterb}\,.
\ee
To write the equations of motion for CDM, one can use the invariance of  the matter action $S_{c}$ under arbitrary diffeomorphisms, $x^\mu \to x^\mu + \xi^\mu$. This gives an expression of the form\footnote{The explicit expression for $Q_c$ in a FLRW background is~\citepubli{Gleyzes:2015pma}
\be
\bar{Q}_c= \frac{H \rho_c}{1+\alpha_{\text{D},c}}    \left\{     \alpha_{\text{C},c} +   \alpha_{\text{D},c} \left(3 +  \frac{\dot\rho_c}{H \rho_c}  \right)
+ \frac{\dot\alpha_{\text{D},c}}{2H (1+\alpha_{\text{D},c})}\right\}   \,. \label{barQ}
\ee}
\begin{align}
&\nabla_\mu T_{(c)}{}_{ \ \nu}^\mu + Q_c \partial_\nu \phi =0 \;, \label{evol_matter} \\ 
&Q_c\equiv - \frac{1}{\sqrt{-g}} \frac{\delta S_c}{\delta \phi} = - \frac{ C_c' }{2 C_c} T_{(c)} - \frac{D_c'}{2C_c} T_{(c)}^{\mu \nu} \partial_\mu \phi \partial_\nu \phi + \nabla_\mu \left(T_{(c)}^{\mu \nu} \partial_\nu \phi \frac{D_c}{C_c} \right) \;,
\end{align}
where a prime denotes a derivative with respect to $\phi$. The explicit form of these equations for baryons and CDM in Newtonian gauge can be found in Eqns. (4.8)-(4.9) of ~\cite{Gleyzes:2015pma}, while those for a generic fluid in Eqns. (3.16)-(3.17) of the same Article.
Finally, the evolution equation for $\phi$ can be obtained by variation of the total action with respect to $\phi$, $\delta S /\delta \phi =0$. One obtains
\be
\label{pi_evol}
\frac{1}{\sqrt{-g}} \frac{\delta S_{\rm g}}{\delta \phi} -  Q_c  =0 \;.
\ee
The explicit form of the above equation is in Eqn. (C.7) of ~\cite{Gleyzes:2015pma}.
\subsection{Background evolution}\label{sec:bgIDE}
On the background, the evolution equations~\eqref{evol_matterb}-\eqref{evol_matter} written in terms of the baryons and CDM energy fractions $\Omega_{b,c}\equiv \rho_{b,c}/(3H^2M^2) $, are
\begin{align}
\dot\Omega_b &=-H \bigg( 3 + 2 \frac{\dot H}{H^2} + \alphaM\bigg) \Omega_b \,, \label{Omegab}\\
\dot  \Omega_c&= - H \bigg( 3 + 2 \frac{\dot H}{H^2} -3\gammac+ \alphaM\bigg) \Omega_c  \,. \label{Omegac}
\end{align}
All the information about the non minimal coupling is encoded in the parameter $\gammac$\footnote{Defined as $ \bar Q_c =3H \rho_c \gamma_c$.},
\be
\gammac=\frac13 \alphaCc+\frac{\dot \alpha_{\text{D},c}}{6H(1+\alphaDc)}\,.
\ee
The presence of the coefficient $\alphaM$ is due to the fact that the mass $M$ can be time-dependent.
As already mentioned, at the background level the dark energy can be defined by giving a specific time evolution for the Hubble parameter. 
I assume that  the expansion history corresponds to that  of $w$CDM, so that $H$ is given by
\begin{align}
H^2 (a) = H_0^2 \left[ \Omega_{\rm m,0} a^{-3} +(1-\Omega_{\rm m,0}) a ^{-3(1+\wDE)} \right] \;,  \label{Hparam}
\end{align}
where $w$ is a constant parameter.\footnote{This choice of parametrisation for the background is motivated by the fact that observations suggest that the recent cosmology is very close to $\Lambda$CDM, which corresponds to $w=-1$, and deviations from $\Lambda$CDM in the expansion history are usually parametrised in terms of  $\wDE \neq -1$.} In the absence of modifications of gravity and non minimal couplings, i.e.~for $\alphaM = \gammac=0$, $\wDE$ coincides with the equation of state of dark energy.
With this parametrisation and for $w\sim-1$, the background expansion remains close to $\Lambda$CDM, even when $\alphaM$ or $\gammac$ are switched on and  matter does not scale as $a^{-3}$ (see eqs.~\eqref{Omegab} and \eqref{Omegac}).  

\subsection{Perturbations in the quasi static regime}\label{pertQSWEP}
In this section, I discuss the phenomenology of perturbations on scales where the so-called ``quasi-static approximation'' holds. Roughly speaking, this corresponds to considering scales where the time derivatives in the Einstein equations can be neglected with respect to the spatial ones. This argument can be made rigorous and it can be shown that it is justified for spatial scales smaller than the sound horizon of dark energy, i.e.~$k \gg aH/c_s$ \cite{Sawicki:2015zya,Lombriser:2015cla}. 
In this regime, all the scalar perturbations $\Phi$, $\Psi$, $\pi$ obey Poisson-like equations. 
One obtains a system of equations for six independent variables: the two metric potentials $\Phi$ and $\Psi$, and the density contrasts ($\delta_b$, $\delta_c$) and velocities ($v_b$, $v_c$) for baryons and CDM. 

\begin{figure}[t]
\centering
\includegraphics[width=0.8\textwidth]{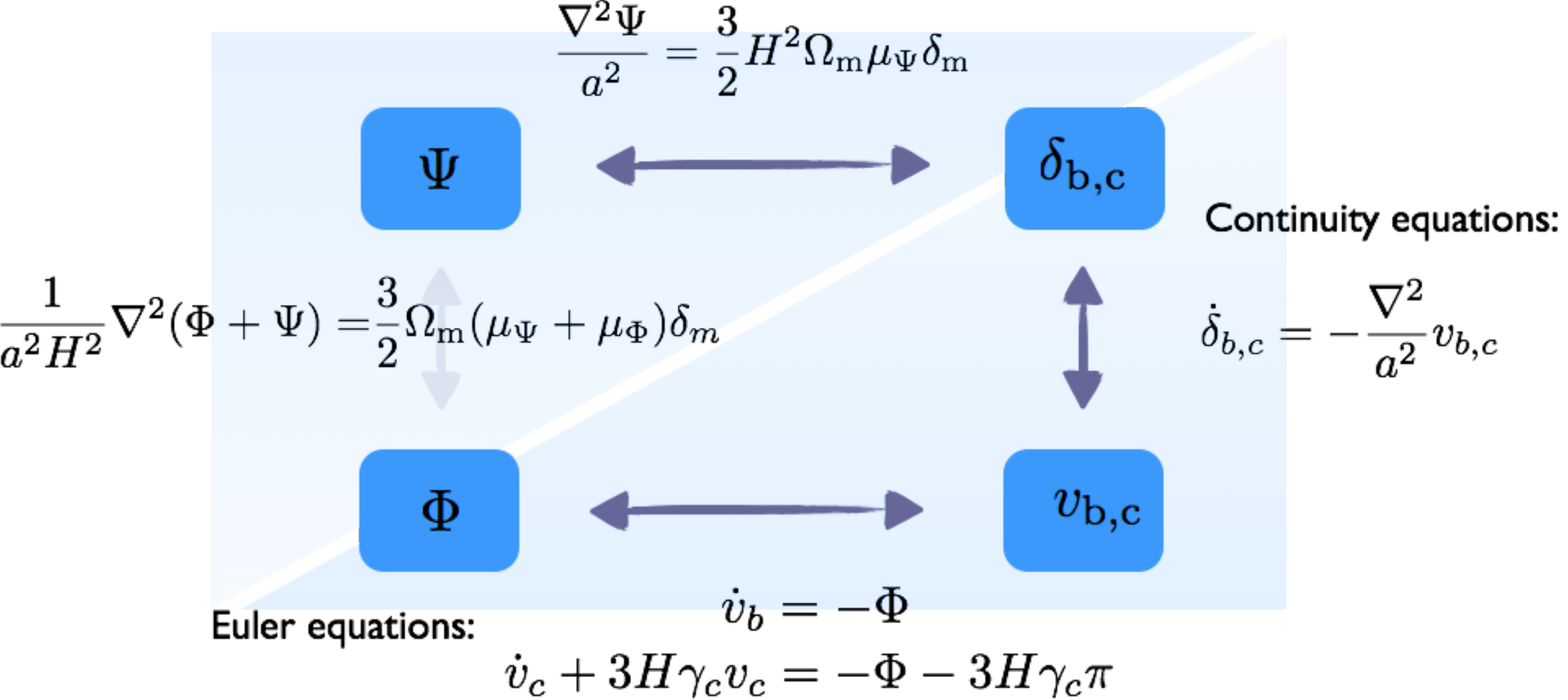}
\caption{Relation between the matter and gravitational perturbations in the interacting CDM model.}
\label{fig:potentialsMG}
\end{figure}

Let me define the total matter density contrast $\delta_{\rm m}=\omega_{\rm c} \delta_c+\omega_{\rm b}  \delta_b$, where $\omega_{\rm b,c}\equiv \Omega_{\rm b,c} / \Omega_{\rm m} $. An analogous definition holds for the velocity potential: $v_{\rm m}=\omega_{\rm c} v_c+\omega_{\rm b}  v_b$. The equations for the scalar field and the metric potentials can be written as:
\begin{align}
& \frac{\nabla^2}{a^2} \Phi = \frac32 H^2 \Omega_{\rm m} \mu_{\Phi} \delta_{\rm m} \, , \label{phipoisson}\\
& \frac{\nabla^2}{a^2} \Psi = \frac32 H^2 \Omega_{\rm m} \mu_{\Psi} \delta_{\rm m} \, , \label{psipoisson}\\
& \frac{\nabla^2}{a^2} \pi = 3H \Omega_{\rm m}\frac{ \tbeta_\xi \omega_b \delta_b +  (\tbeta_\xi + \tbeta_\gamma)  \omega_c \delta_c }{\sqrt{2} c_s \alpha^{1/2} }  \;, \label{pipoisson}
\end{align}
while the continuity and Euler equations take the form:
\begin{align}
& \dot{\delta}_{b}=-\frac{\nabla^2}{a^2} v_{b} \, , \label{contb}\\
& \dot{\delta}_{c}=-\frac{\nabla^2}{a^2} v_{c} \, , \label{contc}\\
& \dot{v}_{b}=-\Phi\, , \label{eulerb}\\
& \dot{v}_{c}+3H\gamma_c v_c=-\Phi-3 H \gamma_c \pi\, . \label{eulerc}
\end{align}
The relations between these quantities are summarised in Figure~\ref{fig:potentialsMG}.
The functions $\mu_{\Phi}$ and $\mu_{\Psi}$ introduced in Eqns.~\eqref{phipoisson}-\eqref{psipoisson} have the explicit expressions
\begin{align}
\mu_\Phi  &= 1+\alphaT +   \tbeta_\xi \big( \tbeta_\xi  +  \tbeta_\gamma  \omega_c b_c  \big)\;, \label{muphi}\\
\mu_\Psi  &= 1+  \tbeta_{\rm B}  \big(  \tbeta_\xi +    \tbeta_\gamma  \omega_c b_c  \big) \;.  \label{mupsi}
\end{align}
I introduced the parameters\footnote{The parameter $\beta_\gamma$ generalizes the parameter $\beta$ defined for coupled quintessence in Sec.~5.3.4 of \cite{Ade:2015rim}. In this case, the relation between the two parameters is $\beta_\gamma = - \sqrt{2} \beta$.}
\be
\label{parameters}
\tbeta_{\gamma}  \equiv \frac{3  \sqrt{2}   }{ c_s \alpha^{1/2}  } \gamma_c\, ,  \quad \tbeta_{\xi}  \equiv \frac{\sqrt{2} }{  c_s \alpha^{1/2} }\xi \equiv  \frac{\sqrt{2} }{  c_s \alpha^{1/2} }\left[ \alphaB (1+\alphaT) + \alphaT - \alphaM\right] \, ,\quad \tbeta_{\rm B}  \equiv \frac{\sqrt{2}  \alphaB }{ c_s \alpha^{1/2} } \;,
\ee
$\alpha$ and $c_s$ were introduced in equations~\eqref{alpha_def} and~\eqref{csHornd}, and $b_c \equiv \delta_c/\delta_{\rm m}$ is a time-dependent bias between the CDM and the total matter density contrast\footnote{In the quasi-static limit, the evolution equations are scale independent so that the ratios $\delta_b/\delta_{\rm m}$ and $\delta_c/\delta_{\rm m}$ do not depend on scales. Note also that the bias parameter introduced here is different from the bias between the total matter density and the galaxy density, that I am going to introduced later.}.
As one can see, modified gravity and a non minimal coupling affect the equations for the two potentials in the following way:
\begin{itemize}
\item The coupling between the potentials and the metric is altered with respect to GR. The modification is encoded in the function $\mu_{\Psi}$ in Eqn.~\eqref{psipoisson}, defined so that in GR $\mu_{\Psi}=1$. An analogous quantity $\mu_{\Phi}$ can be defined for the Poisson equation for $\Phi$, Eqn.~\eqref{phipoisson}. In the absence of nonminimal coupling of CDM, the gravitational coupling $\mu_{\Phi}$ is given by $\mu_{\Phi}=1+ \alphaT + \beta_\xi^2$. If $\alphaT\ge0$
 , this quantity is always larger than one, which tends to enhance the growth of structure.
\item The relation between the two metric potentials is non trivial. In GR, one simply has $\Phi=\Psi$ (and $\mu_{\Psi}=\mu_{\Phi}=1$). In general, one can combine Eqns~\eqref{phipoisson} and~\eqref{psipoisson} to get another Poisson-like equation for the sum of the two potentials:
\be \label{weylpoisson}
\frac{\nabla^2}{a^2} \left(\Phi +\Psi \right) = \frac32 H^2 \Omega_{\rm m} \left(\mu_{\Phi} +\mu_{\Psi} \right)\delta_{\rm m}\, .
\ee
\item The non minimal coupling introduces extra friction and an additional ``fifth-force'' term in the Euler equation for CDM, Eqn~\eqref{eulerc}. This is the result obtained in the context of coupled dark energy (see e.g. \cite{Amendola:2003wa}).
If there is a non trivial coupling of CDM but  gravity itself is not modified, than the Newton constant is not modified, $\mu_\Phi = \mu_\Psi =1$, and $\Phi$ and $\Psi$ are the same as in GR, {\em even if} CDM is nonminimally coupled. Note that in the equations for matter, all the modifications are encoded in the single parameter $\gamma_c$. Therefore, it is not possible to disentangle the conformal and disformal effects. This is due to the fact that the non minimally coupled species is pressureless and that we are in the quasi-static regime.
\end{itemize}

One can combine Eqns.~\eqref{phipoisson}-\eqref{eulerc} to obtain two coupled second-order differential equations for the two density contrasts:
\begin{align}
\ddot \delta_b + 2 H \dot \delta_b &= \frac32  H^2  \Omega_{\rm m} \left\{ (1+\alphaT+ \tbeta_\xi^2 ) \omega_b \delta_b +  \left[ 1+\alphaT+ \tbeta_\xi ( \tbeta_\xi  +  \tbeta_\gamma) \right] \omega_c \delta_c \right\}  \;, \label{eqmatterb}\\
\ddot \delta_c + (2 - 3 \gammac) H \dot \delta_c &= \frac32  H^2 \Omega_{\rm m}  \left\{  \left[ 1+\alphaT+ \tbeta_\xi ( \tbeta_\xi  +  \tbeta_\gamma) \right] \omega_b \delta_b +  \left[ 1+\alphaT+  ( \tbeta_\xi + \tbeta_\gamma )^2 \right]   \omega_c \delta_c  \right\}   \;.\label{eqmatterc}
\end{align}

Since equations (\ref{eqmatterb})--(\ref{eqmatterc}) are independent of the wavenumber $k$, one can factorize the time dependence from the $k$ dependence of the initial conditions and write the solutions in the form
\be
\label{solMD}
\delta_c (t, \vec k) = \GG_c  (t) \, \delta_{c,0} (\vec k)\;, \qquad 
\delta_b(t,\vec k) = \GG_b (t)  \, \delta_{b,0}(\vec k) \;,
\ee
where $\d_{c,0}$ and $\d_{b,0}$ represent the initial density contrasts for CDM and baryons respectively,
defined at some earlier time in the matter dominated era (the choice of initial conditions is described in the next section).
The two functions of time $\GG_c  (t)$ and $\GG_b  (t)$ are the growth factors for CDM and baryons, respectively.

I will solve equations  (\ref{eqmatterb})--(\ref{eqmatterc}) to analyse linear perturbations in the quasi static regime. Let me conclude with some remarks.
Modifications of gravity exchanged by $\pi$ are parametrized by $\beta_\xi$ and the nonminimal coupling of CDM is parametrized by $\beta_\gamma$. This separation of effects is not physical and depends on the choice of frame.
The modification of gravity associated with the parameter $\alphaT$ does not depend on the exchange of $\pi$~\cite{Perenon:2015sla} (see also \cite{Jimenez:2015bwa} for a discussion on local constraints of this effect), and does not mix with the other two effects under change of frame.
Finally, note that the parameter $\alpha$ always appears multiplied by $c_s^2$. From the definition of the sound speed, eq.~\eqref{csHornd}, $c_s^2 \alpha$ is independent of $\alphaK$. This is a consequence of dropping time derivatives in the fluctuations of $\pi$ to reach the quasi-static regime, so $\alphaK$ cannot be constrained by observations in this regime (\cite{Piazza:2013pua}, \citepubli{Gleyzes:2015rua}).\\
In summary, the phenomenology in the quasi-static limit for baryons and nonminimally coupled CDM is captured by the reduced set of parameters:
\be
\alphaK , \, \alphaB, \,  \alphaM , \, \alphaT, \,  \alpha_{\text{C},c} , \,  \alpha_{\text{D},c} \, \mapsto  \alphaB, \,  \alphaM , \, \alphaT, \,  \gamma_c \, .
\ee

\section{Solving the equations: parametrization and initial conditions}

In the effective descriptions, the free functions are time dependent, so one has to choose a parametrisation in order to solve Eqns.~\eqref{eqmatterb}-\eqref{eqmatterc}. I will assume that the functions $\alphaB$, $\alphaM$ and $\alphaT$ share the same time dependence :
\be\label{TimeDepAlphas}
\alpha_{\rm A}=\alpha_{A,0}\, \frac{1-\Omega_{\rm m} (t)}{1-\Omega_{\rm m,0}}
\ee
$A=B,M,T$ and $\alpha_{A,0}$ denote the current values of these parameters. These are the free parameters of the effective description which I will constrain.
The time dependence of $\gammac$ is chosen by assuming that the parameter $\beta_\gamma$, defined in eq.~\eqref{parameters}, is time-independent, so that 
\be
\gamma_c   (t) = \frac{  \tbeta_{\gamma} }{3  \sqrt{2}   }   \,  c_s (t) \alpha^{1/2}  (t) \; . \label{gamma_evol} 
\ee
This choice of parametrisation allows to include coupled quintessence \cite{Amendola:2011ie} as a special case, or more generally other cases where the nonminimal coupling of CDM remains active also when the dark energy density becomes negligibly small, since one can have $c_s \alpha^{1/2} =0$ while  $\tbeta_{\gamma}\neq0$. Moreover, $c_s \alpha^{1/2}$ vanishes in matter domination.
Therefore, when $\Omega_{\rm m} \to 1$, then $\alpha_{\rm A} \to0$ and $\gamma_c \to 0$, which corresponds to the standard matter dominated phase for the background evolution. However, while modifications of gravity switch off in this limit (i.e. $\alphaB, \alphaM, \alphaT \to 0$), the nonminimal  coupling parametrised by $\beta_\gamma$ remains active.
The details of the parametrisation and of the time-dependence are discussed in Section 4 and App. A of~\cite{Gleyzes:2015rua}.

In particular, let me briefly discuss the time dependence of $c_{s}^2\alpha$. Its expression is given in general by Eqn.~\eqref{csHornd}, where one should sum over baryons and CDM. Explicitly, we can write it as
\be
c_s^2 \DDt = (1+\alphaB) (3 - 3\eta - 2 \xi) - 3 \Omega_{\rm m}-2\frac{\dot \alpha_\text{B} }{H}  \, , \label{csalpha}
\ee
where $\xi=\alphaB (1+\alphaT) + \alphaT - \alphaM$, and
\be
\eta \equiv  \frac{1}{3} \left(3+ 2 \frac{\dot H}{H^2 } \right) =-\wDE \, \frac{(1-\Omega_{\rm m,0})a^{-3\wDE}}{\Omega_{\rm m,0}+(1-\Omega_{\rm m,0}) a^{-3\wDE}} \;, \label{eta_def}
\ee

By using Eqn.~\eqref{TimeDepAlphas} and the background evolution equations \eqref{Omegab} and \eqref{Omegac} to evaluate $\dot \alpha_{\rm B}$ in~\eqref{csalpha}, 
this can be written as
\be
\label{eqtheta}
\begin{split}
{c_s^2 \alpha} = 3 (1-\Omega_{\text{m}} -\eta) +   &  \alphaB  \bigg[ 1- 3 \eta \bigg(1 + 2  \frac{\Omega_{\rm m}}{1 - \Omega_{\text{m}}} \bigg)    -  2 ( \alphaM - 3  \gammac\,  \omega_c  )  \frac{\Omega_{\rm m}}{1-\Omega_{\text{m}}} \bigg] \\
&- 2 \alphaB^2 -2 \alphaT  \big(1+ \alphaB  \big)^2 + 2 \alphaM  ( 1 + \alphaB )     \, ,
\end{split}
\ee
Finally, one can replace $\gammac$ by its expression \eqref{gamma_evol} given in terms of ${c_s \DDt^{1/2}}$. The equation \eqref{eqtheta} is thus a quadratic equation for $X\equiv c_s\alpha^{1/2}$. One can extract the relevant solution. 
This is done explicitly in App. A of~\cite{Gleyzes:2015rua}.

The background evolution has been discussed in Sec.~\ref{sec:bgIDE} and it is given by equation~\eqref{Hparam}. 

Let me also comment on the initial conditions needed to solve Eqns.~\eqref{eqmatterb}-\eqref{eqmatterc}. I start the evolution during matter domination, where $\Omega_{\rm m}\simeq 1$ and $\alpha_{\rm A}\simeq 0$ ($A=B,M,T$). This also implies that $\gamma_c\simeq0$ and $\beta_{\xi}\simeq 0$. Thus, at the background level there are no deviations from $\Lambda$CDM, while the perturbations equations~\eqref{eqmatterb}-\eqref{eqmatterc} in this limit are
\begin{align}
\ddot \delta_b + 2 H \dot   \delta_b &\simeq \frac32  H^2 \left[ \omega_b \delta_b +  \omega_c \delta_c \right]  \;, \label{eqmatterb2} \\
 \ddot \delta_c  + 2 H   \dot \delta_c  &\simeq \frac32   H^2  \left[  \omega_b \delta_b +  \left( 1+ \tbeta_\gamma^2 \right)   \omega_c \delta_c  \right]   \;,\label{eqmatterc2}
\end{align}
where $\omega_{b,c}$ are constant. The solutions of the above system can written as 
\be
\d_b=\bbi \, \d_{\rm m}\,, \qquad \d_c=\bci\,  \d_{\rm m},
\ee
with constant and scale-independent bias parameters given by 
\be
\label{bias}
\bbi = \frac{1+\beta_\gamma^2 \omega_c - \sqrt{ 4 \beta_\gamma^2 \omega_c^2 + (1-\beta_\gamma^2 \omega_c)^2 }}{2 \beta_\gamma^2 \omega_c \omega_b } \;, \qquad 
\bci = \frac{-1+\beta_\gamma^2 \omega_c + \sqrt{ 4 \beta_\gamma^2 \omega_c^2 + (1-\beta_\gamma^2 \omega_c)^2 }}{2 \beta_\gamma^2 \omega_c^2  } \,.
\ee
The respective growth functions $\GG_c$ and $\GG_b$ are identical, solutions of  the  equation 
\be
\ddot \GG + 2 H \dot \GG - \frac32 H^2 \left(  1 + \beta_\gamma^2 \omega_c^2 \bci   \right) \GG =0 \;. \label{D_evol} 
\ee
I set initial conditions on the growing mode, $\GG_{+}$. This analysis also shows that baryons and CDM possess spectra that are initially proportional and then grow similarly.

\section{Observables}\label{sec:obs}
In the next section, I will present constraints based on a Fisher matrix analysis applied to three observables that are targets of future surveys: the galaxy and weak lensing power spectra \cite{Tegmark:1996bz,Tegmark:1997rp} and the correlation between the ISW effect in the CMB and the galaxy distribution \cite{Crittenden:1995ak}. Here, I give an analytical understanding of the effects of modifications of gravity on these observables and the expression for their Fisher matrices.

\subsection{Galaxy clustering}
The observed number density of galaxies in redshift space can be related to the one in real space by a term that depends on the line-of-sight component of the galaxy's peculiar velocity, $v_{\text{g},z}$ (see e.g.~\cite{Bernardeau:2001qr}),
\be
\label{RSD}
\delta_{ \text{g},s} =\delta_{\text{g}}-  \frac{1}{aH}\nabla_z v_{\text{g},z}  \,.
\ee
To compute the above quantity, we need to obtain an expression for the peculiar velocity of the galaxy, $ v_\text{g}$. Let me show how to do so. The idea is to relate the peculiar velocity $v_\text{g}$ to the CDM and baryon fluid velocities ${v}_b$, ${v}_c$ that satisfy the Euler equations~\eqref{eulerb}-\eqref{eulerc}.
\begin{enumerate}
\item I shall effectively treat galaxies as test particles moving in the Hubble flow (see e.g.~\cite{Chan:2012jj}). They are composed by baryon and CDM  mass fractions $x_b \equiv M_b/M_\text{g}$ and $x_c \equiv M_c/M_\text{g}$ ($M_\text{g} \equiv M_b+M_c$), respectively\footnote{In the following I assume the same baryon-to-CDM ratio for each galaxy and I set this to be the background value, i.e.~$x_c = \omega_c$ and $x_b=\omega_b$. However, one could also consider different populations of galaxies with different baryon-to-CDM ratios and study the effects of equivalence principle violations on large scales between these different populations (see e.g.~\cite{Creminelli:2013nua}).}. 
\begin{figure}[t]
\centering
\includegraphics[width=0.8\textwidth]{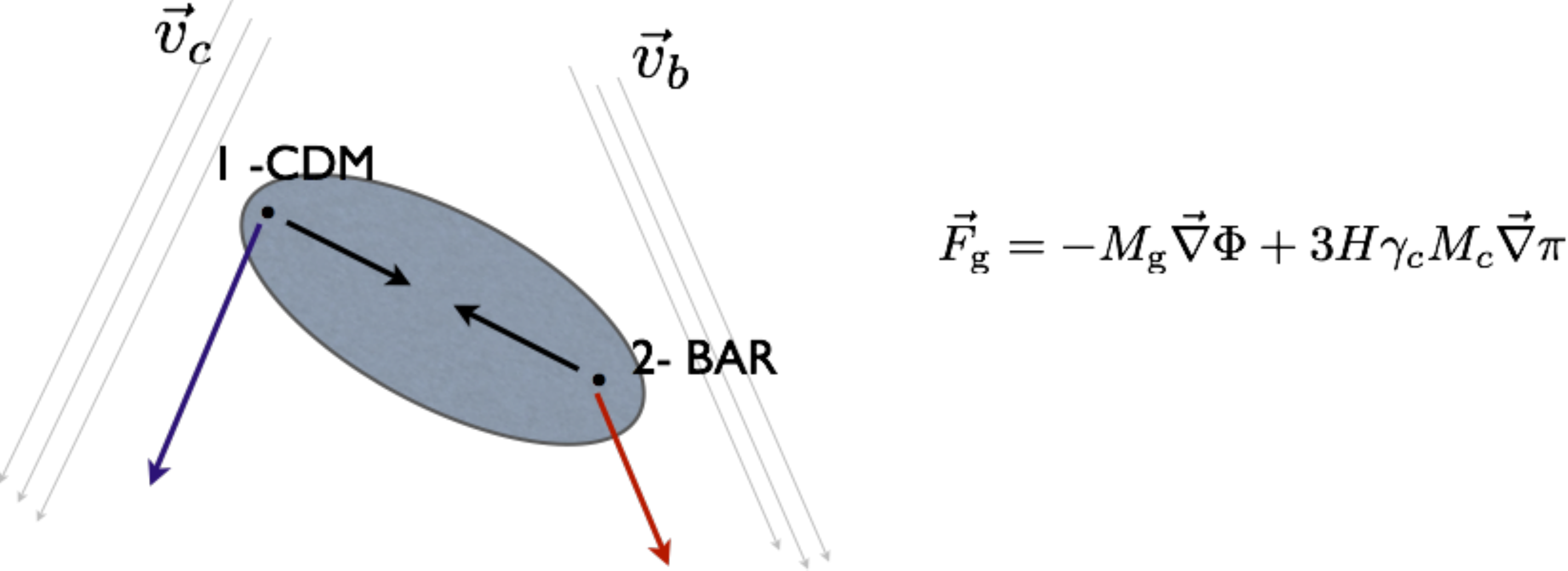}
\caption[Toy-model of a galaxy made by baryons and CDM.]{Toy-model of a galaxy made by baryons and CDM. The centres of mass of the two components feel different forces due to the non minimal coupling of CDM particles, and this result in an additional force felt by the galaxy.}
\label{fig:galaxy}
\end{figure}
A representation of this toy model is given in figure~\ref{fig:galaxy}. Newton's law for the galaxy, including the fifth force on the CDM component, can be written as ($\vec{v}_{g}=\vec{\nabla} v_{g}$):
\be
\frac{d}{dt}\left(M_\text{g}\vec{v}_\text{g}\right)=M_\text{g} \dot{\vec{v}}_\text{g}+3H\gamma_c M_c \vec{v}_\text{g}=\vec{F}_\text{g}=- M_\text{g} \vec \nabla \Phi + 3 H \gamma_c M_c \vec \nabla \pi . 
\ee
In the first equality, I used the fact that in absence of screening the mass of the CDM component in the galaxy is not conserved and obeys
$
\dot{M}_c=3H\gamma_c M_c\,. \label{changeMc} 
$
The last term on the right hand side can be rewritten in terms of the baryon and CDM velocities using the Euler equations~\eqref{eulerb}-\eqref{eulerc} with $\vec{v}_{b,c}=\vec{\nabla} v_{b,c}$. Doing so, one obtains that the above equation is solved by
\be \label{vg}
\vec{v}_\text{g}=x_c\vec{v}_c+x_b \vec{v}_b \;.
\ee

\item Then, one can use the continuity equations~\eqref{contb}-\eqref{contc} in Fourier space to define a growth rate for the CDM and baryons. This gives:
\beq
v_{b,c}=\frac{a^2 H}{k^2} f_{b,c} \, \delta_{b,c} \, ,\quad f_{b,c}\equiv\frac{1}{\delta_{b,c}}\frac{\text{d}\delta_{b,c}}{\text{d} \ln a} \, ,
\eeq
In such a way, one can finally express the peculiar velocity of the galaxy given by Eqn.~\eqref{vg} as a function of the density contrasts of the baryons and CDM given by the solutions of the system~\eqref{eqmatterb}-\eqref{eqmatterc}:
\be
v_\text{g}=\frac{a^2H}{k^2}\left(x_c f_c \, \delta_{c}\, +x_b f_b\,  \delta_{b}\right)\, .
\ee 
\end{enumerate}
One can then proceed as in the standard calculation and compute the galaxy power spectrum in redshift space from the galaxy number density in real space, Eqn.~\eqref{RSD}. This is given by 
\be \label{galps}
P_{\text{g},s} (z, \vec k)  =\,  {\big( b_{\rm g}(z)^2+\mu^2  {\fg(z)} \big)}^2 P_{\rm m} (z,k)   \;,
\ee
where $\mu\equiv k_z/k$, and I have introduced the \emph{effective growth rate} of the galaxy distribution as
\be \label{feff}
\fg \equiv {x_c f_c \, b_c\,+x_b f_b \, b_b} \,, 
\ee
and the galaxy bias $b_\text{g}$, defined as $\delta_\text{g}=b_\text{g} \,  \delta_{\rm m}$.
The matter power spectrum $P_{\rm m} (z, k) $ can be written in terms of the growth functions of CDM and baryons using Eqn.~\eqref{solMD}. Since $\delta_{\rm m}=\omega_{\rm c} \delta_c+\omega_{\rm b}  \delta_b$, we have 
\be
P_{\rm m} (z,k) = T_{\rm m}^2 (z)  P_{0} (k)\;, 
\ee
where 
\be
\label{Tm}
T_{\rm m} (z) \equiv \omega_b (z) \, \bbi  \, \GG_b (z) + \omega_c (z) \, \bci  \, \GG_c(z)  \; 
\ee
is the matter transfer function, $P_{0} (k)$ is the initial power spectrum of matter fluctuations, $\d_{\rm m,0}$, during matter domination and $\bbi$, $\bci$ are defined in eq.~\eqref{bias}. As the effects of dark energy and modified gravity intervene at late times, the initial spectrum is independent on modifications of gravity.

Finally, I include the corrections due to the Alcock-Paczynski effect. 
The observed power spectrum reads \cite{Seo:2003pu} 
\be
P_\text{obs}(z;k,\mu)=  {\cal N} (z)   \left[b_{\rm g}(z) + \fg(z) \mu^2 \right]^2\, P_{\rm m}(z,k) \;,
\ee 
where the normalization factor  ${\cal N} (z)$  is given by 
\be
{\cal N} (z) \equiv \frac{H(z) \hat D_{\rm A}^2(z)}{\hat H(z) D_A^2(z)} \;, \qquad  D_A (z) \equiv \frac{1}{ 1+z} \int_0^z \frac{d \tilde z}{H(\tilde z)} \;,
\ee
$D_A $ is the  angular diameter distance, and a hat denotes that the corresponding quantity is evaluated on the background.

I assume a spectroscopic redshift survey with Euclid-like characteristics \cite{Laureijs:2011gra}. I particular, I assume a $15 \, 000$ squared degrees sky coverage, sliced in eight equally-populated redshift bins\footnote{the galaxy distribution is taken as the one given by \cite{Geach:2009tm} with a limiting flux  placed at $4 \times 10^{-16} \, \text{erg} \, \text{s}^{-1} \, \text{cm}^{-2}$} between $z=0.5$ and $z=2.1$.\\
The corresponding Fisher matrix for a set of parameters $\boldsymbol \theta$ reads
\be
F^{\rm LSS}_{ab}(z)= \sum_{\rm bins} \, \frac{V}{2{(2\pi)}^3}\int_{k_{\rm min}}^{k_{\rm max}} 2\pi  k^2 dk \int_{-1}^{1} d\mu\,   \frac{\partial  \ln {P_\text{obs} (z;k, \mu)}} {\partial \theta^a} \frac{\partial  \ln {P_\text{obs} (z;k, \mu)}} {\partial \theta^b}\, ,
\label{Fisher_galaxy}
\ee
where $V$, $k_{\rm min}$ and $k_{\rm max}$ are, respectively, the comoving volume and the minimum and maximum wavenumbers of the bin. 
In this formula I neglected the intrinsic statistical error associated with the white shot noise from the Poisson sampling of the density field \cite{Feldman:1993ky}. However, to be conservative, I choose the maximum wavenumber $k_{\rm max}$ such  that the galaxy power spectrum  dominates over the shot noise and we are well within the linear regime\footnote{More specifically, for each redshift bin I take  $k_{\rm max}$ as the minimum between $\pi/(2  R)$,  where $R$ is chosen such that the r.m.s.~linear density fluctuation of the matter field in a sphere with radius $R$ is 0.5, and the value of $k$ such that $\bar n_i P_g(k) =1$, where $\bar n_i$ is the number density of galaxies inside the bin. 
These values of $k_{\rm max}$ are always smaller than $H/(\sigma_g (1+z))$, with $\sigma_g = 400\, \text{km} \, \text{s}^{-1}$, i.e.~the scale where the peculiar velocity of galaxies due to their virialized motion becomes important. 
}. 
For the minimum wavenumber, I assume $k_{\rm min}=10^{-3} h$ Mpc$^{-1}$.

\subsection{Weak Lensing}
A powerful cosmological probe for dark energy is weak lensing, which depends on the so-called scalar Weyl potential, i.e.~the sum of the two gravitational potentials $\Phi$ and $\Psi$.
In particular, I consider lensing tomography \cite{Hu:1999ek}. 

I assume a photometric survey of $15 \, 000$ squared degrees in the redshift range $0 < z < 2.5$, with a redshift uncertainty $\sigma_z (z) = 0.05 (1+z)$, and a galaxy distribution \cite{Smail:1994sx}
\be
n (z) \propto  z^2 \exp\left[ - \left( \frac{z}{z_{ 0}} \right)^{1.5} \right]  \; ,
\ee
where $z_0 = z_{m}/1.412$ and $z_{m}$ is the median redshift, assumed to be $z_{m}=0.9$ \cite{Amara:2006kp,Amendola:2012ys}.
I divide the galaxy distribution in 8 equally populated redshift bins. For each bin $i$, I define the distribution $n_i(z)$ by convolving $n(z)$ with a Gaussian whose dispersion is equal to the photometric redshift uncertainty $\sigma_z(z_i) $, $z_i$ being the center of the $i$th bin (see also \cite{Giannantonio:2011ya,Amendola:2011ie}). Each distribution $n_i(z)$ is normalised to unity, $\int_0^\infty dz \, n_i(z) = 1$.

The angular cross-correlation spectra of the lensing cosmic shear is given by
\be \label{lenspowspec}
C^{\rm WL}_{ij} (\ell)=  \frac{\ell}{4} \int_0^{\infty} \frac{dz}{H(z)}\, \frac{W_i(z) W_j(z) }{ \chi^3(z)}\, k_\ell^3 (z) P_{\Phi + \Psi} [ z, k_\ell(z)] \;, 
\ee
where $\chi(z) \equiv \int_0^z dz/H (z)$ is the comoving distance and $k_\ell (z) \equiv \ell /\chi(z)$ is the wavenumber which projects into the angular scale $\ell$. I also used the lensing efficiency in each bin, defined as
\be
W_i (z) \equiv \chi(z) \int_z^{\infty} d \tilde z \, n_i(\tilde z)  \frac{\chi(\tilde z) - \chi(z)}{\chi(\tilde z) } \;.
\ee
$P_{\Phi + \Psi} $ in Eqn.~\eqref{lenspowspec} is the  power spectrum of $\Phi + \Psi$. Using Eqn.~\eqref{weylpoisson} in Fourier space, we can relate it to the matter power spectrum $P_{\rm m}$:
\be
P_{\Phi + \Psi} (z,k)  =  {\left[-\frac{3a^2H^2}{2k^2}\Omega_{\rm m} \left(\mu_\Psi +\mu_\Phi  \right) \right]}^2 P_{\rm m} (z,k) \;.
\ee
Similarly to the matter case, we can define a transfer function for $\Phi+\Psi$,
\be
P_{\Phi + \Psi} (k)  =  T_{\Phi+\Psi} ^2 (z,k) P_{0} (k) \;,
\ee
where
\be
\label{TPhiPsi}
T_{\Phi+\Psi} (z,k) \equiv -\frac{3a^2H^2}{2k^2}\Omega_{\rm m} \left(\mu_\Psi +\mu_\Phi \right) T_{\rm m} (z) \;.
\ee
From the above equation, we see that the lensing is sensitive to the combination
\be
\mu_\Psi +\mu_\Phi = 2+\alphaT +  (\tbeta_{\rm B}  + \tbeta_\xi ) \big(  \tbeta_\xi +    \tbeta_\gamma  \omega_c b_c  \big).
\ee
Neglecting the shot noise error due to the intrinsic ellipticity of galaxies, 
the Fisher matrix for the cross-correlation spectra in eq.~\eqref{lenspowspec} is given by  \cite{Hu:1998az,Hu:2003pt}
\be
F^{\rm WL}_{ab}= f_{\rm sky} \sum_{\ell = \ell_{\rm min}}^{\ell_{\rm max}} \frac{2\ell+1}{2} \, \text{Tr} \left\{ \frac{\partial  C^{\rm WL}_{ij} (\ell)} {\partial \theta^a}  \big[C^{\rm WL}_{jk}(\ell) \big]^{-1}\frac{ \partial C^{\rm WL}_{km}(\ell)} {\partial \theta^b} \big[C^{\rm WL}_{mi}(\ell) \big]^{-1} \right\} \, ,
\ee
where I choose $\ell_{\rm min} = 10$ and $\ell_{\rm max} =300$. Assuming Euclid-like characteristics \cite{Laureijs:2011gra} for the galaxy density and intrinsic ellipticity noise, the chosen $\ell_{\rm max}$ corresponds to scales where the shot noise is negligible and perturbations are only mildly beyond the linear regime at small redshift.

\subsection{ISW-Galaxy correlation}
As a third probe, I consider the cross-correlation between the ISW effect of the CMB photons and the galaxy distribution, which is a valuable probe of dark energy and of its clustering properties in the late-time universe (see e.g.~\cite{Hu:2004yd,Corasaniti:2005pq}). The galaxy distribution is assumed to come from the same photometric survey as for weak lensing, described in the previous section.

The angular power spectra of the ISW effect and the cross-correlation spectrum depend on the time evolution of the gravitational potentials. The ISW term is 
\be\label{ISWterm}
\frac{\Delta T}{T}^{\rm ISW} (\hat n)=-\int_0^{\infty}  {dz} \frac{\partial}{\partial z} \big(  \Phi+  \Psi \big) [z , \hat n \chi (z)] \; .
\ee
As for galaxies, the projected galaxy overdensity in the bin $i$ is given by~\cite{Ho:2008bz}
\be
g_i (\hat n) =\int_0^{\infty} dz\; n_i(z) b_{\rm g}(z) \delta_{\rm m}[z , \hat n \chi (z)]\;, 
\ee
With these definitions, the angular power spectra of the projected galaxy overdensity and of the ISW effect are respectively given by
\begin{align}
C_{ij}^{\rm gal} (\ell) &=\int_0^{\infty} dz\; \frac{H(z)}{\chi^2(z)} n_i(z) n_j(z)  {b_{\rm g}^2(z)} \,  P_{\rm m} [z,k_\ell (z)] \; , \\
C^{\rm ISW} (\ell) &=  \int_0^{\infty} dz\;  \frac{H(z)}{\chi^2(z)} \bigg[{\bigg(  \frac{\partial T_{\Phi+\Psi} }{\partial z}  (z,k)   \bigg)}^2 \!   \, P_0 (k ) \bigg]_{k=k_\ell (z)}  \; .
\end{align}
Analogously, the angular cross-correlation spectrum between the ISW effect and galaxies reads
\be
C^{\text{ISW-gal}}_i (\ell)=  - \int_0^{\infty} dz\;  \frac{H(z)}{\chi^2(z)} n_i(z) b_{\rm g}(z) T_{\rm m}(z) \bigg[ \frac{\partial T_{\Phi+\Psi} }{\partial z}  (z,k)    P_0 (k) \bigg]_{k=k_\ell (z)} \;.
\ee
The Fisher matrix for the ISW-galaxy correlation is given by (see e.g.~\cite{Douspis:2008xv,Majerotto:2015bra})
\be
F^{\text{ ISW-gal}}_{ab}= f_{\rm sky} \sum_{\ell = \ell_{\rm min}}^{\ell_{\rm max}} (2\ell+1) \, \frac{\partial  C_{ j}^{\text{ISW-gal}} (\ell) } {\partial \theta^a}  \big[\text{Cov}_{jk}(\ell) \big]^{-1}\frac{ \partial C_{ k}^{\text{ISW-gal}} (\ell) } {\partial \theta^b} \, ,
\ee
where I use $\ell_{\rm min}=10$ and $\ell_{\rm max}=300$ and the covariance matrix is given by
\be
\text{Cov}_{jk}(\ell) =C_{ j}^{\text{ISW-gal}} (\ell) C_{ k}^{\text{ISW-gal}}(\ell) +C^{\rm CMB} (\ell)  C_{ jk}^{\rm gal} (\ell) \; ,
\ee
where $C^{\rm CMB}(\ell)$ is the full CMB angular power spectrum.

\section{Forecasts}
\begin{figure}[t]
\begin{center}
{\includegraphics[scale=0.4]{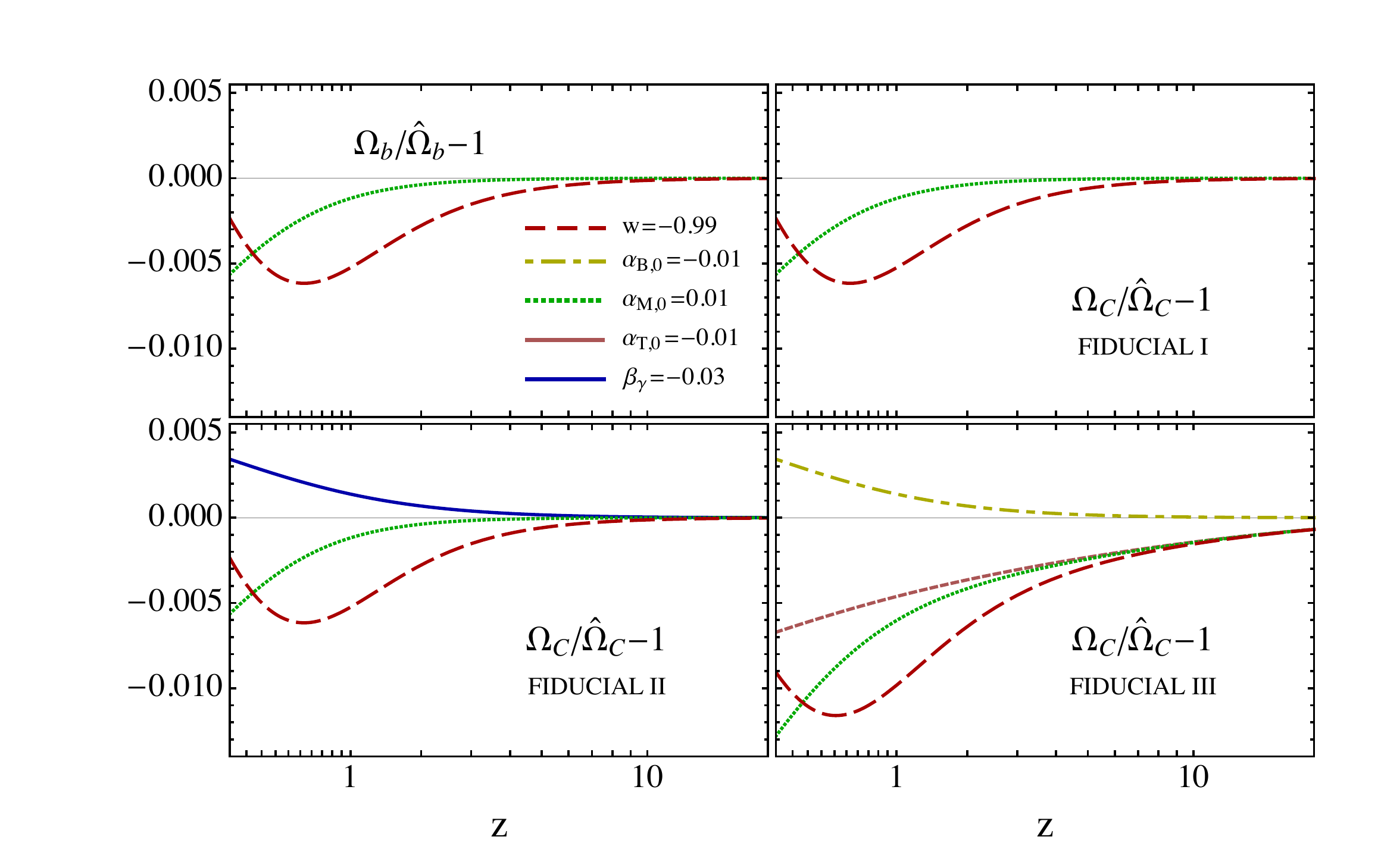}}
\caption[Modification of the background in interacting dark energy models]{ 
Relative change of the baryon and CDM density fractions,  with respect to their fiducial values, as a function of the redshift $z$, depending on the values of the parameters $w$, $\alpha_{\text{B},0}$, $\alpha_{\text{M},0}$, $\alpha_{\text{T},0}$ and $\beta_\gamma$.}
\label{fig:back}
\end{center}
\end{figure}

\begin{figure}[h!]
\begin{center}
{\includegraphics[scale=0.35]{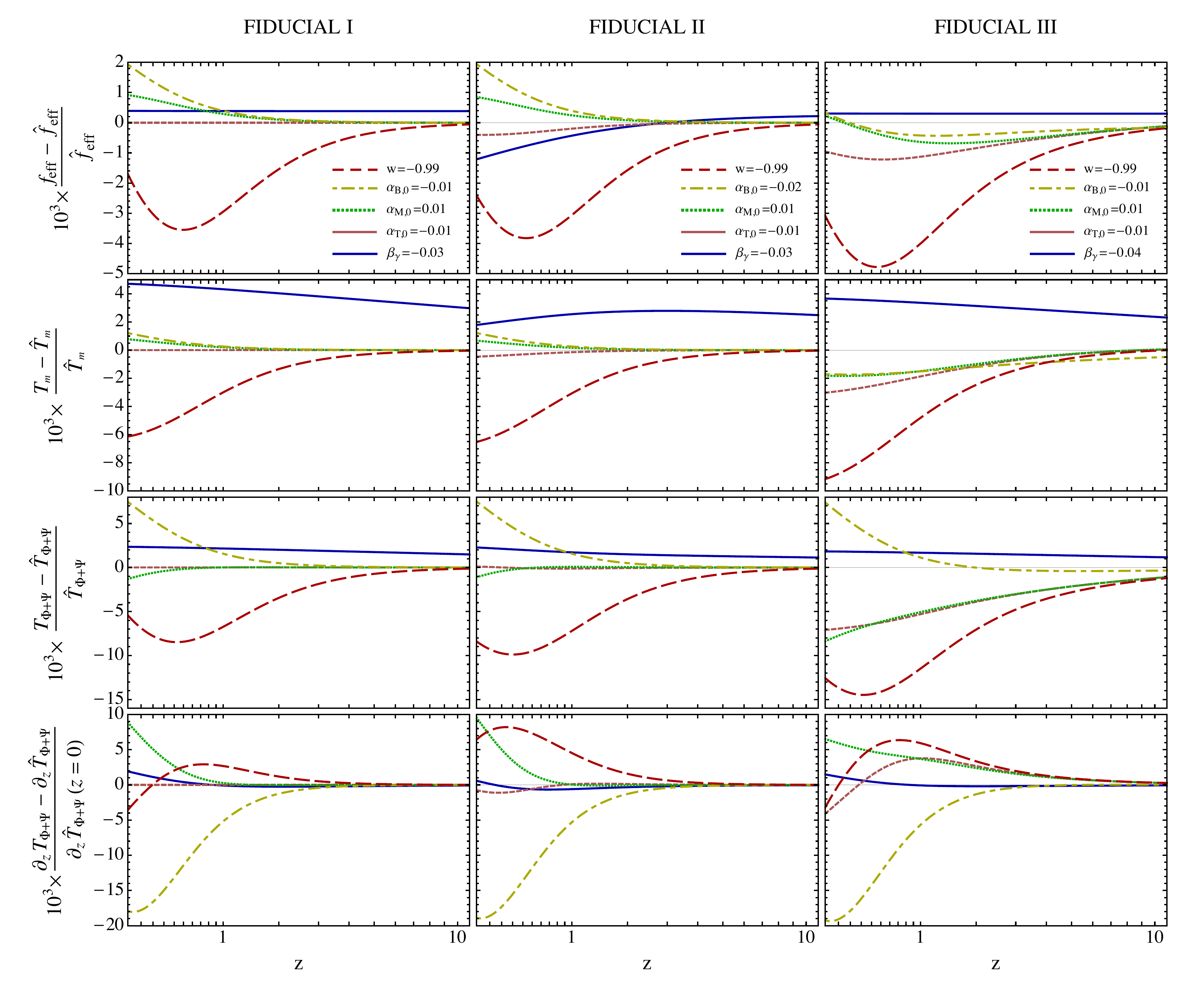}}
\caption[Modification of perturbations in interacting dark energy models]{ Modifications of the evolution of perturbations from their fiducial values, as a function of redshift, for the different parameters $w$, $\alphaBz$, $\alphaMz$, $\alphaTz$ and $\beta_\gamma$. From top to bottom,  relative variation of the effective growth factor $\fg$, eq.~\eqref{feff}, the matter transfer function $T_{\rm m}$, eq.~\eqref{Tm}, the Weyl potential transfer function $T_{\Phi +\Psi}$, eq.~\eqref{TPhiPsi} and its derivative with respect to  redshift, $\partial_z T_{\Phi +\Psi}$, for the three different fiducial models (respectively I, II and III, from left to right). As $\partial_z T_{\Phi +\Psi}$ vanishes in matter domination, we have normalized it to its value at $z=0$ instead of its value as a function of the redshift.}
\label{fig:pert}
\end{center}
\end{figure}

To concentrate on the effects of modifications of gravity and to simplify the analysis I fix the background cosmological parameters to their Planck estimated values. For $w=-1$ these are given by \cite{Ade:2015xua} $h = 0.6731$, $h^2 \Omega_{b,0} = 0.0222 $ and $h^2  \Omega_{c,0} = 0.1197$, while for $w\neq -1$ I choose the values of $\Omega_{b,0}$ and $\Omega_{c,0}$ so to maintain the same angular diameter distance as in the $w=-1$ case \cite{Ade:2015xua}.  The details are in the App.~A.1 of~\cite{Gleyzes:2015rua}.
In summary, the parameters I am going to constrain are:\footnote{In the fiducial models 
I and 
III 
$\gamma_c$ vanishes when varying along $\beta_\gamma$ (since $c_s\alpha^{1/2}=0$) and thus, since $\beta_\xi =0$ (see eqs.~\eqref{eqmatterb} and \eqref{eqmatterc}), $\beta_\gamma$  only appears quadratically in the perturbation equations. For the fiducial 
II, observables depend only mildly on $\gamma_c$. 
Thus, we chose $\beta_\gamma^2$ rather than $\beta_\gamma$ as the independent variable in the analysis.}
\be
\boldsymbol \theta \equiv 
\{\wDE \, , \, \alphaBz \, , \, \alphaMz \, , \, \alphaTz \, , \, \beta_\gamma^2 \} \;.
\ee
For the background, I take as fiducial evolution of the Hubble parameter the function
\beq
\label{H_fid}
\hat{H} (a) = H_0 \sqrt{ \Omega_{\rm m,0} a^{-3} +1-\Omega_{\rm m,0} } \;, \qquad ({\rm Fiducial})
\eeq
which corresponds to the $\Lambda$CDM evolution, i.e.~$\wDE = -1$ in eq.~\eqref{Hparam} and a quantity evaluated on the fiducial model is denoted by a hat. 
The fiducial value for two of the parameters is zero,
\beq
\label{parameters_fid}
\hat{\alpha}_{\text{M},0} = \hat{\alpha}_{\text{T},0} =0 \;, \qquad ({\rm Fiducial})
\eeq
but I consider several options for the parameters $\beta_\gamma$ and $\alphaBz$:
In particular, I will  distinguish three fiducial models:
\begin{description}
\item[I) $\Lambda$CDM:]  $\hat{\alpha}_{\text{B},0}= \hat{\beta}_\gamma=0$,
\item[II) Braiding:] $\hat{\beta}_\gamma=0$, $\hat{\alpha}_{\text{B},0}= - 0.01$,
\item[III) Interacting:] $\hat{\alpha}_{\text{B},0}=0$, $\hat{\beta}_\gamma=-0.03$,
\end{description}
In Fig.~\ref{fig:back} and~\ref{fig:pert} I show the effects of the different operators on the background and on perturbations, for the three different fiducials. These are useful to understand the results of the Fisher analysis. For the background, in Fig.~\ref{fig:back} I show the relative difference between $\Omega_{b,c}$ and their respective fiducial values.  For the perturbations, I plot the quantities that are relevant for the three observables introduced in Sec.~\ref{sec:obs}: the effective growth rate $\fg$ (see Eqn.~\eqref{feff}), the matter transfer function, defined in Eqn.~\eqref{Tm}, the transfer function for $\Phi+\Psi$, Eqn.~\eqref{TPhiPsi}, and its derivative with respect to the redshift $z$.

The unmarginalized  errors on the parameters are summarized in Tab.~\ref{taberrors} while in Tab.~\ref{tabfiducial} I report, for each Fisher matrix, the eigenvector associated to the maximal eigenvalue (called here maximal eigenvector), which provides the direction maximally constrained in parameter space, i.e.~the one that minimizes the degeneracy between parameters. The two-dimensional  contours are presented in Fig.~\ref{fig:FM1},~\ref{fig:FM1} and~\ref{fig:FM3} for the three fiducials\footnote{For each observable, the Fisher matrix  including all the parameters is ill-conditioned and cannot be inverted. This means that the observables do not have the constraining power to resolve the degeneracies (see e.g. \cite{Vallisneri:2007ev}). Thus, when plotting the two-dimensional contours I do not marginalise over the other parameters but I fix them to their fiducial values. }. The shaded blue regions in the plots correspond to instability regions, where $c_s^2 \alpha < 0$\footnote{Here I conservatively exclude the instability region from the allowed parameter space. A more refined 
treatment would require multiplying the likelihood function by a theoretical prior that excludes the  forbidden region, which is impossible to achieve with a Fisher matrix analysis (our priors cannot be represented with an invertible matrix).   }. Let me now comment on the results for the three fiducials.

\subsection{ Fiducial I: $\Lambda$CDM}
\begin{figure}[t]
\includegraphics[scale=0.35]{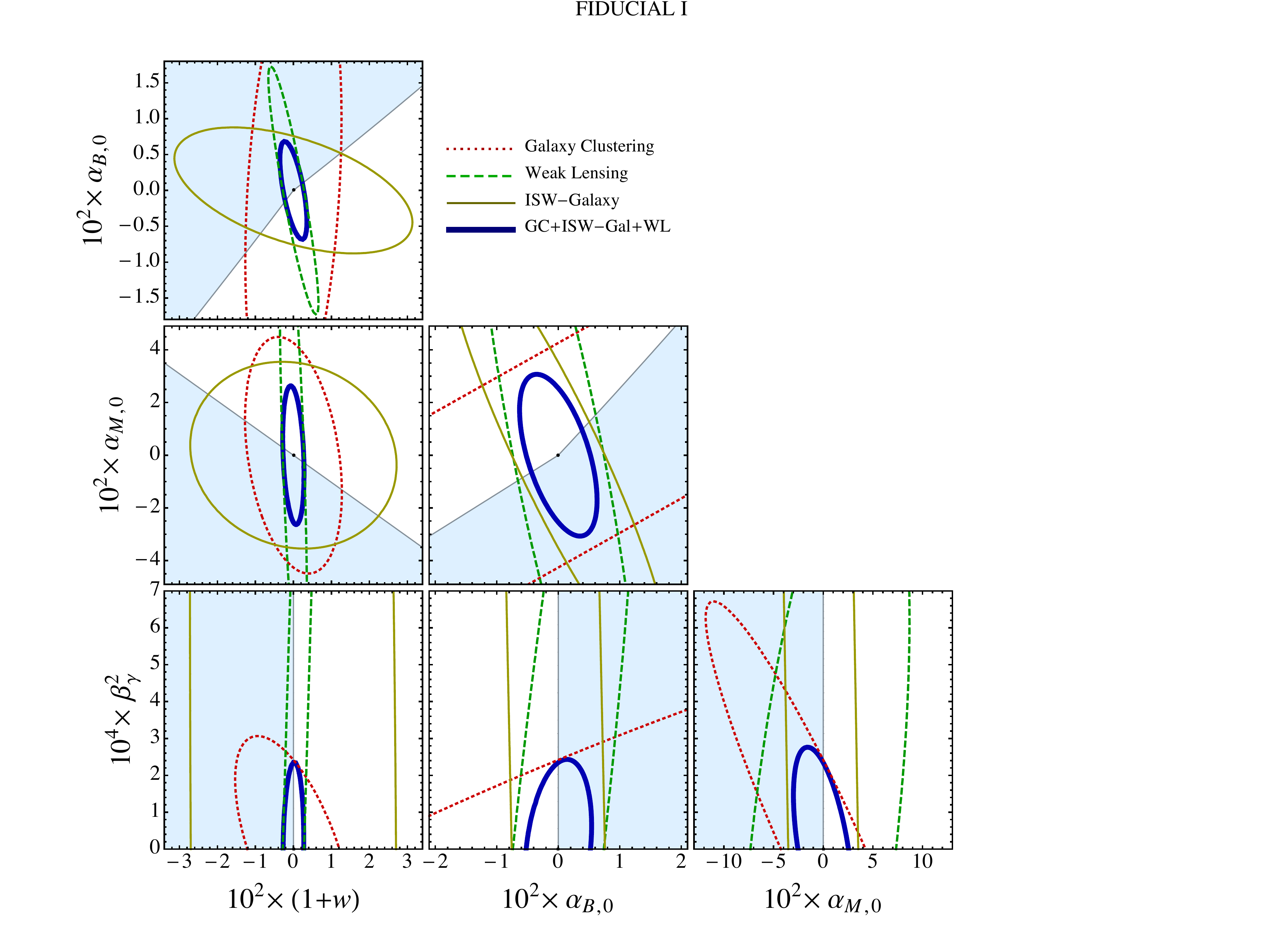}
\caption[Two-dimensional $68\%$ CL contours for interacting dark energy fiducial model I.]{Two-dimensional $68\%$ CL contours for the fiducial model I ($\Lambda$CDM model), obtained by fixing all the other parameters to their fiducial values. The parameter $\alphaTz$ is absent for $\Lambda$CDM, as it is unconstrained on this fiducial model. Shaded blue regions correspond to theoretically forbidden parameter space where $c_s^2 \alpha<0$. Note that the axis range is different for different parameter planes.}
\label{fig:FM1}
\end{figure}
This fiducial gives the usual  $\Lambda$CDM  for the perturbations. In this case the generalised Einstein equations and the modified continuity and Euler equations reduce to the standard ones. The two-dimensional contours are presented in Fig.~\ref{fig:FM1}. 
Let me comment on the effects of the different functions:
\begin{itemize}
\item The parameter $w$ mainly affects the background. In particular, it changes the function $H(z)$, thus also the evolution of the baryons and CDM energy densities, $\rho_b$ and $\rho_c$ (thus also $\Omega_b$ and $\Omega_c$). Since on this fiducial $\beta_{\gamma}$ vanishes, the changes in $\Omega_b$ and $\Omega_c$ are the same. This can be seen in the upper panels of Fig.~\ref{fig:back}.
\item The parameter $\alphaTz$ is unconstrained in this fiducial model, as when we vary it fixing all the other on the fiducial it disappears from the equations. Indeed, when $w=-1$ and $\beta_\gamma=0$, one finds that $\mu_{\Phi}=1+\alphaT+\beta_{\xi}^2$, $\mu_{\Psi}+\mu_{\Phi}=2+\alphaT+\beta_{\xi}^2$. Moreover, we have $c_s^2\alpha=-2\xi$. From the definition in Eqn.~\eqref{parameters} it follows that $\beta_{\xi}^2=-\alphaT$, and any dependence on $\alphaT$ disappears.
\item Switching on $\alphaB$ gives $\mu_{\Phi}=1+2\alphaB^2/(c_s^2 \alpha)$,  $\mu_{\Phi}+\mu_{\Psi}=2+4\alphaB^2/(c_s^2 \alpha)$, with $c_s^2 \alpha=-(2+3\Omega_{\rm m}) \alphaB-2\alphaB^2$. For small $\alphaB$, this can be approximated as
\be\label{f1_aB}
\mu_{\Phi}\simeq 1-\frac{2}{2+3\Omega_{\rm m}} \alphaB\,,\qquad \mu_{\Phi}+\mu_{\Psi}\simeq 2-\frac{4}{2+3\Omega_{\rm m}} \alphaB\,.
\ee
This shows that the effect of $\alphaB$ is larger when $\Omega_{\rm m}$ decreases, in agreement with the plots in Fig.~\ref{fig:pert}. 
\item The parameter $\alphaM$ affects both the background and the perturbations. At the background level, it does not change the evolution of $\rho_b$ and $\rho_c$, but it changes  
$\Omega_b$ and $\Omega_c$, according to equations~\eqref{Omegab}-\eqref{Omegac}. This can be seen in Fig.~\ref{fig:back}. As in the case of $w$, $\Omega_b$ and $\Omega_c$ are affected the same way since on the fiducial $\beta_{\gamma}=0$.
For perturbations, changing only $\alphaM$ from the fiducial gives $\mu_{\Phi}=1+\alphaM$ and $\mu_{\Phi}+\mu_{\Psi}=2+\alphaM$.
\end{itemize}
Note that the effect of $\alphaM$ is approximately equal in magnitude and opposite in sign to the one of $\alphaB$, Eqn.~\eqref{f1_aB}, as it can also be seen in Fig.~\ref{fig:pert}. In the $\alphaBz$--$\alphaMz$ panel of Fig.~\ref{fig:FM1} for galaxy clustering, one can see a degeneracy that is qualitatively explained by this result. A similar argument does not hold for weak lensing. In this case, a non vanishing $\alphaM$ also changes the background, and this effects has a non negligible impact on the transfer function $T_{\Phi+\Psi}$. 

Finally, a large region of the observationally constrained parameter space is forbidden by the stability requirements. This shows the importance of the analysis of stability conditions presented in Chapter~\ref{chap:EFToDE_lin}.
\subsection{ Fiducial II: Braiding}
\begin{figure}[h]
\includegraphics[scale=0.35]{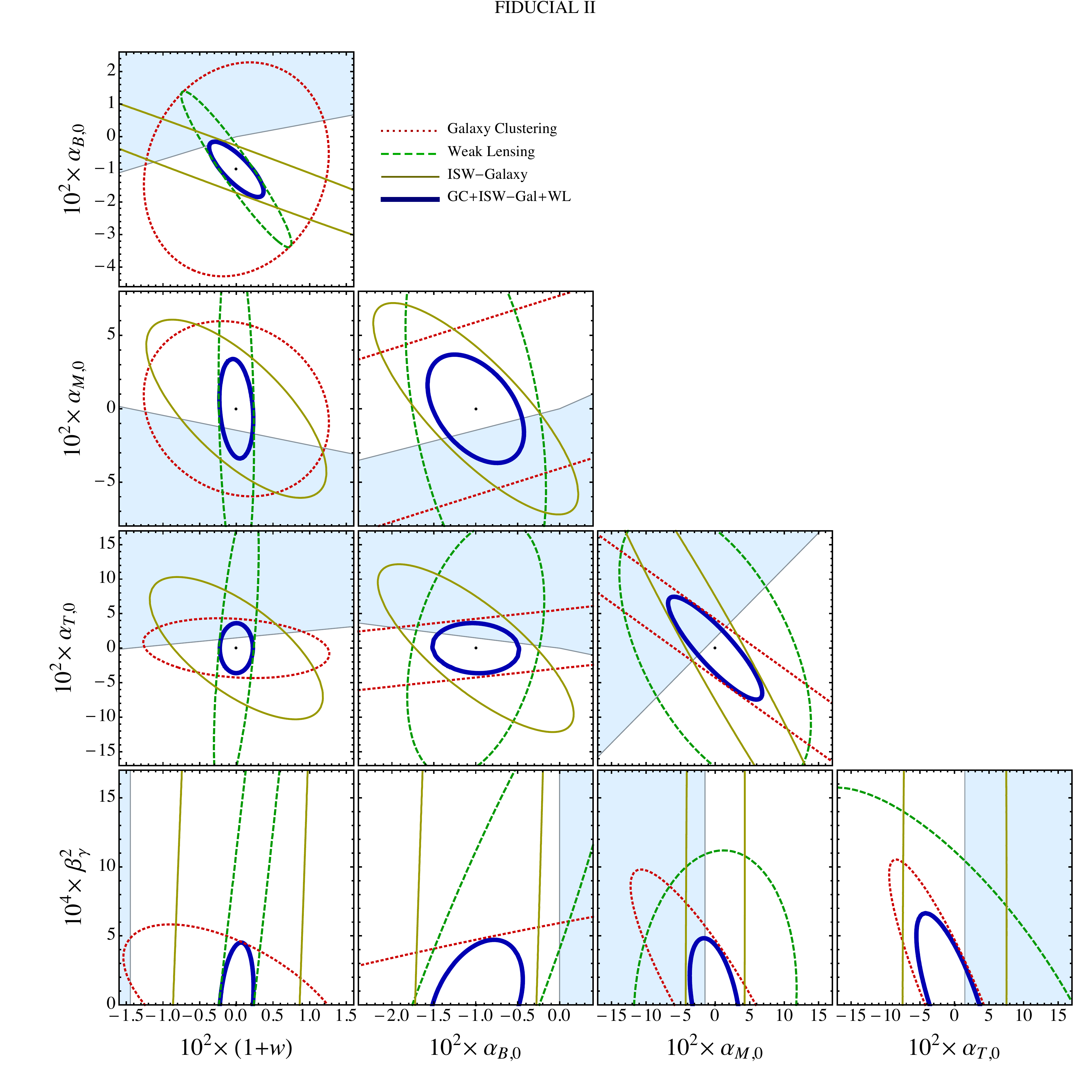}
\caption[Two-dimensional $68\%$ CL contours for interacting dark energy fiducial model II.]{Two-dimensional $68\%$ CL contours for the fiducial model II 
(braiding model with $\alphaBz=-0.01$), obtained by fixing all the other parameters to their fiducial values. Shaded blue regions correspond to theoretically forbidden parameter space where $c_s^2 \alpha<0$. Note that the axis range is different for different parameter planes.}
\label{fig:FM2}
\end{figure}
This fiducial corresponds to a mixing between the dark energy and gravity kinetic terms at the level of the perturbations. 
The two-dimensional contours are presented in Fig.~\ref{fig:FM2}.
The allowed parameter space is larger than in the previous fiducial because for $\alphaBz \neq 0$ the null energy condition can be violated without  instabilities \cite{Creminelli:2006xe}. Let me again comment on some effects and degeneracies.
\begin{itemize}
\item The effect of $w$ and $\alphaM$ on the background is the same as for $\Lambda$CDM.
\item $\alphaTz$ has how to be included in the analysis. In particular, the plane $\alphaBz$--$\alphaTz$ in Fig.~\ref{fig:FM3} has the same background evolution as $\Lambda$CDM.  Therefore, all the effects are controlled by $\mu_{\Phi}$ and $\mu_{\Phi}+\mu_{\Psi}$. This allows to explain some degeneracies analytically.  For small $\alphaBz$ and $\alphaTz$ one finds
\be
\mu_{\Phi}  \simeq 1+ \frac{3   \alpha_{\text{B},0} \left(\Omega _{\rm m}-1\right) \left(2 \alpha_{\text{B},0}+\left(2-3 \Omega _{\rm m}\right) \alpha_{\text{T},0}\right)}{\alpha_{\text{B},0} \left(6 \Omega _{\rm m}+4\right)+4 \alpha _{\text{T},0}}  \simeq  (1-\Omega_{\text{m}}) \left(0.54 \alpha_{\text{T},0}-0.6 \Delta \alpha_{\text{B},0}\right) \;, 
\ee
where in the last equality I expanded at linear order for small $1-\Omega_{\rm m}$ and used $\alphaBz = -0.01 + \Delta \alphaBz$. This explains the degeneracy between $\Delta \alphaBz$ and $\alphaTz$ observed in the growth. By the same procedure one finds $\mu_{\Phi}+\mu_{\Psi} \simeq (1-\Omega _{\text{m}}) \left(0.18 \alpha _{\text{T},0}-1.2 \Delta \alpha _{\text{B},0}\right) $, which explains why $\Delta \alpha _{\text{B},0}$ is more constrained than $ \alphaTz$ by lensing observations.
\item For the $\alphaBz$-$\alphaMz$ plane, the situation is similar to the one of $\Lambda$CDM. The two functions have effects opposite in sign and of the same magnitude. This explains the degeneracy in the growth of structures, while for weak lensing background effects are more relevant. 
\end{itemize}
 %
 \subsection{ Fiducial III: Interacting}
\begin{figure}[t]
\includegraphics[scale=0.35]{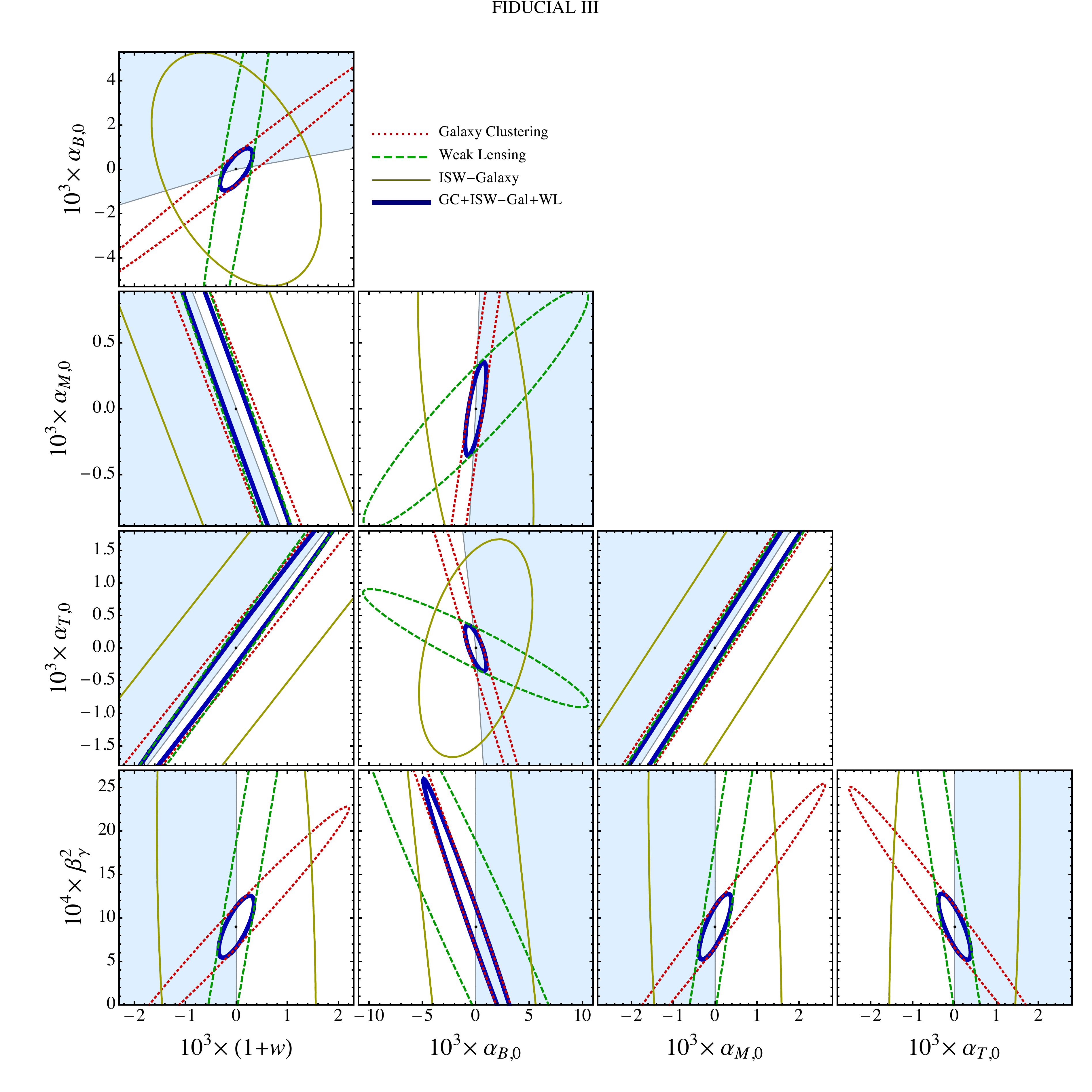}
\caption[Two-dimensional $68\%$ CL contours for interacting dark energy fiducial model III.]{Two-dimensional $68\%$ CL contours for the fiducial model III 
(interacting model with $\hat{\beta}_\gamma = -0.03$), obtained by fixing all the other parameters to their fiducial values. Shaded blue regions correspond to theoretically forbidden parameter space where $c_s^2 \alpha<0$. Note that the axis range is different for different parameter planes.}
\label{fig:FM3}
\end{figure}

In this fiducial, a non vanishing interaction between dark energy and CDM is present, which is active for perturbations but does not affect the background because $c_s \alpha^{1/2} =0$, and thus $\gammac = 0$. The two-dimensional contours are presented in Fig.~\ref{fig:FM3}.
The constraints for this fiducial model are generally stronger than those for models I and II. This is due to the enhancement of the effects on the observables, caused by the nonminimal coupling. In this case, the term $\beta_\xi\beta_\gamma$ in eqs.~\eqref{eqmatterb} and \eqref{eqmatterc} encodes the new effects that arise when both modifications of gravity and nonminimal couplings are considered. 
These effects explain the qualitative difference, in the size and shape, between the contours of fiducial 
III and those of the other two fiducial models. Not only are the constraints tighter by an order of magnitude in this case, but also the maximal eigenvectors of the Fisher matrices point in different directions, see Tab.~\ref{tabfiducial}. 

An analytical understanding of the degeneracies is complicated by the fact that the background evolution of the CDM density contrast $\Omega_c$ is changed by a change in  any of the parameters. For $w$ and $\alphaM$, the effect is amplified with respect to the other fiducials. Moreover, in this case, a non vanishing coupling $\gamma_c$ is present even when $\alphaT$ or $\alphaB$ are nonzero, since since both $\beta_{\gamma}\neq0$ and $c_s^2 \alpha \neq 0$.

One can try to partially understand analytically some noticeable degeneracies:
\begin{itemize}
\item When $\alphaTz$ and $\alphaMz$ are switched on, we have 
\be
\mu_{\Phi}=\alphaM - \sqrt{\alphaM- \alphaT} \, \beta_\gamma \omega_c b_c \, , \quad \mu_{\Phi}+\mu_{\Psi}=2+ \alphaM -  \sqrt{\alphaM- \alphaT} \, \beta_\gamma  b_c \omega_c \,.
\ee
Still, this does not completely explain the degeneracy since in this case the background is changed also for a non vanishing $\alphaT$.
\item Another strong degeneracy is present between $w$ and the parameters $-\alphaTz$ or $\alphaMz$. This can be partially understood  from the fact that $w$ appears in the combination  
\be
\label{wxi}
c_s^2 \alpha \simeq 3 (1+w) (1-\Omega_{\rm m})- 2 (\alphaM - \alphaT)  \simeq 3 (1-\Omega_{\rm m}) \left(1+w  - \alphaMz + \alphaTz \right)  \;,
\ee
where I used $\eta \simeq - w (1-\Omega_{\rm m})$ in  eq.~\eqref{csalpha} for the first equality and $\Omega_{\rm m,0} \simeq 1/3$ in the last one.
\end{itemize}
\renewcommand{\arraystretch}{1.4}
\begin{table}[t]
\small
\begin{center}
\begin{adjustbox}{max width=\textwidth}
\begin{tabular}{|c||c|c|c|c|c|c|c|cl}
  \hline
  Obs. & Fiducial I &  Fiducial II &  Fiducial III    \\
 \hline\hline
 GC &  $   (0.012 , - 0.007, 0.005, 0 ,1)
$  
& $     ( 0.022   , - 0.013, 0.007 ,  0.01, 1 )
$ 
& $  (-0.626 , 0.348  ,-0.629 ,0.64 , 1) 
$  \\
  \hline
 WL &  $   (-0.345 , - 0.115, - 0.007, 0 ,1)  
$  
& $    ( -0.463, - 0.136 , - 0.001 ,  0.004, 1) 
$ 
& $  (1, -0.074 , 0.910 , -0.914 ,-0.293 ) 
$  \\
  \hline 
  ISW-g &  $   (0.053 , 0.7, 0.154, 0 , 1) 
$  
& $    ( 0.856,  1,  0.117 , 0.063 , -0.609) 
$ 
& $    (-0.997 ,- 0.138 , -0.989 , 1 , -0.068)  
$  \\
  \hline
Comb. &  $   (-0.008 , - 0.012, 0.005, 0 , 1) 
$  
& $    (-0.055 , -  0.034 , 0.006 , 0.009 , 1)
$ 
& $    (1,- 0.285 , 0.953 ,  -0.964 , -0.867 )  
$  \\
  \hline
  \end{tabular}
\end{adjustbox}
\end{center}
\caption[First  eigenvector of the Fisher matrices in interacting dark energy models.]{First  eigenvector of the Fisher matrices, for the basis $\{ w, \alphaBz, \alphaMz, \alphaTz, \beta_\gamma^2\}$, with the maximum eigenvalue, corresponding to the combinations of parameters that are maximally constrained by experiments. The coefficients are normalized by the maximum component and rounded to three significant digits. }
\label{tabfiducial}
\end{table}
\renewcommand{\arraystretch}{1.4}
\begin{table}[h!]
\small
\begin{center}
\begin{adjustbox}{max width=\textwidth}
\begin{tabular}{|l|l||c|c|c|c|c|c|c|cl}
  \hline
Fid. & Obs. & $ 10^3 \times \sigma (1+w)$  & $ 10^3 \times \sigma( \alphaBz) $ & $ 10^3 \times \sigma( \alphaMz )$  &   $10^3 \times \sigma(\alphaTz)$  & $10^4 \times \sigma (\beta_\gamma^2)$ \\
 \hline \hline
 I &  GC &  $   7.0 $  & $ 18.6   $ & $   24.5 $  & --  & $1.4 $ \\
  \cline{2-7}
& WL &  $   1.6 $  & $ 4.3   $ & $   42.1 $  & --  &  $5.7  $ \\
  \cline{2-7} 
 & ISW-g &  $   15.5 $  & $ 4.4   $ & $   20.2 $  & --  & $31.3$ \\
  \cline{2-7}
&Comb &  $   1.6 $  & $ 3.0   $ & $   14.6 $  & --  & $1.35$ \\
  \hline
   II &  GC &  $   7.2 $  & $ 18.6   $ & $   33.8 $  & $ 24.4$  & $2.7$ \\
  \cline{2-7}
& WL &  $   1.4 $  & $  4.4  $ & $   67.4 $  & $ 98.9 $  & $ 6.4$ \\
  \cline{2-7} 
 & ISW-g &  $  5.0 $  & $ 4.2   $ & $  24.5 $  & $ 43.2$  &  $56.0$ \\
  \cline{2-7}
&Comb &  $   1.3 $  & $ 3.0   $ & $   19.0 $  & $ 20.8 $  & $2.5$  \\
\hline
   III &  GC &  $   0.22 $  & $  0.40  $ & $   0.22 $  & $ 0.22$  & $1.4$ \\
  \cline{2-7}
& WL &  $   0.17 $  & $ 2.12   $ & $   0.18 $  & $ 0.18$  &  $5.7$\\
  \cline{2-7} 
 & ISW-g &  $   0.88 $  & $ 2.78   $ & $   0.88 $  & $ 0.87$  & $31.3$   \\
  \cline{2-7}
&Comb &  $   0.13 $  & $ 0.39   $ & $   0.14 $  & $ 0.14$  & $1.4$ \\
  \hline
  \end{tabular}
\end{adjustbox}
\end{center}
\caption[$68\%$ confidence level (CL) errors in Fisher forcast for interacting dark energy.]{ $68\%$ confidence level (CL) errors on each individual parameter, assuming that the others take their fiducial values, for each fiducial model and observable: galaxy clustering (GC), weak lensing (WL), ISW-galaxy correlation (ISW-g) and the combination of the three (Comb). }
\label{taberrors}
\end{table}
\subsection{Comments}
For the current values of $\alphaB$, $\alphaM$ and $\alphaT$, the errors are of the order of  $ 10^{-2}$--$10^{-3}$  for fiducial models I and 
II and an order of magnitude better for the fiducial model III, while the error on  $\beta_\gamma^2$ is of the order of $10^{-4}$ for all fiducial models. For all the models, strong degeneracies are present. While some of them can be understood analytically, other result from a non-trivial combination of background and perturbations effects. In general, a combination of different probes such as the three considered can help substantially in breaking these degeneracies. One should recall also that the background cosmological parameters should be included in the analysis as nuisance parameters. In this case, it is important to take as well into account other cosmological data such as the CMB, the baryon acoustic oscillations and the supernovae Type Ia\footnote{An analysis similar to the one presented here has been extended to include these probes~\cite{Leung:2016xli}; in this case, the authors were also able to marginalise over the nuisance parameters. Where comparable, their results agree with those discussed here.}. One can hope to reduce degeneracies going beyond the quasi-static approximation, 
even if in this case, at least one more parameter, $\alphaK$, must be considered in the analysis.
For the case of Horndeski theories without a nonminimal coupling, the forecasts above have been recently extended~\cite{Alonso:2016suf} and the parameters of the effective description constrained~\cite{Bellini:2015xja}.




\chapter{Kinetic Matter Mixing}\label{chap:KMM}
\lhead{\emph{Kinetic Matter Mixing}} 

In this chapter, I analyse the second case introduced in Chapter~\ref{chap:EFToDE_m}, i.e. a gravitational sector described by a ``beyond-Horndeski'' theory and matter having a conformal coupling that depends on the scalar field only and a disformal one that depends on the gradient of the field as well, see Eqn.~\eqref{cdgenKMM}. 
I also assume that the WEP holds. In this case, only one function $C_{\rm m}(\phi)$ and one function $D_{\rm m}(\phi,X)$ are sufficient to characterise the matter coupling. At the linear level, according to the discussion in Chapter~\ref{chap:EFToDE_m}, we then have three functions of time $\alphaCm$, $\alphaDm$ and $\alphaXm$ characterising the matter sector for linear perturbations.

I also explained how the new physical effect present in these theories, Kinetic Matter Mixing, can be interpreted either as a modification of gravity (in the Jordan frame where $\alphaH\neq0$ and $\alphaXm=0$), or as a particular type of disformal coupling (in a frame where $\alphaH=0$ and $\alphaXm\neq0$). I work in the frame where all matter species are minimally coupled, hence I consider the former case where $\alphaXm=0$ and all the effects of KMM are encoded in the function $\alphaH$. I consider the late universe in presence of CDM, i.e. a non-relativistic fluid with vanishing pressure and speed of sound.
On the other hand, I generalise the study of the phenomenology in two ways with respect to Chapter~\ref{chap:Pheno}, i.e. showing the peculiar effects of the function $\alphaH$ associated to linear perturbations in this class of theories, and using results obtained with a Boltzmann code without resorting to the quasi-static approximation. 

\section{Analytical results}
In this subsection I derive some analytical results that are useful to interpret the numerical ones and the mixing of the propagating degrees of freedom. In particular, one can consider two regimes.
\begin{itemize}
\item \emph{Oscillatory regime}. 
On short scales, the gradients of the scalar field $\phi$ support an oscillatory regime, and in presence of KMM the oscillations are also shared by matter, even when it is made of nonrelativistic species with no pressure gradients.
To study the oscillations it is useful to consider the kinetic limit, i.e.~the limit where the spatial and time derivatives are larger than the expansion rate $H$. In this case, it is possible to find a redefinition of the metric perturbations  that  de-mixes  the new metric variables from the scalar field $\pi$ and removes the higher derivative term from the gravitational action~\cite{Gleyzes:2014qga} (see equation 3.5 of~\cite{DAmico:2016ntq}). In the kinetic limit, the dynamics of the relevant dynamical variables, i.e. the scalar field $\pi$ and the matter velocity potential $v_{\rm m}$  
is decoupled from that of the metric potentials in the new frame and we can study them separately.
Allowing for the moment for a non-vanishing speed of sound and pressure for matter, $c_{\rm m}^2$ and $p_{\rm m}$ respectively, the normalised fields
\be
\pi_{\rm c}\equiv \frac{H M \alpha^{1/2} }{1+\alphaH}\, \pi\, , \qquad v_{\rm c}\equiv \left(\frac{\rho_{\rm m} +p_{\rm m}}{c_{\rm m}^2 }\right)^{1/2} \, v_{\rm m} \; ,
\ee
have dynamics described by the Lagrangian
\be
\label{PiKinet2}
{\cal L} = \frac{1}{2}  \bigg\{  \bigg(1+ \frac{c_s^2}{c_{\rm m}^2} \lambda^2 \bigg) \dot \pi_{\rm c}^2 -  c_{s}^2  {(\nabla \pi_{\rm c})^2} + {\dot{v}_{\rm c}}^2- c_{\rm m}^2 {{(\nabla v_{\rm c})}^2}  + 2 \frac{c_s}{c_{\rm m}} \lambda  \; \dot{v}_{\rm c}\,\dot {\pi}_{\rm c}\bigg\}\,,
\ee
where $\lambda$, defined in Eqn.~\eqref{gpar}, is the frame-independent parameter quantifying KMM, and $c_s$ is the sound speed given in Eqn.~\eqref{SSBH}.
So, in presence of KMM, $\lambda\neq0$, it can be seen from the Lagrangian~\eqref{PiKinet2} that there is a kinetic coupling between $\pi_{\rm c}$ and $v_{\rm c}$.
One can find the normal modes of the system,
\be
\begin{pmatrix}
 c_{s}^3 \lambda/c_{\rm m}  &  c_{-}^2-c_{s}^2 \\
- c_s^3 \lambda /c_{\rm m} & c_s^2 - c_+^2
\end{pmatrix} \begin{pmatrix}
\pi_c \\
v_c
\end{pmatrix} \;,
\ee
where $c^2_\pm$ are the eigenvalues of the system, given by the two solutions of eq.~\eqref{km}. 
As I anticipated in the introduction to this Chapter, I am interested in studying the late universe in presence of a pressureless CDM component. So, I now take the $c_{\rm m}^2 =0$ limit. Going back to standard normalisation, the eigenmodes and respective eigenvalues of the system are
\begin{align}
\label{eigenm}
X_-= \ &v_{\rm m}+\pi \,\frac{\alphaH}{1+\alphaH}\; , \qquad c_-^2=  c_{\rm m}^2 =0  \;, \\
X_+ = \ &\pi - v_{\rm m} \, \lambda^2 \frac{1+\alphaH}{ \alphaH} \;, \qquad c_+^2=c_s^2 (1+\lambda^2)  \; , \label{Xp}
\end{align}
with $\lambda^2 =  3 \alphaH^2  \Omega_{\rm m} / ({ \alpha c_s^2}) $.
While $X_+$ displays oscillations with frequency $\omega = \pm i c_+ k$, the speed of the fluctuations of $X_-$ vanishes as that of matter. 
\item\emph{Quasi-static regime}. 
When including the Hubble expansion, we expect  the oscillations of $X_+$ to get damped~\cite{Sawicki:2015zya}.
In the absence of the oscillatory mode $X_+$, the time evolution is dominated by the Hubble friction and  time derivatives are of the order of the Hubble rate $H$. In this case, we can consider the short-scale limit $k \gg k_+$, where $k_+$ denotes the sound horizon scale of the oscillating mode,
\be
\label{ks}
k_+ \equiv \frac{aH}{c_+} =  \frac{a H}{c_s \sqrt{1+\lambda^2}} \;.
\ee
This is the quasi-static regime (in Appendix C of~\cite{DAmico:2016ntq} one can find a discussion of how this regime is reached in the cosmological evolution.).
\begin{figure}[t]
\centering
\includegraphics[width=0.8\textwidth]{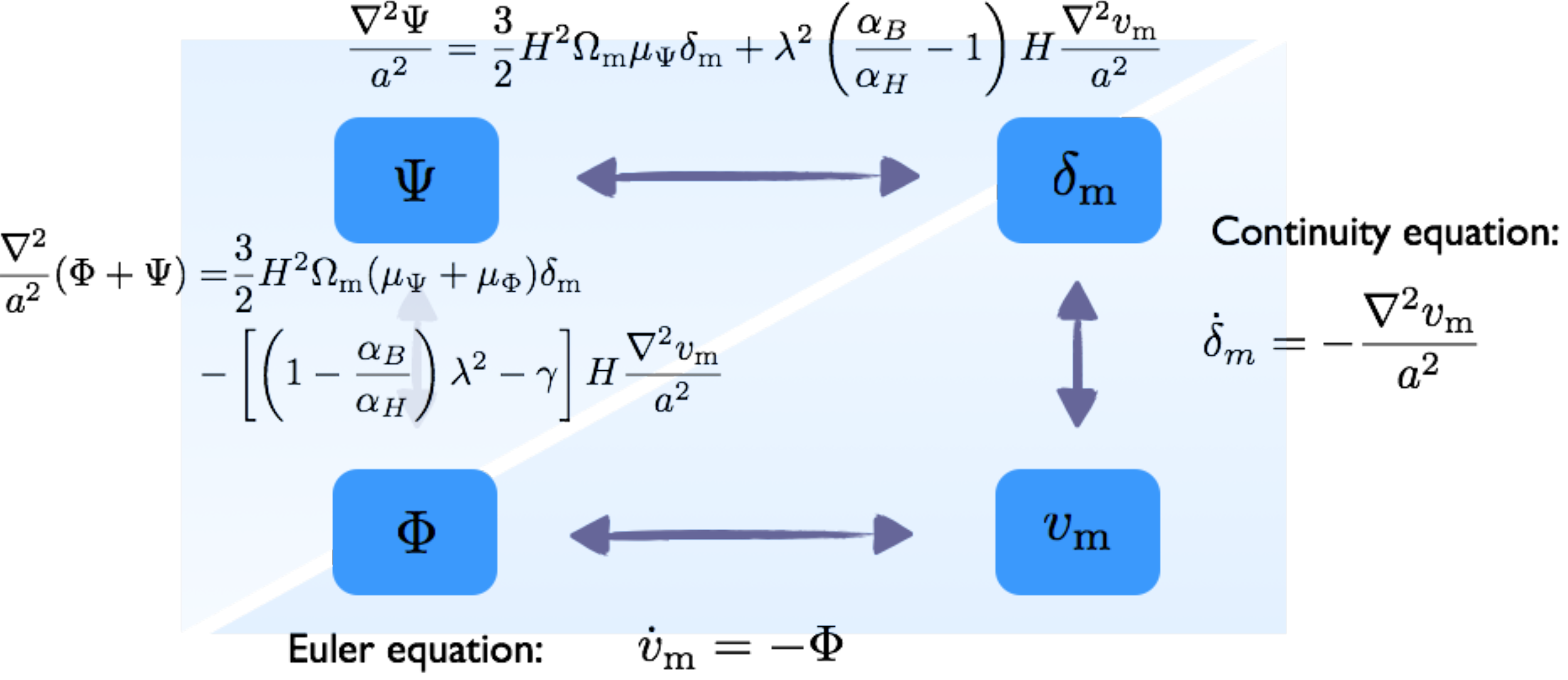}
\caption{Relation between the matter and gravitational perturbations in presence of KMM.}
\label{fig:potentialsKMM}
\end{figure}
Analogously to the case treated in the previous chapter, this time we have a system of only four variables $\Phi$, $\Psi$, $\delta_{\rm m}$, $v_{\rm m}$:
\begin{align}
\label{nablapsi}
  \frac{\nabla^2 \Psi}{a^2}
&= \frac{3}{2}  H^2 \Omega_{\rm m}    \mu_{\Psi} \delta_{\rm m} +  \lambda^2 \left( \frac{\alphaB }{\alphaH } - 1 \right)  H \frac{\nabla^2 v_{\rm m}}{a^2}    \;, \\
\label{nablaphi}
  \frac{\nabla^2 \Phi}{a^2}
&=  \frac32 H^2 \Omega_{\rm m} \mu_{\Phi} \delta_{\rm m} + \gamma  H \frac{\nabla^2 v_{\rm m}}{a^2 } \;,\\
\dot{\delta}_{\rm m} &= - \frac{\nabla^2 v_{\rm m}}{a^2} \label{conti} \;, \\
\dot {v}_{\rm m} &= - \Phi \label{Euler}\;.
\end{align}
The above equations are summarised in Figure~\ref{fig:potentialsKMM}. 
Being matter minimally coupled, the energy-momentum conservation equations take the standard form, while in the gravitational sector we see peculiar modifications in the equations that are characterised by the presence of the laplacian of the matter velocity potential $\nabla^2 v_{\rm m}$. The functions  $\mu_{\Psi}$ and $\mu_\Phi$ in~\eqref{nablapsi}-\eqref{nablaphi}, analogously to Eqs.~\eqref{muphi}-\eqref{mupsi}, are defined as
\begin{align}
\label{muphiKMM}
\mu_\Psi & \equiv \frac{1}{ 1+ \alphaH} \bigg[    1+ \frac{2(\alphaB - \alphaH)}{c_s^2 \alpha}   \left(\xi - \frac{\dot \alpha_{\rm H}}{H} \right)
  \bigg] \;, \\
\label{GeffKMM}
\mu_\Phi &\equiv  \frac{1}{(1+  \lambda^2 ) (1+ \alphaH)^2} \left\{ c_T^2 + \frac{2\xi}{c_s^2 \alpha} \left( \xi - \frac{ \dot \alpha_{\rm H} }{H} \right) +
a M^2 \alphaH (1+\alphaH)\right. \nonumber\\
&\left. \left[ \frac{2 }{a H M^2 c_s^2 \alpha } \left(\xi - \frac{ \dot \alpha_{\rm H} }{H} \right)  \right]^{\hbox{$\cdot$}}  \right\} \;,
\end{align}
and in Eqn.~\eqref{nablaphi} I introduced the parameter
\begin{align}
\label{gamma}
\gamma  \equiv  \frac{d \ln \left(1+ \lambda^2 \right)}{d \ln a}  \;.
\end{align}
Again, these equations can be combined in a single second-order differential equation for the density contrast:
\be
\label{deltaevol}
\ddot{\delta}_m + (2 + \gamma) H \dot{\delta}_m =  \frac32 H^2 \Omega_{\rm m} \mu_\Phi \delta_m   \;.
\ee

Finally, summing eqs.~\eqref{nablapsi} and \eqref{nablaphi}, one can obtain an equation for the Weyl potential, 
\be
\label{Weyl}
\begin{split}
 &  \frac{1}{a^2 H^2 } \nabla^2 ( \Phi+\Psi ) =  \frac32 \Omega_{\rm m}( \mu_\Psi  + \mu_\Phi   ) \delta_{\rm m} +  \left[  \left( 1 -\frac{\alphaB}{\alphaH} \right) \lambda^2 -{\gamma}   \right] \frac{\dot \delta_{\rm m}}{ H }  \;,
\end{split}
\ee
where I used the continuity equation to replace the velocity $v_{\rm m}$ by $\dot \delta_{\rm m}$.

Let me summarise the most important features of the equations in presence of KMM comparing them to the case studied in the previous chapter:
\begin{itemize}
\item In absence of KMM, we recover the result of Eqn.~\eqref{mupsi}, $\mu_{\Phi}=1+ \alphaT + \beta_\xi^2$. As already noticed in Sec.~\ref{pertQSWEP}, this means that the exchange of the fifth force tends to enhance gravity on small scales~\citepubli{Gleyzes:2015pma,Gleyzes:2015rua}-\cite{Piazza:2013pua,Pogosian:2016pwr}. On the contrary, in the presence of KMM $\mu_\Phi - (1+\alphaT)$ can be negative, corresponding to a repulsive scalar fifth-force, thus weakening gravity.
\item The modifications of the Poisson equations for $\Psi$ and $\Phi$ are qualitatively different in presence of KMM, and they include contributions depending on the laplacian of the matter velocity. In particular, the last term on the right-hand side of the Poisson equation for $\Phi$ gives extra friction $\gamma$ in Eqn.~\eqref{deltaevol}. 
\end{itemize}

\end{itemize}
\section{Observational effects}
As I showed in the previous section, KMM can lead to a repulsive scalar fifth force.
 This can leave peculiar signatures on structure formation with respect to the other effective theory operators. In this Section I will show these signatures on the matter power spectrum and on the CMB.
 
To go beyond the quasi-static limit, one should solve the full equations for linear perturbations. The minimal non-redundand set of equations is given by a second-order differential equation for one of the gravitational potentials, e.g. $\Psi$, another second-order differential equation for the scalar fluctuation $\pi$, and the equations for matter perturbations. In presence of $\alphaH$, the full equations can be found in Ref.~\cite{Gleyzes:2014rba}.\\
Moreover, in order to fully capture the properties of matter, CDM and baryons can be treated as collisionless and collisional fluids respectively. To treat properly photons and neutrinos, on the other hand, one has to resort to a phase-space description and solve the Boltzmann transport equations. The distribution functions are expanded in Legendre polynomials $P_{\ell}$, where $\ell$ is the multipole. The expansion up to order $\ell$ depends on terms of order $\ell+1$, so one obtains an infinite hierarchy of moment equations and can truncate the expansion at some given order $\ell_{\rm max}$ depending on the accuracy needed. The system of coupled differential equations can then be solved numerically to compute observables.\\
Two main Einstein-Boltzmann solvers exist for perturbations in $\Lambda$CDM. These are CAMB~\cite{Lewis:1999bs,Lewis:2002ah} and CLASS~\cite{Lesgourgues:2011re}.
Recently, the interest in testing alternatives to $\Lambda$CDM motivated to implement Boltzmann codes for scalar-tensor models. Noticeably, these are based on an effective description of dark energy. This is another point that shows the importance of this topic. \\
In particular, MGCAMB~\cite{Zhao:2008bn,Hojjati:2011ix} and EFTCAMB~\cite{Hu:2013twa,Raveri:2014cka} are two codes based on CAMB, while \emph{hi\_class} is an extension of CLASS~\cite{Bellini:2015xja,Zumalacarregui:2016pph}. The latter uses the parameters of the effective description described in this thesis (in particular, the equations are taken from Ref.~\cite{Bellini:2014fua}). These codes implement models within the Horndeski class. Finally, the code COOP~\cite{Huang:2015srv,zqhuang_2016_61166} (see also \url{http://www.cita.utoronto.ca/~zqhuang/} for documentation) includes also the beyond Horndeski function $\alphaH$ and uses the equations given in Ref.~\cite{Gleyzes:2014rba}.\\
In the rest of this Chapter, I will show results using the Boltzmann solver COOP. At the beginning of Sec.~4 of~\cite{DAmico:2016ntq} more details are given on the procedure used in the code to solve the equations. The detailed algorithm can be found in Ref.~\cite{Huang:2012mt}.

I assume the same parametrisation as in Chapter~\ref{chap:Pheno}, Eqn.~\eqref{TimeDepAlphas}, but focussing on the effects of KMM only, i.e. with $\alphaB= \alphaM=\alphaT=0 $, and
\be
\alphaK =  \alpha_{\rm K,0} \frac{1-\Omega_{\rm m} (t)}{1-\Omega_{\rm m,0}}\;, \qquad
\alphaH =  \alpha_{\rm H,0} \frac{1-\Omega_{\rm m} (t)}{1-\Omega_{\rm m,0}}\;.
\ee
The background expansion history is fixed to $\Lambda$CDM, Eqn.~\eqref{Hparam} with $w=-1$, which is the simplest assumption that one can make to focus on the effects of KMM at the level of linear perturbations. In the following I set the current value of $\alphaK$  to unity, $\alpha_{\rm K,0} =1$ and I plot the effect of $\alphaH$ in terms of  four different values of this parameter today, i.e.~$\alpha_{\rm H,0} = 0.06$, $0.12$, $0.24$ and $0.48$. These values have been chosen in order to comply with the stability conditions~\eqref{alpha_def_bH} and~\eqref{cpmgeq0}, that in this case read\footnote{To avoid that scalar fluctuations become superluminal in the past we must also require
\be
\alpha_{\rm H} \le \frac15 \alpha_{\rm K} \; . \label{superlum}
\ee
Just for the purpose of illustration, in the next two subsections I ignore constraints from superluminality, as using large values of $\alphaH$ allows to better visualise the effects on the observables.}: 
\be \label{stabWin}
\alphaK \ge 0\,, \qquad 0 \le \alphaH \le 1 +\frac{2}{3 \Omega_{\rm m}}\; .
\ee
\begin{figure}[t]
\includegraphics[width=0.504\textwidth]{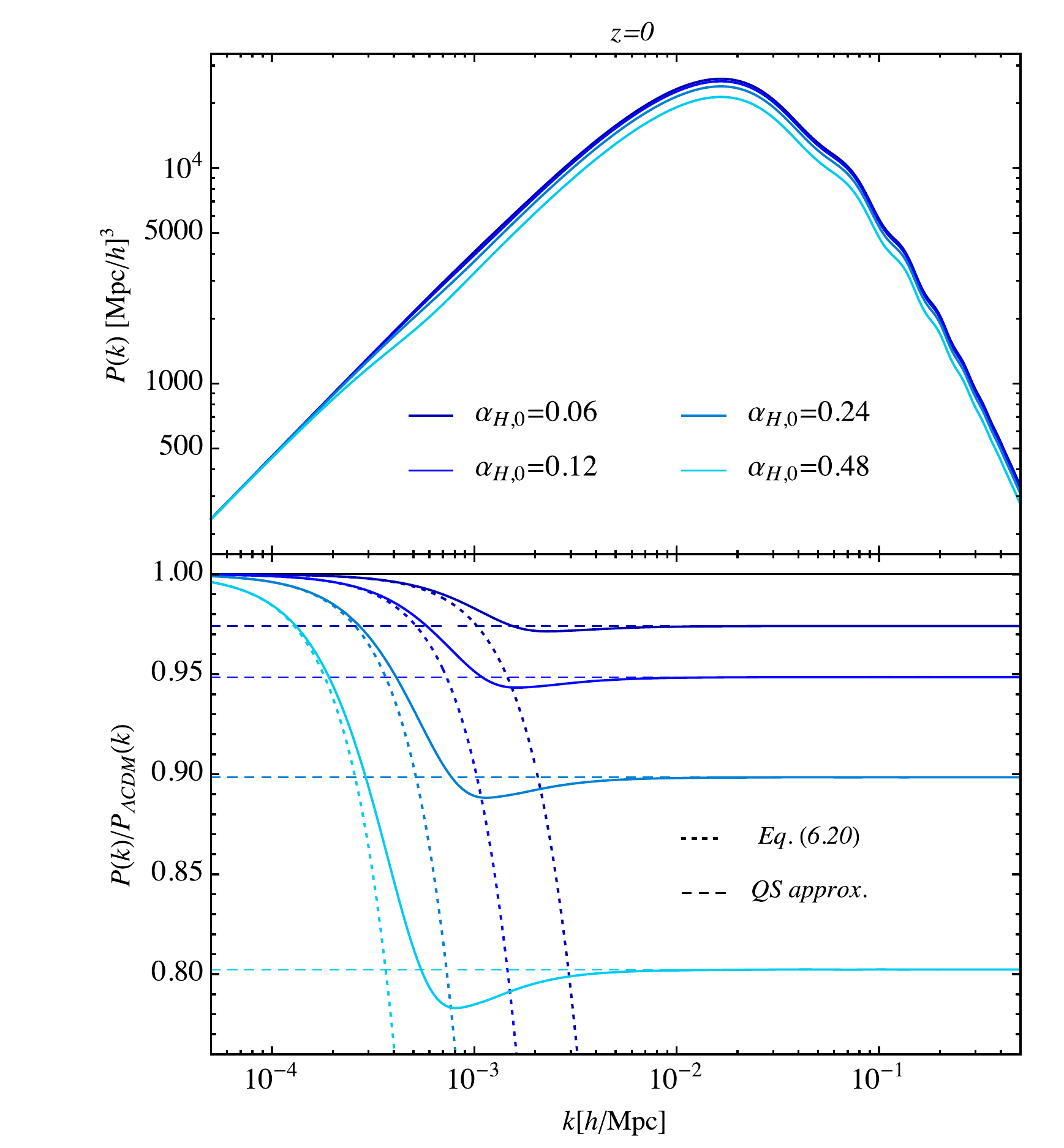}
\includegraphics[width=0.5\textwidth]{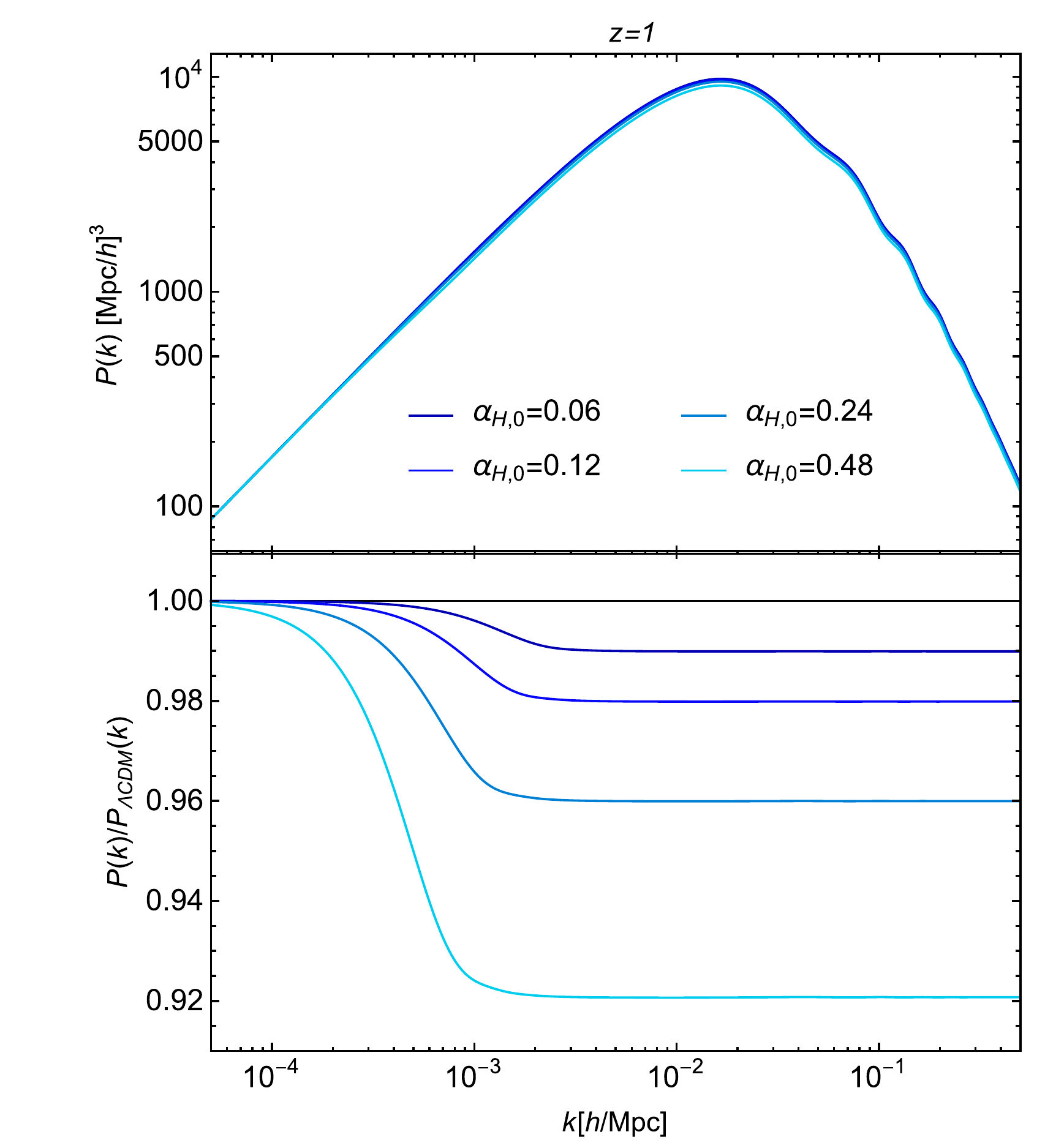}
\caption[Matter Power spectrum in presence of KMM]{Effect of KMM on the matter power spectrum for four different values of $\alphaH$ today, i.e.~$\alpha_{\rm H,0} = 0.06$, $0.12$, $0.24$ and $0.48$, at redshift $z=0$ (left panel) and $z=1$ (right panel). The lower plots display the ratio of these power spectra with the respective spectra for $\alphaH=0$. For comparison, the dashed and dotted lines in the left lower panel respectively show the quasi-static approximation and the perturbative solution~\eqref{Deltam2nd}. }
\label{fig:alphaH_PS}
\end{figure}
\subsection{Matter power spectrum}
\begin{figure}[t]
\centerline{\includegraphics[width=0.55\textwidth]{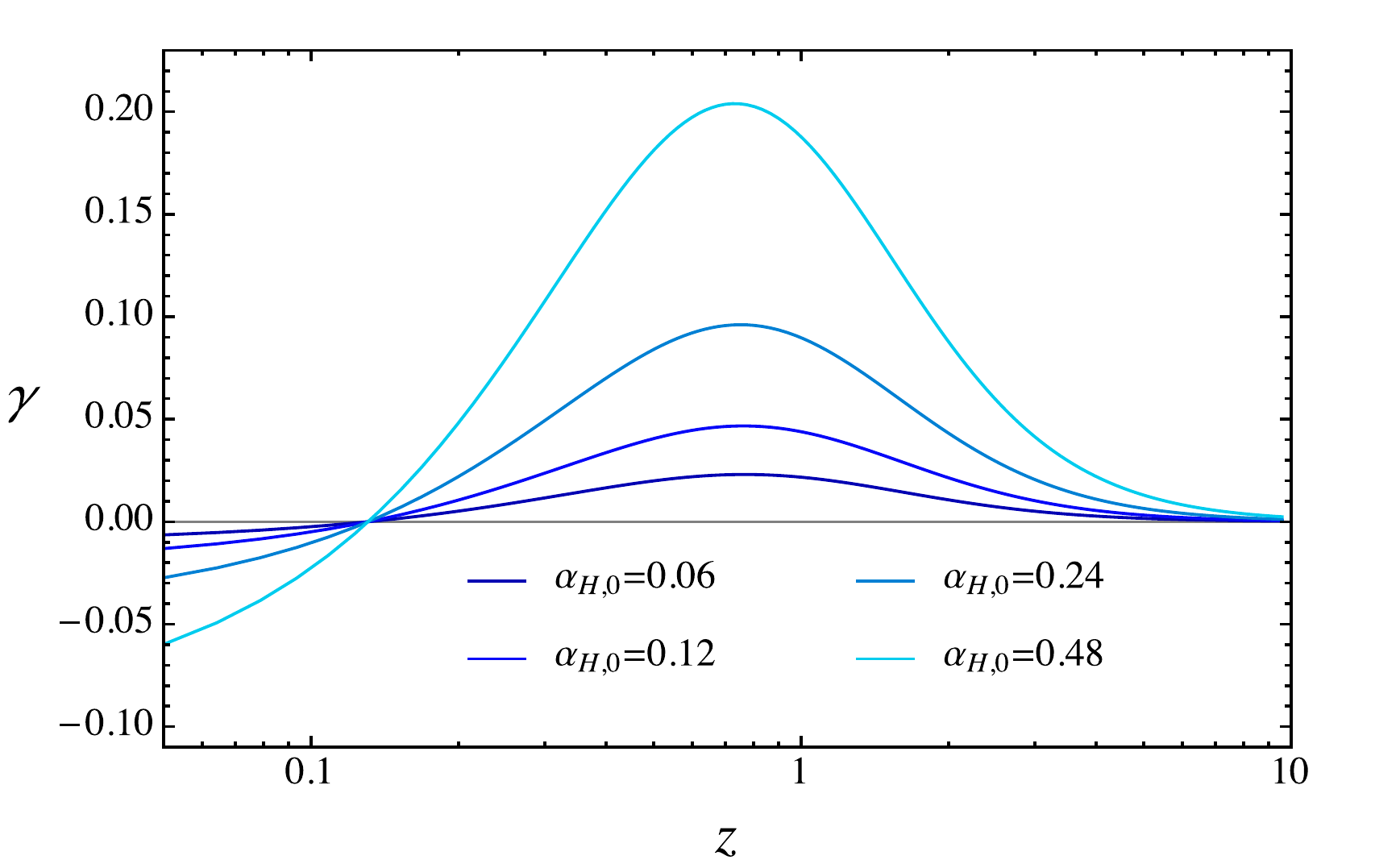}}
\caption{Friction term $\gamma$ given in Eqn.~\eqref{gamma}, as a function of redshift.}
\label{fig:gamma}
\end{figure}

The matter power spectrum is shown as a function of $k$ in Fig.~\ref{fig:alphaH_PS} for $z=0$ (left panel) and $z=1$ (right panel). From this plot we see that increasing $\alpha_{\rm H,0}$ suppresses the growth of structures. On small scales we can understand the power suppression applying the quasi-static approximation, i.e.~eq.~\eqref{deltaevol}. Two effects contribute to this result: the presence of $\gamma$, which is positive in matter domination and provides extra friction, and $\mu_{\Psi}$ which is smaller than unity, which means that the scalar force exchanged by $\pi$ in the presence of KMM is always repulsive.\\
Indeed, with only nonvanishing $\alphaK$ and $\alphaH$ and for the time parametrisation chosen, $\gamma$ and $\mu_{\Psi}$ are related by $\mu_{\Psi}=1-\gamma$. $\gamma$ as a function of redshift is plotted in Fig.~\ref{fig:gamma}. It starts positive and changes sign only recently. In particular, during matter domination (i.e. $\Omega_{\rm m}\approx 1$) it behaves as
\be \label{gammaMD}
\gamma = \frac95 \alphaH  +{\cal O} (1-\Omega_{\rm m})^2  \;.
\ee
Given that  $\mu_{\Psi}=1-\gamma$, this also means that $\mu_\Phi$ starts smaller than unity decreasing the strength of gravity, and gets larger than one only when $\gamma$ changes sign. This has the cumulative effect of suppressing the  power spectrum  with respect to the $\Lambda$CDM case. 

Note that the sign of $\alpha_{\rm H}$ is fixed by the stability condition, and so is the one of $\gamma$ according to Eqn.~\eqref{gammaMD}. This means that the weakening of gravity is a well defined prediction for stable theories under the assumptions made in this section.
Finally, a comment on the quasi-static approximation is in order. On the right panel of Fig.~\ref{fig:alphaH_PS}, I show the comparison between the quasi-static solution (dotted) and the full solution. On the scales where the former is valid, the agreement is excellent.

On the other hand, as expected, the quasi-static approximation fails on scales that become comparable to the sound horizon scale $k_+ $ defined in eq.~\eqref{ks}. Corrections  are expected to be of the order ${\cal O} ( k_+^2 / k^2 )$. Interestingly, one can still find an integral solution for the matter density perturbation that agrees with the numerical one. This is done by solving the Einstein and scalar field equations perturbatively in $\alphaH$ (while keeping the exact dependence on $\alphaK$ to avoid inconsistencies \cite{Iglesias:2007nv}). In particular, deviations from $\Lambda$CDM arise  at second-order in $\alphaH$, as the backreaction
effect
of $\pi$ on gravity. The detailed calculation can be found in Sec.~4.1 of~\cite{DAmico:2016ntq}.  At the end, one obtains a solution for the comoving matter density contrast $\Delta_{\rm m} \equiv \delta_{\rm m} - 3 H v_{\rm m} $:
\be
\Delta_{\rm m} =  \Delta_{{\rm m}}^{(0)} \left[ 1 - 2 a H^2  \frac{\alphaH^2}{\alphaK} \frac{k^2}{a^2}  \left( \int \frac{dt}{a H^3} - H \int \frac{dt }{aH^4} \right) \left(1 - \frac{H}{a} \int a dt  \right)^{-1}  +  {\cal O} (\alphaH^3) \right] \;. \label{Deltam2nd}
\ee
Notice that this solution breaks down on small scales because the quasi-static limit assumes $\alphaH \neq 0$.\\
On very large scales, i.e.~for
\be
\begin{split}
k \lesssim k_* & \equiv  \frac{\sqrt{\alphaK}}{ \sqrt{2} \, \alphaH}\frac{a}{H}   \left(1- \frac{H}{a} \int a dt \right)^{1/2 }  \left(  a \int \frac{dt}{a H^3} -   {Ha} \int \frac{dt }{aH^4} \right)^{-1/2} \\
& \simeq  \frac{\sqrt{\alpha_{\rm K,0}}}{\alpha_{\rm H,0}} \times 5.4 \times 10^{-4} h /\text{Mpc} \;,
\end{split}
\ee
the power spectrum is unmodified by KMM, although this restricts only to the case where the background expansion is that of $\Lambda$CDM. On intermediate scales, $k_* \lesssim k \lesssim k_+ $, the power spectrum drops as $k^2$ due to the second term on the right-hand side of eq.~\eqref{Deltam2nd}. The perturbative solution~\eqref{Deltam2nd} is shown in the left panel of Fig.~\ref{fig:alphaH_PS}.

\subsection{Cosmic Microwave Background}
\begin{figure}[t]
\centerline{\includegraphics[width=0.5\textwidth]{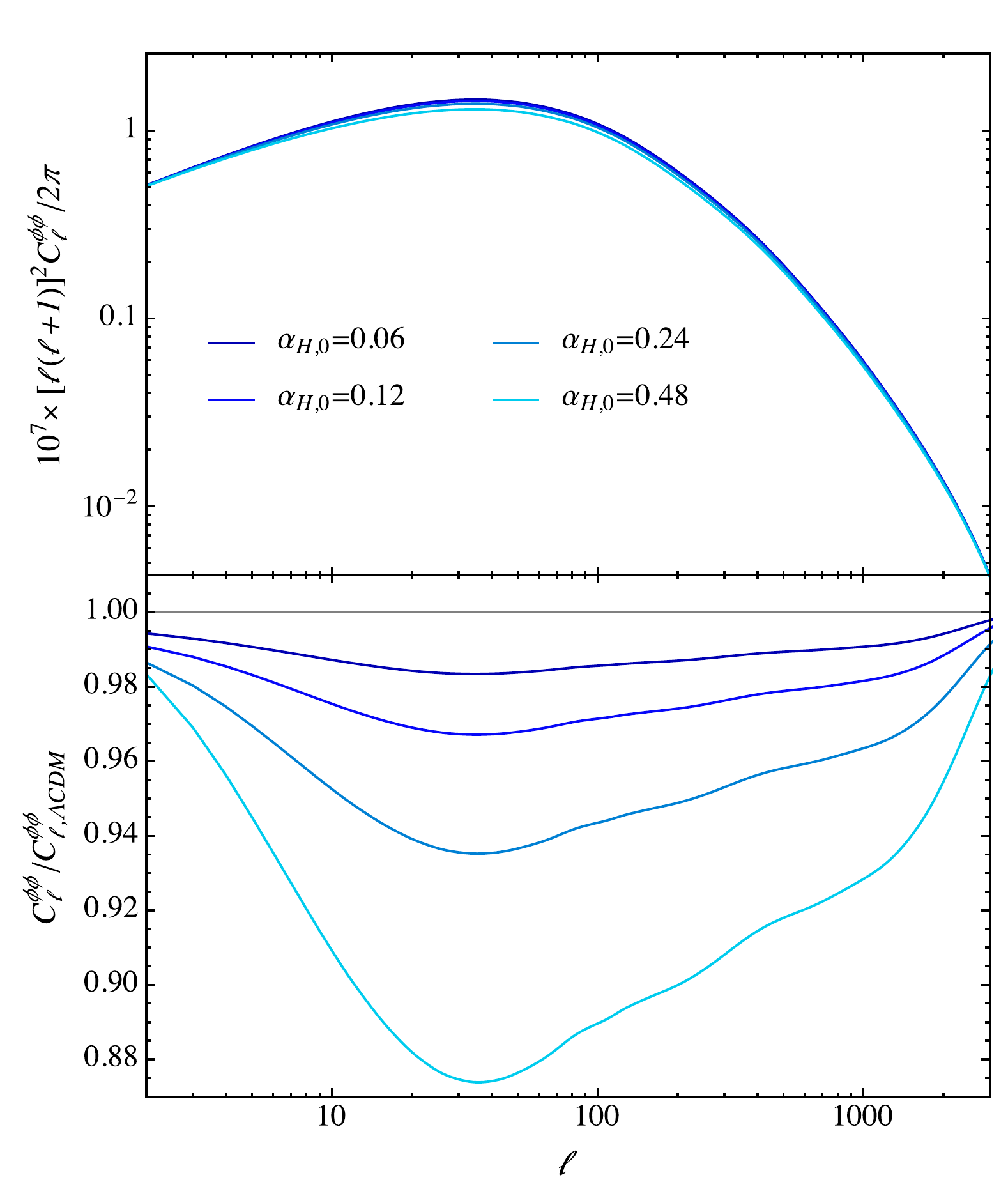}
\includegraphics[width=0.5\textwidth]{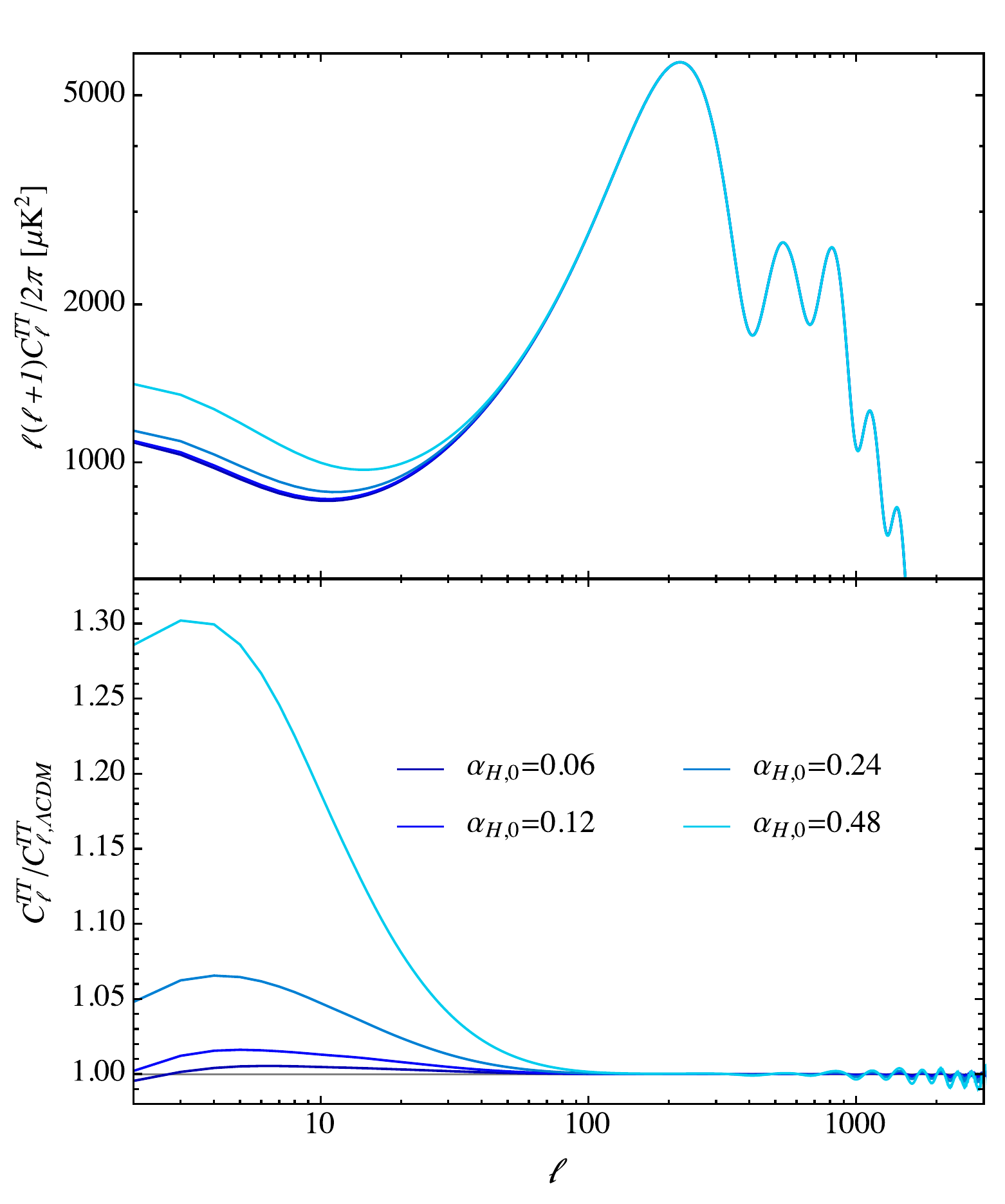}}
\caption[CMB anisotropies in presence of KMM]{Effect of KMM ($\alphaH$) on the CMB lensing potential  (left panel) and on the CMB anisotropies (right panel) angular power spectra. The lower plots display the ratio of these angular spectra with the respective spectra for $\alphaH=0$.}
\label{fig:alphaH_CMB}
\end{figure}

The effects on the CMB lensing potential are shown in the left panels of Fig.~\ref{fig:alphaH_CMB}, while those on the angular power spectrum of the CMB anisotropies in the right panels.

In presence of KMM, the CMB lensing potential is suppressed. 
In the previous Chapter, I showed that lensing effects are sensitive to the combination $\mu_{\Phi}+\mu_{\Psi}$. 
In particular, the CMB lensing potential is
defined as \cite{Lewis:2006fu}
\be
\phi (\hat n) = -\int_0^{z_*} \frac{d z}{H(z)}  \frac{\chi (z_*) - \chi (z)}{\chi (z_*)  \chi (z) } \big[ \Phi (\chi  \hat n, z) + \Psi (\chi  \hat n, z) \big] \;,
\ee
where $\chi \equiv \int_0^z dz/H(z)$ is the comoving distance and $z_*$ denotes the redshift of last scattering.\\
As I discussed in the previous Chapter, lensing effect are thus sensitive to the Weyl potential $\Phi+\Psi$. 
We can understand the effect in the quasi-static approximation. Indeed, the bulk of the CMB lensing kernel is at $ 0. 5\lesssim z  \lesssim 6$ \cite{Lewis:2006fu}, where deviations from  this approximation are below $\sim 5\%$ for the values of $\alpha_{\rm H,0}$ that I considered.\\
When KMM is present, we see from Eqn.~\eqref{Weyl} that the combination $\mu_{\Phi}+\mu_{\Psi}$ does not fully encode deviations from GR, because of the presence of the terms proportional to $\dot \delta_{\rm m}$ on the right-hand side of this equation. One can define the  quantity \citepubli{Gleyzes:2015pma}
\be
\label{lensingmu}
\mu_\text{\rm WL} \equiv  \frac{2 \nabla^2 ( \Phi+\Psi )}{3 a^2 H^2 \Omega_{\rm m} \delta_{\rm m}}  \;,
\ee
that can be used in general to characterise the deviations in weak lensing observables from the $\Lambda$CDM case. When KMM is absent,  $\mu_\text{\rm WL}=\mu_{\Phi}+\mu_{\Psi}$. In presence of KMM, this definition cannot be directly applied to eq.~\eqref{Weyl}, because of the presence of the terms proportional to $\dot \delta_{\rm m}$ on the right-hand side of this equation.
We can still simplify the discussion replacing 
 $\dot \delta_{\rm m}$ by its expression in matter domination, $\dot \delta_{\rm m} \simeq H   \delta_{\rm m}$.
Setting $\alphaB=\alphaM=\alphaT=0$
and employing the approximation above in eq.~\eqref{Weyl}, the effect of $\alphaH$ in weak lensing observables can be rewritten as
\be
\label{mulensingH}
\mu_\text{\rm WL}-    2  =   \alphaH \frac{8 - 9 \Omega_{\rm m} (1+ \Omega_{\rm m})}{2 + 3(1-\alphaH) \Omega_{\rm m}}  \;.
\ee
One can verify that this quantity is negative for $z \gtrsim 0.5$, i.e.~inside the bulk of the CMB lensing kernel. Therefore, the lensing potential is suppressed by the modification of gravity induced by $\alphaH$.
For small $\Omega_{\rm DE}$, in matter domination this suppression is roughly proportional to $\alphaH$.
%

As for the CMB anisotropies, at large $\ell$, we don't see any signature because they are generated at recombination,\footnote{Because of this, polarisation is also unaffected.} when $\alphaH$ vanishes. The only visible effect is an oscillating pattern observed at high $l$, due to the change in the CMB lensing. Indeed, lensing smears the CMB acoustic peaks; for larger values of $\alpha_{\rm H,0}$ the smearing  is suppressed and CMB peaks  enhanced.\\
At low $l$, the deviations from the $\Lambda$CDM case are dominated by the ISW effect, which is enhanced by KMM. 
For these multipoles, the deviations from the $\Lambda$CDM case are dominated by the ISW effect, introduced in Eqn.~\eqref{ISWterm}. Again, we can understand the effect by using the quasi static limit. Taking the derivative of eq.~\eqref{mulensingH} with respect to the $e$-foldings,  one obtains the following relation:
\be
\left. \frac{d \ln (\Phi+\Psi) }{ d \ln a} \right|_{\rm QS}  =  f_{\rm QS} -1 + \frac{d \ln \mu_{\rm WL}}{d \ln a} \;,
\label{ISW}
\ee
where
\be
f_{\rm QS} \equiv \left. \frac{ d \ln \delta_{\rm m} }{d \ln a} \right|_{\rm QS} \;,
\label{fQS}
\ee
is the growth rate computed using the quasi-static approximation.
In $\Lambda$CDM, $\mu_{\rm WL} = 2$ and the time variation of $\Phi+\Psi$ is given by the first two terms on the right-hand side, i.e.~the deviation of the matter growth rate from unity, which is negative. When gravity is modified, the last term on the right-hand side does not vanish. In the case of KMM, it  contributes with the same sign as the first term, enhancing the ISW effect. For example, assuming matter domination and expanding in $\alphaH$ one finds
\be
\frac{d \ln \mu_{\rm WL}}{d \ln a} =  - 3 \alphaH + {\cal O}(\Omega_{\rm DE}^2)\;,
\ee
which explains the enhancement in the ISW effect observed in the right panel of Fig.~\eqref{fig:alphaH_CMB}, roughly proportional to $\alphaH$.
\subsection{Growth rate of matter}
\begin{figure}[t]
\centering
\includegraphics[width=1.\textwidth]{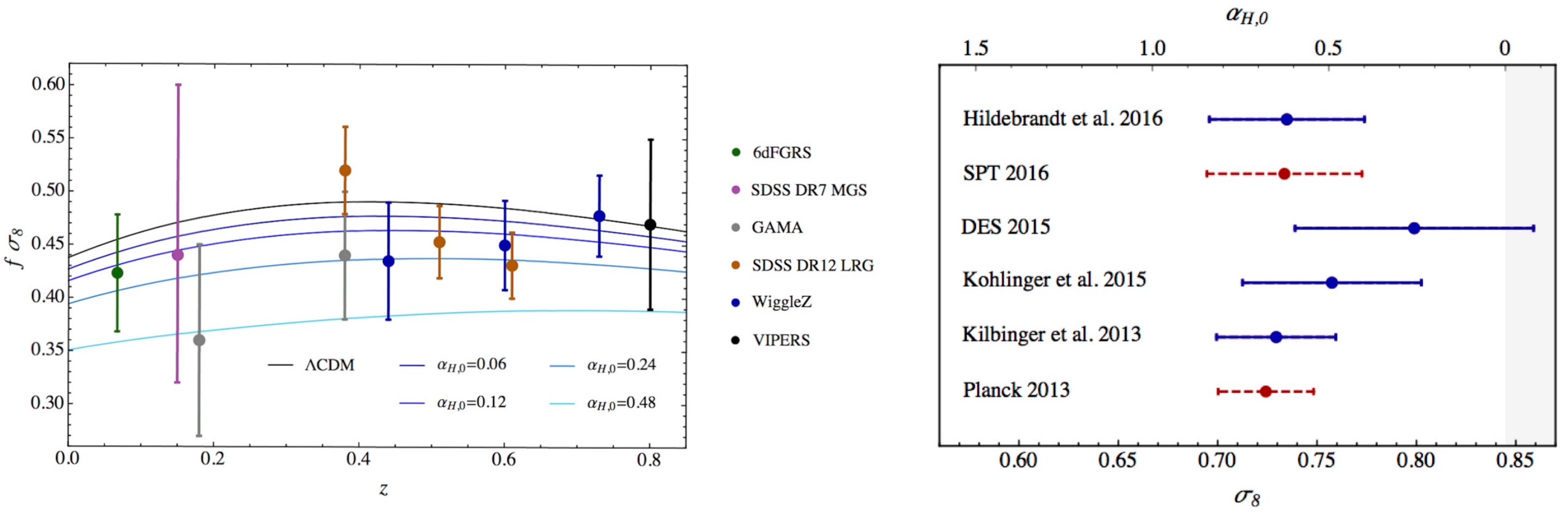}
\caption[$f \sigma_8$ and $f \sigma_8$ in presence of KMM, compared with measurements]{Left: the quantity $f \sigma_8$ as a function of redshift for different values of $\alpha_{\rm H,0}$.
The plot also shows the measurements of $f \sigma_8$ and their respective 1-$\sigma$ errors from several redshift surveys: 6dF GRS \cite{Beutler:2012px}, SDSS DR7 MGS  \cite{Howlett:2014opa}, GAMA\cite{Blake:2013nif}, SDSS DR12 LRG \cite{Alam:2016hwk}, WiggleZ \cite{Blake:2012pj} and VIPERS  \cite{delaTorre:2013rpa}.\protect\footnotemark~\\
Right: relation between $\alpha_{\rm H,0}$ and the corresponding $\sigma_8$ at redshift $z=0$, respectively in the top and bottom $x$-axes. The $\alpha_{\rm H,0} = 0$ line
corresponds
to $\Lambda$CDM and the region $\alpha_{\rm H,0} < 0$ is shaded because it is out of the stability window.
The plot also shows the measurements of $\sigma_8$ and their respective 1-$\sigma$ errors from several
collaborations.
In particular, the constraints based on cluster counts (red dashed lines) are from
Planck 2013~\cite{Ade:2013lmv} and SPT 2016 \cite{deHaan:2016qvy}.
The constraints based on weak lensing observations (blue solid lines) are from several analysis of the CFHTLens, by Kilbinger et al.~2013 \cite{Kilbinger:2012qz}, K\"ohlinger et al.~2015 \cite{Kohlinger:2015tza} and Hildebrandt et al.~2016 \cite{Hildebrandt:2016iqg}, and from the cosmic shear study of DES 2015 \cite{Abbott:2015swa}.}
\label{fig:fsigma8}
\end{figure}

In order to illustrate the effect of KMM on the growth rate, in the left panel of Fig.~\ref{fig:fsigma8} I plot the combination $f \sigma_8$ --- where $f \equiv d \ln \delta_{\rm m} / d \ln a$ is the growth factor and $\sigma_8$ is defined as the rms of the fractional density fluctuation in a sphere of $8h^{-1}$Mpc--- as a function of redshift for different values of $\alpha_{\rm H,0}$. In the right panel of the same figure, I plot $\sigma_8$  at redshift $z=0$ for different values of $\alpha_{\rm H,0}$ and show this relation together with a set of large scale structure (weak lensing and cluster counts) measurements. $\sigma_8$ scales linearly with $\alpha_{\rm H,0}$,
 \be
\sigma_8 \simeq  (0.84 - 0.18 \, \alpha_{\rm H,0} ) \cdot \frac{A_s}{2.2 \times 10^{-9}} \;,
\label{sigma8}
\ee
where  $A_s$ is the amplitude of scalar primordial fluctuations as measured by Planck. The figure shows that there is some tension between weak lensing and cluster counts measurements and the Planck best-fit $\Lambda$CDM model, which corresponds to the $\alpha_{\rm H,0}$=0 line, and it seems to suggest that  a value $\alpha_{\rm H,0} \sim\text{few} \times 0.1$ would provide the suppression needed to alleviate this tension.

\footnotetext{When possible, I plotted conditional constraints assuming a $\Lambda$CDM background cosmology with Planck 2015 parameters. In particular, the WiggleZ constraints were taken from Fig.~16 of~\cite{Ade:2015xua}.}

Let me make some remarks about the above results, in particular concerning the suppression of power in presence of KMM.
Recently, some tension has been found between the value of  $\sigma_8$ inferred from the CMB anisotropies~\cite{Ade:2013zuv,Ade:2015xua}, and the one measured with the large scale structures at low redshift (in weak lensing~\cite{Kilbinger:2012qz, Heymans:2013fya, Kitching:2014dtq,Kohlinger:2015tza,Hildebrandt:2016iqg} and cluster counts~\cite{Ade:2013lmv,Ade:2015fva,deHaan:2016qvy}). A similar tension is reflected in redshift space distortion measurements~\cite{Macaulay:2013swa} for the combination of $f \sigma_8$ (see the left panel of Fig.~\ref{fig:fsigma8}) which seem to be lower than the one predicted by the Planck best-fit model.
In light of these tensions, it is indeed interesting that the effect of KMM points in the direction of weakening gravity. Of course, one cannot claim those tensions to be highly significative at the current state of the art, and it must be kept in mind that the amount of tension can depend on aspects related to data analysis, such as  the modelling of non-linear scales and of the galaxy bias or other systematic effects.

As for the effects of KMM presented here, notice that the constraints on $\sigma_8$  reported from the respective articles have been extracted from data {\em assuming standard gravity}, hence we cannot infer from Fig.~\ref{fig:fsigma8} a ``best-fit value'' for $\alpha_{\rm H,0}$.
Moreover, as pointed out above, a large $\alpha_{\rm H,0}$ could lead to a too large ISW effect, even if this could be compensated by a variation in other parameters. The message is that one can draw conclusions only after a global fit to data. One should also keep in mind that possibilities different from modification of gravity of the scalar-tensor type can be put forward, see Sec. 4.3 of~\cite{DAmico:2016ntq} for a discussion.

For what concerns the class of theories considered in this thesis, there are anyway some interesting remarks to make. 
For theories within the Horndeski class with the same expansion history as $\Lambda$CDM, one can obtain a suppression of the growth rate around redshift $0.5 \lesssim z \lesssim1$ in self-accelerating models~\cite{Perenon:2015sla,Tsujikawa:2015mga}. This is due to the fact that $\Omega_{\rm m}$  on the right-hand side of Eqn.~\eqref{deltaevol}, contains the time-dependent effective Planck mass $M^2$ at the denominator. The enhancement of the latter due to self-acceleration lowers $\Omega_{\rm m}$ with respect to the standard $\Lambda$CDM case at intermediate redshifts. The scalar fifth-force on the other hand remains attractive, $\mu_{\Phi} >1$, but this effect can be subdominant with respect to the suppression due to self-acceleration.

On the contrary,
 the effect of KMM stands out as the unique leading to a repulsive force mediated by $\pi$: in this case, $\Omega_{\rm m}$ remains the standard one, but $\mu_{\Phi} <1$, which can be considered the distinct signature of KMM for this class of models.


\chapter*{Conclusions}
\addcontentsline{toc}{chapter}{Conclusions} 
\lhead{\emph{Conclusions}}  

This thesis contains results of my work on the so-called ``Effective Theory of Dark Energy''. As I explained, this approach allows to describe linear perturbations around a flat FLRW background in scalar-tensor theories of gravity. Deviations from the cosmological standard model, $\Lambda$CDM, are encoded in a few functions of time only. This approach is based on the symmetries of a homogeneous and isotropic FLRW universe and for this reason it is very general. One can map any model formulated in terms of a covariant Lagrangian to this description. However, the most useful feature of an effective approach is that it can be used independently of any fundamental theory to gain information about deviations from $\Lambda$CDM. I adopted this second strategy in the second part of the thesis, Chapters~\ref{chap:Pheno}-\ref{chap:KMM}, where I studied some phenomenological aspects resorting to a parametrisation of the free functions of the effective theory.

Adopting this strategy, in principle one could start directly from the effective theory for linear perturbations. 
Of course, a fully nonlinear formulation of modifications of GR is important for different reasons in phenomenology. The most important one is that it allows to describe physics also at scales where the linear approximation breaks down. In this work I did not study nonlinear aspects, but I showed that there is at least a second reason why having a fully nonlinear understanding of the theory can be very important even for the phenomenology at the linear level. This reason is related to degeneracy, introduced in Chapter~\ref{chap:DHOST}. 
The full degeneracy conditions $\CI$~\eqref{Ia}-$\CII$~\eqref{IIa} that one has to impose on DHOST Lagrangians at the linear level to get a healthy theory are obtained from a covariant, nonlinear analysis. If we restricted to linear perturbations in unitary gauge, we would conclude that the condition $\CU$~\eqref{C_U} is enough to avoid the presence of an additional degree of freedom.
This results into a complicated dispertion relation for the propagating mode, where $\omega(k)$ is a ratio of polynomials in $k^2$. Even if imposing that this dispertion relation takes its standard form $\omega=c_s k $ we recover the full degeneracy conditions, in principle one has no reason to do so. Thus, if we had to constrain the free functions basing on a linear analysis in unitary gauge, imposing only the condition $\CU$, we could in principle explore regions of the parameter space that are actually excluded by the full set of degeneracy conditions. 

So, to give a more complete understanding I chose to dedicate the first chapter to the nonlinear, covariant formulation of the most general class of scalar-tensor theories currently known, called DHOST or EST theories. First of all their study addresses a very interesting field theoretical question: is it possible, and under which conditions, to introduce higher-order derivatives in a Lagrangian without introducing also additional propagating modes? As I explained, the answer is not trivial and for long time having second order dynamics was considered a necessary condition to get a healthy theory. 
Besides this aspect, DHOST/EST theories can prove very interesting candidates to test against $\Lambda$CDM. This is the main reason why I studied them. Indeed, when they were discovered, we realised that the introduction of operators built with time derivatives of the lapse function in the Effective Theory of Dark Energy naturally describes all DHOST theories. This leads to extend the effective description with respect to its original formulation that covers Horndeski models, and its earlier extension to ``beyond Horndeski'' theories.

In Chapter~\ref{chap:EFToDE} I gave an introduction to the Effective Theory of Dark Energy. I chose to include directly the results of my work rather than proceeding in chronological order. In particular, the effective description was originally developed for Horndeski theories~\eqref{Hornd} with minimally and universally coupled matter fields. In this case, four functions of time ($\alphaM$, $\alphaT$, $\alphaB$, $\alphaK$) are enough to describe linear perturbations. One additional function $\alphaH$ has to be introduced for the theories ``beyond Horndeski''\eqref{L4bH}-\eqref{L5bH}. The study of DHOST theories is part of my contribution. In this case, four additional functions have to be introduced. We called them $\aL$, $\bun$, $\bdeux$, $\btrois$. However, these are not independent but subject to three degeneracy conditions that leave only one of them free. All the functions of the effective description can be given a physical interpretation that I summarised in Sec.~\ref{sec:AlphaInterpr}. \\
A second aspect I studied in detail in my work is the coupling to matter. This was the object of Chapter~\ref{chap:EFToDE_m}. When a minimal coupling is adopted, our description of the physics in different frames can be very different.
One can be more general and couple matter to a metric which is conformally and disformally related to the gravitational one. This gives equivalent frames and the fact that the structure of the theory is preserved by the aforementioned transformations reduces the number of free functions. 
The effective description can be further generalised to include the possibility that different species couple differently to the gravitational sector. This allows to study violations of the Weak Equivalence Principle. 
In the most general case, the conformal/disformal coupling is characterised, at linear level, by four functions of time for each species, $\alphaCI$, $\alphaDI$, $\alphaXI$, $\alphaYI$, introduced in Eqn.~\eqref{defalphasM}.\\
The coupling to matter is also relevant for the viability of the theory and for the possible mixing between matter and the scalar perturbations. 
In Horndeski theories, the two are decoupled and propagate with their respective speeds of sound, as I discussed in Sec.~\ref{sec:HorndDOF}. In theories beyond Horndeski, the propagating modes are on the contrary mixed states of matter and the scalar, see Sec.~\ref{KMMsec}. I showed that the mixing can be quantified in a frame-independend way. For theories beyond Horndeski, it can be seen either as a modification of gravity due to the operator $\alphaH$ or as an X-dependent disformal coupling to matter quantified by the function $\alphaXm$. Similar considerations can be made for DHOST theories satisfying the degeneracy conditions $\CI$.\\
Theories satisfying the degeneracy conditions $\CII$ can be instead ruled out from an analysis of linear stability in both the tensorial and scalar sector. In this case, I showed in Sec.~\ref{sec:CosmonoMatterDHOST} that a gradient instability necessarily arises in one of the two sectors, which makes these theories phenomenologically unviable. Moreover, these theories fail in recovering a Poisson equation on a Minkowski background at linear level, as I showed in Sec.~\ref{sec:DHOSTmatter}. \\
Let me point out here that the above results are an example of the ``effectiveness'' of the effective description. Among DHOST theories, there exist seven classes of purely quadratic theories, nine of purely cubic, and 25 combinations of quadratic and cubic. These all reduce to just two classes at linear level, among which one could be ruled out by stability, and the other is equivalent to Horndeski+beyond Horndeski theories with matter conformally and disformally coupled. These results thus remarkably reduce the class of allowed theories and simplify the study of their phenomenology.

Phenomenological aspects were the subject of the second part of the thesis. I believe this is the side where most progress has been made by the community since the beginning of my PhD, and where many questions are still open. Given the expectations that we have to get constraints on cosmological perturbations from next generation surveys, a general and natural question to ask is what their constraining power will be on the effective description. First, this requires to solve the evolution equations. To this extent, an intense activity led to the development of three Boltzmann codes that use the effective description introduced in this work~\cite{Hu:2013twa,Raveri:2014cka,Bellini:2015xja,Zumalacarregui:2016pph,Huang:2015srv,zqhuang_2016_61166}. Their exploitation is only at the beginning and even some cases treated in this thesis are not yet included. One example is interacting dark energy treated in Chapter~\ref{chap:Pheno}. The forecasts presented there were obtained resorting to the quasi-static limit, where a full Boltzmann code is not needed to solve the equations. Even this way, we were able to get interesting results. In general, we could be able to get constraints on deviations from $\Lambda$CDM at the $10^{-2}-10^{-3}$ level. The combination of different probes can constrain different combinations of the parameters and it is thus very important in order to achieve this precision. 
In Chapter~\ref{chap:KMM} I studied a case where the equations are solved employing the Boltzmann code COOP. This captures the effects of modifications of gravity at all linear scales. It also allows to verify numerically the consistency of the quasi static approximation. Besides this, this code is the only one publicly available that includes effects of the operator $\alphaH$ characterising theories beyond Horndeski. In its presence, differently from the Horndeski case, the extra force mediated by the scalar field can be repulsive when stability conditions are imposed. It is also interesting that this effect goes in the direction of alleviating the tensions between different measurements of the amplitude of fluctuations, $\sigma_8$.

The message is that the Effective Theory of Dark Energy presented in this thesis is a very useful and flexible tool to constrain deviations from $\Lambda$CDM for different reasons, and it opens different directions to follow. Not only it is very general and covers the most studied scalar-tensor theories, from the oldest ones to their most recent generalisations. Different couplings with matter can also be included in the description. It also provides a general insight on the phenomenological aspects of entire classes of theories, capturing their common features at the level of linear perturbations. I already recalled the dramatic reduction that happens in DHOST theories. Another example is the result that all Horndeski theories give an attractive fifth force in contrast to their extensions beyond Horndeski. This last case is also an example of a potentially phenomenologically relevant aspect discovered via the effective description.
Finally, the equations can be implemented once and for all in numerical codes.\\
Let me spend a few more word on this point. As I underlined several times, the price to pay to have a model-independent effective description is that the free functions have to be given a time dependence in order to solve the equations. This means that we have to parametrise them if we don't want to commit to any specific model. In this work, I used a parametrisation where the free functions are proportional to the fractional energy density of dark energy $\Omega_{DE}(t)$. This derives from the assumption of associating the onset of deviations from $\Lambda$CDM at the level of perturbations with the beginning of the dark energy dominated phase on the background. Indeed, our initial goal was to get general indications on the possibility to constrain the free functions and to understand degeneracies, and this simple parametrisation allows to do that. On the other hand, one can wonder if this approach actually captures accurately enough the time evolution of the $\alpha$ and $\beta$ in \emph{all} the theories under consideration~\cite{Linder:2016wqw}. Put at the level of comparison with data, we should ask how much the constraints would change under a change in the time evolution of these functions. Some recent studies indicate that the impact can be non negligible~\cite{Alonso:2016suf}. This remains a very interesting direction to follow.

Let me conclude by saying that constraining deviations from $\Lambda$CDM with the Effective Theory of Dark Energy is a program to which the community has started to dedicate increasing attention. The Planck collaboration included in the analysis the parametrisation described in this thesis~\cite{Ade:2015rim}. Besides the results presented here, a considerable amount of work has been spent to investigate the constraining power of future surveys using different parameterisations~\cite{Perenon:2016blf,Perenon:2015sla,Leung:2016xli,Alonso:2016suf}.
Constraints in the case of Horndeski theories using complementary datasets were studied in~\cite{Bellini:2015xja} and extended to include neutrinos~\cite{Bellomo:2016xhl}. 
Another open direction is the impact of stability conditions. I showed that they can reduce substantially the parameter space allowed. Several works have started to address this issue at the level of comparison with data~\cite{Peirone:2017lgi,Salvatelli:2016mgy,Raveri:2017qvt}. 
The fact that complementary observations are needed to break degeneracies led also to look for other ways to constrain the parameters. The most interesting one is the tensor speed excess $\alphaT$. 
Cosmic rays observations put a very stringent ($10^{-15}$) lower bound on the propagation speed of gravitons \cite{Moore:2001bv}.
At lower energies, an order~$\sim 1\%$ constraint come from  binary pulsar orbital periods~\cite{Jimenez:2015bwa}, while the arrival timing of GW150914~\cite{Abbott:2016blz} between the two LIGO detectors set an upper bound, $c_T < 1.7  c$~\cite{Blas:2016qmn}. Ref.~\cite{Bettoni:2016mij} discusses current and future bounds. 

The above discussion indicates that several interesting directions are open. In particular, I believe that the impact of the parametrisation of the free functions on constraints is a question that should be answered. The available codes allow to study modifications of gravity numerically, and several operators introduced in this work can still be added to them, such as the beyond Horndeski function $\alphaH$ and the conformal and disformal couplings to matter.
With the advent of next generation surveys, our forecasts will be translated in actual constraints that will improve those we already have. This will enable us to test General Relativity on cosmological scales at unprecedented precision, and I believe that the Effective Theory of Dark Energy presented in this work provides a very promising tool to do so.




\newpage
\addcontentsline{toc}{chapter}{Publications related to this thesis}
\bibliographystylepubli{other/utphys1}  
\bibliographypubli{biblio_thesis}  

\addcontentsline{toc}{chapter}{Other publications not related to this thesis}
\bibliographystylepaper{other/utphys1}  
\bibliographypaper{biblio_thesis}  

\lhead{Bibliography}  

\renewcommand{\refname}{Bibliography}
\addcontentsline{toc}{chapter}{Bibliography}
\bibliographystyle{other/utphys2}
\bibliography{biblio_thesis}

\addtocontents{toc}{\vspace{2em}} 




\end{document}